**Article**

# Femtosecond Three-Dimensional Imaging of Single-Protein with Hard X-ray Laser

Paul Lourdu Xavier,[1,2,3*] Andrew J. Morgan,[1,6] Johan Bielecki,[2] Amit K. Samanta,[1] Simon Dold,[2] Chufeng Li,[1] Xinyue Gao,[15] Spencer K. Passmore,[22] Ruslan P. Kurta,[2] Tim B. Berberich,[2] Emiliano De Santis,[17] Wenhui Zhang,[1] Juncheng E,[2] Faisal H. M. Koua,[2] Romain Letrun,[2] Luca Gelisio,[2] Oleksii Turkot,[2] Egor Sobolev,[2] Lukas V. Haas,[1,4] Jannik Lübke,[1,4] Jingxuan He,[1,4] Madeline B. F. Memovich,[1] Kevin Janson,[1] Stefanie Lenzen,[1,4] Safi Rafie-Zinedine,[2] Mansi Butola,[1] Oleksandr Yefanov,[1] Fabian Trost,[2] Chan Kim,[2] Tomas Popelar,[2] Tokushi Sato,[2] Mohammad Vakilli,[2] Yevheniy Ovcharenko,[2] Rebecca Boll,[2] Sergey Usenko,[2] Thomas M. Baumann,[2] Alberto De Fanis,[2] Björn Senftleben,[2] Matteo Porro,[2,21] Andrea Castoldi,[23,24] Ibrahym Dourki,[2] Cyril Danilevski,[2] Ekaterina Round,[2] Christina Schmidt,[2] Huijong Han,[2] Roberto Alvarez,[7] Ervin S.H. Chia,[8] Cornelia Cazey,[25] Carolin Seuring,[25] Ruojie Sha,[9] Maia Azubel,[10] David A. Bushnell,[10] Sébastien Boutet,[11] Sergey Ovchinnikov,[12,13] Adrian P. Mancuso,[2] Michael Meyer,[2] Richard Bean,[2] Anton Barty,[1] Filipe R.N.C. Maia,[14,18] Richard A. Kirian,[7] Thomas D. Grant,[20] N. Duane Loh,[8] Carl Caleman,[1,16] Erik G. Marklund,[15] Nadia Zatsepin,[22] Andrew V. Martin,[19] Saša Bajt,[1,4] Roger D. Kornberg,[10] Jochen Küpper,[1,4,5] and Henry N. Chapman[1,4,5*]

[1]Center for Free-Electron Laser Science CFEL, Deutsches Elektronen Synchrotron DESY, Notkestrasse 85, 22607 Hamburg, Germany
[2]European XFEL, Holzkoppel 4, 22869 Schenefeld, Germany
[3] Max-Planck Institute for the Structure and Dynamics of Matter, Luruper Chausee, 22761 Hamburg, Germany
[4]The Hamburg Centre for Ultrafast Imaging, Universität Hamburg, Luruper Chaussee 149, 22761 Hamburg, Germany
[5]Department of Physics, Universität Hamburg, Luruper Chaussee 149, 22761, Hamburg, Germany
[6] University of Melbourne, Victoria 3010, Australia
[7]Arizona State University, Tempe, Arizona 85287, USA
[8]National University of Singapore, Singapore 117551, Singapore
[9]New York University, New York City, New York 10003, USA
[10]Stanford School of Medicine, Stanford University, CA 94305, USA
[11]Linac Coherent Light Source, SLAC National Accelerator Laboratory, Menlo Park, CA 94025, USA
[12]FAS, Harvard University, Cambridge, MA 02138, USA
[13]Massachusetts Institute of Technology, Cambridge, MA 02139, USA
[14] Laboratory of Molecular Biophysics, Department of Cell and Molecular Biology, Uppsala University, Husargatan 3 (Box 596), SE-75124, Uppsala, Sweden
[15]Department of Chemistry for Life Sciences, Uppsala University, Box 576, SE-75123, Uppsala, Sweden.
[16]Department of Physics and Astronomy, Uppsala University, Box 516, SE-751 05 Uppsala, Sweden
[17]Department of Physics, University of Rome Tor Vergata and INFN, Via della Ricerca Scientifica 1, 00133, Rome, Italy
[18]NERSC, Lawrence Berkeley National Laboratory, Berkeley, CA 94720, USA
[19]Department of Physics, School of Science, RMIT University, Melbourne VIC 3000, Australia
[20]Department of Structural Biology, Jacobs School of Medicine and Biomedical Sciences, University at Buffalo, Buffalo, NY USA
[21]Department of Molecular Sciences and Nanosystems, Ca' Foscari University of Venice, 30172 Venezia, Italy
[22]Optical Sciences Centre, Swinburne University of Technology, Hawthorn VIC 3122, Australia
[23]Dipartimento di Elettronica, Informazione e Bioingegneria (DEIB), Politecnico di Milano, Piazza Leonardo da Vinci 32, Milano, 20133, Italy
[24]INFN, Sezione di Milano, Via Celoria 16, Milano, 20133, Italy
[25]Center for Structural Systems Biology (CSSB), Notkestrasse 85, 22607 Hamburg, Germany

*lourdu.xavier.paulraj@desy.de, henry.chapman@desy.de

**Abstract**

**The extreme intensity of X-ray free-electron laser (XFEL) pulses has enabled imaging of radiation-sensitive samples, such as macromolecular microcrystals, beyond radiation damage limits. These sources have the potential to enable biomolecular single-particle imaging, similar to cryo-electron microscopy but without the need for cryo-fixation and with temporal resolution from femtoseconds to milliseconds. While this possibility was recognized before XFELs were built, the biological single-particle imaging work flow has previously only been demonstrated on large virus particles. Building on improvements in X-ray beam focusing, particle delivery, diffraction detection, and advanced analysis, here we demonstrate imaging of a single molecular complex, the giant-hemoglobin erythrocruorin (Ery) with X-ray laser pulses. Two-dimensional classes of diffraction patterns could be reconstructed to 15 Å resolution, and 3D images to ~20 to 30 Å, even though the 3D merged intensity in reciprocal space extended beyond 20 Å. The resolution discrepancy is likely due to heterogeneity caused by gas-phase compaction of the complexes. With increased throughput, this approach could be used to reveal in-situ structural details during mass spectrometry studies of biomolecules, while improvements in sample delivery may provide ultrafast snapshot imaging of biological single-particles in their native-state beyond the limitations of radiation damage.**

## Main

Imaging the conformational dynamics of proteins has been a long-sought goal in structural biology[1,2]. To achieve this goal, one is confronted with fundamental limitations. Short-wavelength radiation is required to form an atomically resolved image of a single macromolecule, but such an image requires an exposure strong enough for scattering events to occur from a high fraction of atoms of the molecule[3]. Furthermore, a three-dimensional representation requires the molecule to be viewed at many orientations according to the principles of diffraction tomography. Only electrons or x-rays provide high enough scattering cross sections and can be delivered in sufficient flux density to image an unstained single molecule[3], but either of these choices results in a transfer of energy that degrades the structure under investigation. This radiation damage can be overcome by methods that aggregate information from many weaker single-particle measurements that individually only give partial information. Cryo-electron microscopy (cryo-EM)[4,5] achieves this by immobilizing molecules in vitreous ice and forming noisy images of many such molecules in different random orientations that are combined using correlation approaches[4].

With their much lower scattering cross section, and thousand-fold increase of energy deposited per scattering event as compared with electrons, X-rays cannot scatter enough information to provide an orientable scattering pattern without destroying the molecule. Using an intense femtosecond-duration pulse from an X-ray free-electron laser (XFEL), it nevertheless becomes possible to record the diffraction pattern of the molecule before the destruction is manifest[6]. Images are not formed directly, as in cryo-EM, but are recovered by phase retrieval as in crystallography[7–9]. In this approach, dubbed "diffraction before destruction", molecules do not have to be cooled or immobilized since they are frozen in time by the femtosecond-duration exposure, and thus also can be measured in flight without the need of a supporting substrate[10]. The short exposure duration may also afford high time resolution for the study of conformational dynamics that can be triggered in time[11].

The ability to out-run radiation damage, proposed before the introduction of XFEL sources[12–14], has now been well demonstrated in serial femtosecond crystallography (SFX), where single-shot diffraction patterns of crystals of 100 nm to 10 μm size are measured one at a time with XFEL pulses without the need for cryogenic cooling[15–19]. Conformational changes of photoactive macromolecular systems have been obtained by triggering them with an ultrafast optical pulse prior to the X-ray exposure, over delay times of tens of picoseconds while molecules move in synchrony in the crystal. Short-lived intermediate states can also be observed at longer times if a significant population accumulates within the crystal. Single particle imaging (SPI) in the cryo-EM can sort noisy images from an ensemble of molecules into classes representing distinct conformational states[20]. With sufficient images, the free energies of these states could be determined from their relative populations[21,22], and millisecond to microsecond time resolution can be achieved by rapid cryogenic trapping of activated molecules[23]. Imaging of single macromolecules with XFEL pulses has the promise to combine ultrafast triggering with conformation selection, to be able to separate various reaction pathways. Given the bosonic nature of photons, it also has the potential to generate images with more signal per atom than with electrons[2], higher time resolution, and a higher throughput.

## Ultrafast XFEL-SPI and Hemeproteins

The first demonstrations of X-ray SPI were carried out when XFELs came online about 20 years ago[24]. With developments in aerosol particle delivery into vacuum, improvements in X-ray focusing and detection, and analysis techniques, encouraging results were achieved on 3D imaging of viruses[25–27] and inorganic nanoparticles[28,29] at modest resolution,  and recently detecting diffraction of single protein molecules with soft x-rays[30]. However single-protein 3D diffractive imaging, especially with hard x-rays (short wavelength), has remained elusive. The primary hurdles have been to reduce background scattering to enable the recording of diffraction patterns with the necessary signal to background ratios, and to record a high enough number of single-molecule diffraction patterns. Using the nanofocusing SPB/SFX (single particles, biomolecules, and clusters/serial femtosecond crystallography) hard x-ray instrument at the European XFEL[31] and an electrospray-aerosol injection assisted by helium gas[32], we obtained diffraction patterns of an isolated hemeprotein complex, erythrocruorin (Ery)[33]. Beginning with the first protein structures solved by Kendrew and Perutz using x-ray crystallography[34], hemeproteins have long provided insights in structural biology and, lately, for ultrafast biofunctional structural dynamic studies[35]. Heme-containing protein systems play crucial roles in gas and electron transport and enzyme catalysis in living systems. As a model system, Ery has been used extensively in the development of three-dimensional structure determination of proteins with electron microscopy[36,37] and has since been solved with x-ray crystallography and cryo-EM[38–40]. Its large mass (3.6 MDa), high symmetry, complex morphology, and photolabile moieties make it suitable for the development of ultrafast time-resolved single-particle XFEL imaging towards tracking large-scale and local conformational changes[41].

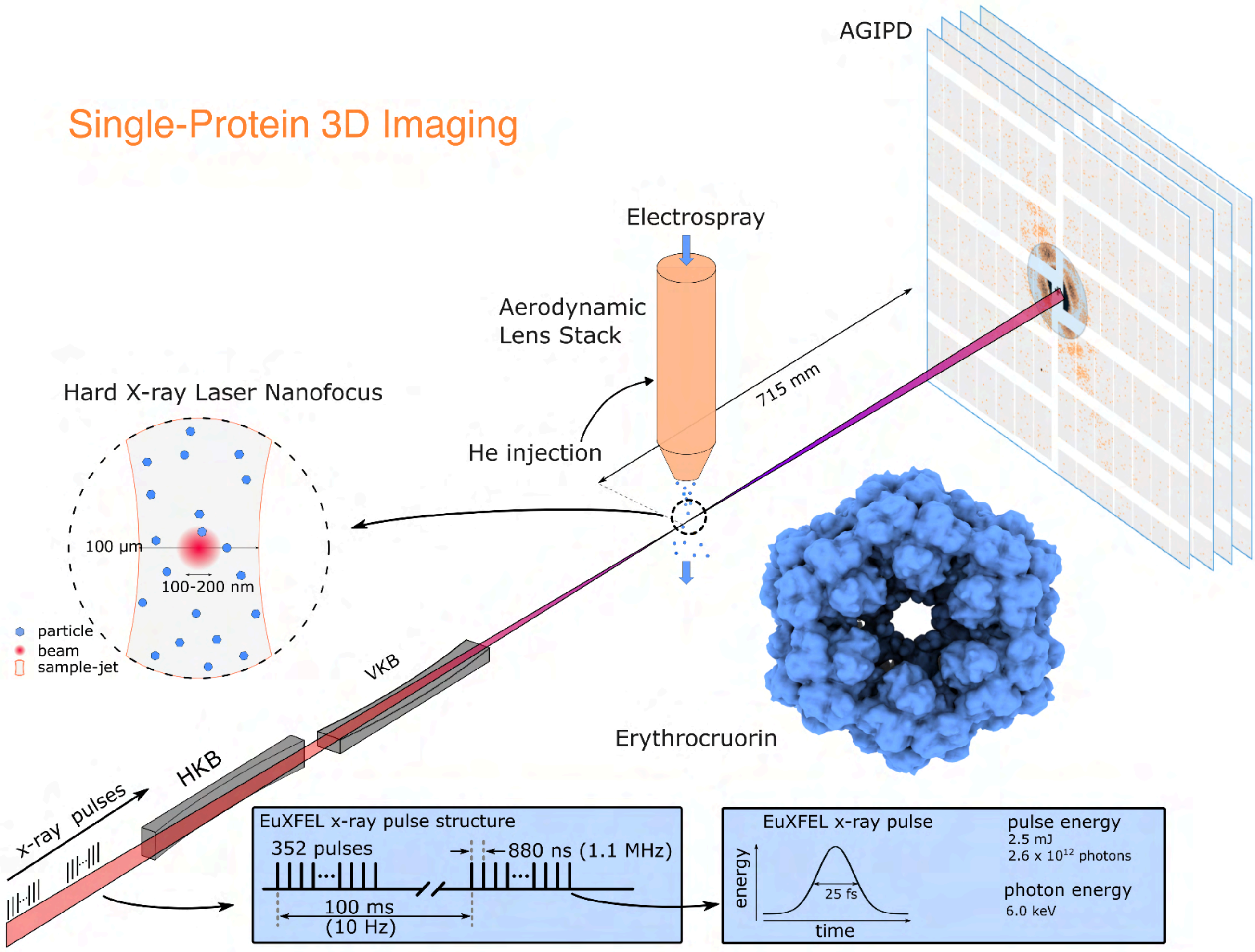

**Fig. 1 | Experimental setup for single-particle diffractive imaging.** Aerosolized single Erythrocruorin (Ery) protein complexes are focused with a gas mixture that provides a low scattering background to intersect XFEL pulses. These, at a photon energy of 6 keV, are focused by Kirkpatrick–Baez mirrors to a 180 × 220 $nm^2$ spot. Scattering of X-rays by a single protein produces a measurable diffraction pattern on the detector. The inset at bottom shows the timing structure of the XFEL pulses at the instrument.

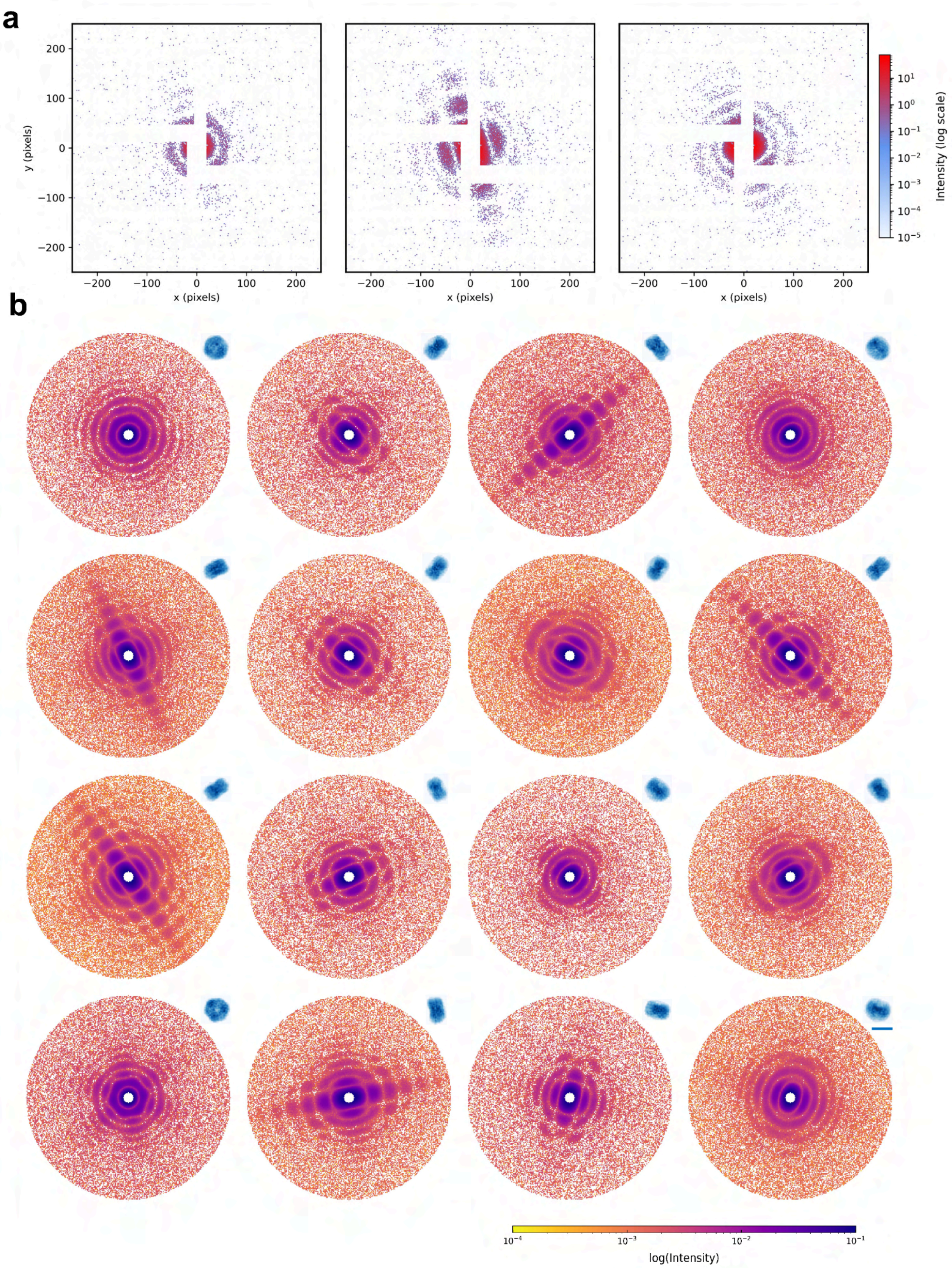

**Fig. 2 | Single-particle Erythrocruorin diffraction patterns.** (a) Example single-shot diffraction patterns of an isolated single Ery complex with nanofocus hard X-ray laser at 6 keV from almost 200,000 collected with XFEL pulses. The central speckle seen here indicates a size of ~170 to 270 Å matching the height and width of the double doughnut

structure of Ery. (b) Selected representative 2D diffraction EMC class averages of Ery out of 144 classes (see Fig. S36-38 for all the 144 class averages). Some classes show well defined diffraction features to the edge of the reconstruction domain at 19 Å  resolution (colorbar: intensity in logscale from 10 - $10^{-4}$), and some show hexagonal faces consistent with D6 symmetry, as expected for Ery. In blue at the top upper right of each class is displayed the corresponding 2D phased real space projection image of the protein. The scale bar (blue) below one such projected protein image (blue) in the last row's rightmost class is 20 nm and it applies to all the projection images displayed .

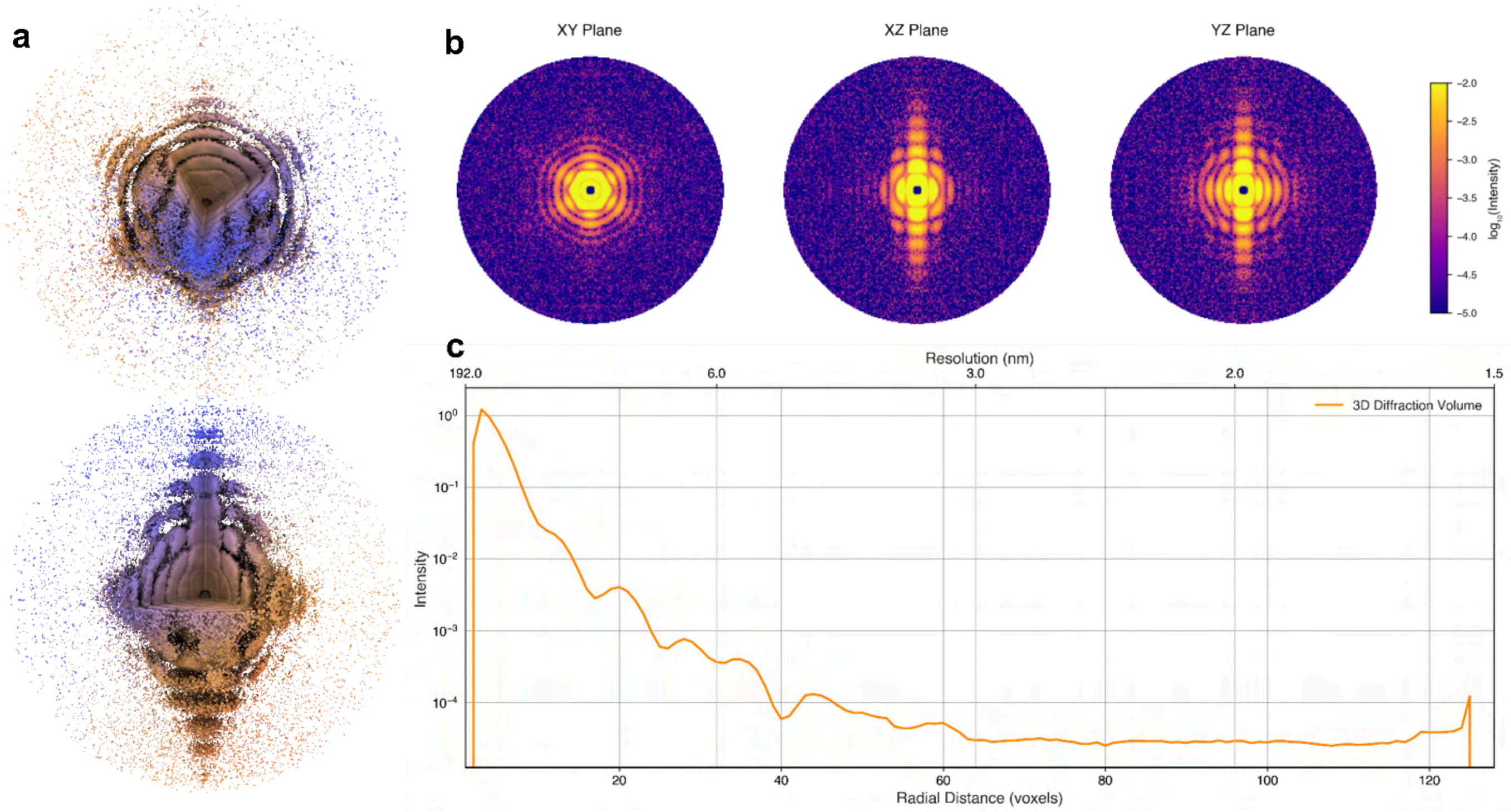

**Fig. 3 | 3D Fourier intensities of single-particle Ery:** (a): Three-dimensional merge of the Fourier intensities of Ery obtained using the background-aware EMC algorithm. Two different orientations of the 3D map are displayed (top and bottom), each with an octant cut. (b) Orthogonal slices of the 3D merged Fourier intensities. Note that seven fringes are observable in all three slices. (c) Profile of the 3D merged Fourier intensities, averaged over resolution shells, as a function of resolution. The signal goes beyond 20 Å resolution in the obtained 3D merged intensity (see Extended Data Fig. 10).

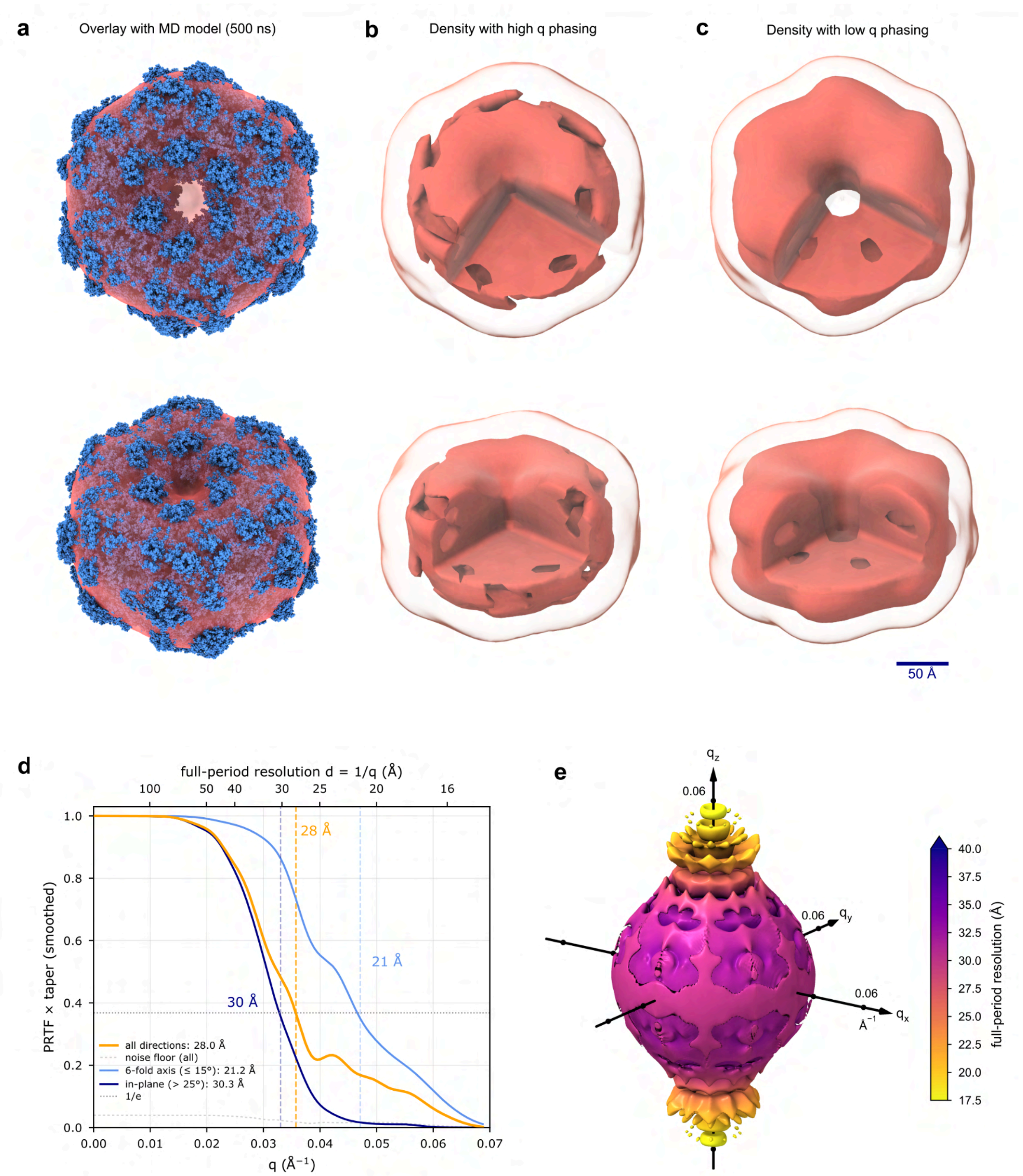

**Fig. 4 | 3D electron density of Ery determined by iterative phase retrieval.** (a) 3D electron density map derived from the 3D merged intensities from one of the classes after 3D classification (Fig. S13-15) (transparent salmon, carved at 80% of maximum density) overlaid on a compacted molecular model (blue) of Ery from MD simulations after 500 ns in a gas environment. (b) and (c) The derived 3D electron density map in two different orientations with octant cuts showing central hole and inner cavities for the high-*q* and low-*q* phased maps respectively (80% and 20% of maximum density level, see Methods). (d) PRTF plot indicating a resolution of 28 Å in all directions, and 21 Å in 6-fold axis, and 30 Å in

in-plane as determined using a threshold of 1/e. The maps in (a-c) were obtained with ChimeraX 1.10.1. (e) Isosurface plot of the 3D PRTF profile with a 1/e threshold indicating anisotropic resolution. Comparison with molmaps with residual solvent too indicates resolution must be around or below 30 Å which gives a range of ~20 to 30 Å effective resolution (see also Fig. S6-9, S19, S25).

## Results

### XFEL-SPI Diffraction Measurements

Diffraction measurements of aerosolised Ery particles were made at a photon energy of 6 keV (2 Å wavelength). XFEL pulses were generated in regular bursts of 352 pulses, 10 times per second, and diffraction patterns were recorded on the Adaptive Gain Integrating Pixel Detector (AGIPD), a 1 megapixel detector specifically designed to keep pace with the EuXFEL pulse rates to generate 3,520 frames per second. The detector was placed at 715 mm downstream of the particle interaction region to give a full-period resolution limit at the centre-edge of the detector of 14 Å (and a corner-edge resolution of 9 Å). The X-ray beam was focused to a size of 180 × 220 $nm^2$ (Fig. 1). With an average energy of about 1.4 mJ in the focused beam (1.4 × $10^{12}$ photons), the fluence in the focus is estimated to reach 3.1 × $10^{13}$ photons/$\mu m^2$ under ideal conditions. The elastic scattering cross section of oxygen atoms (for example) is 1.3 × $10^{-15}$ $\mu m^2$ at 6 keV photon energy, so a well-positioned molecule should scatter a photon from at most about 4% of the atoms. Given that Ery has about 200,000 non-hydrogen atoms, we would expect up to 8000 scattered photons from a single molecule. We note that the cross section for photoabsorption in oxygen is 50 times higher than for scattering, so all atoms will be photoionised within the X-ray pulse. The short pulse duration of 25 fs outruns the subsequent Coulomb explosion (see Methods).

The diffraction measurements were carried out in vacuum in order to maintain a scattering background lower than the expected single-molecule diffraction signal. For this, a 100-μm diameter beam of aerosolised Ery particles, moving perpendicular to the X-ray beam, was focused by an aerodynamics lens and aligned to intersect the X-ray focus (see Methods). The particle focusing was achieved with a gas mixture that was optimised to create the lowest X-ray scattering background yet still maintain a particle beam. The gas scattering without particles was 800 photons per exposure. This is comparable to the expected signal but, since the gas is unstructured, it results in a uniform background compared with the particle's diffraction pattern. Detector noise, after calibration and identification of photons, was almost absent at less than 50 photons per frame.

Over an integrated collection time of 21 hours, 270 million detector frames were recorded. Of these, 188,557, or 0.07%, of the frames were found to contain diffraction signal above the background level, and a significant number of diffraction patterns were deemed to originate from Ery single particles (with size estimation Fig. S35) with the expected diffraction patterns for a hexagonal double-doughnut. Single-protein diffraction patterns were observed in random orientations. Some examples are shown in Fig. 2, including some consistent with D6 symmetry when the 6-fold axis of the Ery molecule was oriented along the beam axis. Candidate single particles were distinguished from foreign objects or multiple particles based on the size of the central speckle (Fig. S35). A comparison of the measured diffraction with calculated patterns of single particles, assuming a structure and atomic scattering factors unaffected by the X-ray pulse, indicate a highest pulse fluence of 1.7 × $10^{13}$ photons/$\mu m^2$ (Extended Data Fig. 6) which is half the maximum achievable by the beamline as stated above. This discrepancy may be due to beamline losses of the pulse energy, a larger X-ray focus, or a reduction of the pulse-integrated scattering due to ionization.

**Classification and Image Reconstruction**

The processing of the diffraction data was carried out in several stages. From the initial selection of 188,557 detector frames, the signal to background ratio (SBR) was estimated from the particle scattering counts in the low-angle region indicated in Extended Data Fig. 7 and from the background counts at high scattering angles. We chose 50,000 frames with the highest SBR (see Methods) for pattern classification and orientation. This was done by applying the Expand–Maximize–Compress (EMC) algorithm in two dimensions (2D) to aggregate patterns into classes of similar particle orientations.[42,43] The algorithm iteratively estimates the class averages while assigning probabilistic weights to individual frames by correlating them with the class averages for each iteration cycle. A selection of 16 out of the full 144 class averages are shown in Fig. 2 (b). The classes each contain about a few hundred diffraction patterns on average, adding signal and averaging away noise to enable real-space projection images to be recovered by iterative phase retrieval (see Methods). The images include a view along the 6-fold axis where low density is observed on axis, as would be expected from the known crystal structure shown in Fig. 1. The resolution of these 2D images is about 15 Å (see Methods and Fig. S36-38, Fig S25).

As in similar analysis in cryo-EM SPI, classifying the diffraction patterns enhances the quality and homogeneity of the data. Many of the classes were obviously not attributable to diffraction from Ery particles, based on the contrast of the diffraction fringes (low contrast implies addition of nonreproducible objects) and by comparison with forward simulations of diffraction of the reference structure (see Fig. S36-38, S25). A selection of 33 classes was chosen, consisting of 26,908 diffraction patterns, as probable diffraction from single Ery particles. This pared selection was then used to build up the 3D diffraction volume of a single particle, using a 3D implementation of the EMC algorithm where diffraction patterns contribute to central spherical cuts (Ewald spheres) through reciprocal space (Fig. 3). It was found that estimates were improved by classifying these into three separate 3D diffraction classes (“conformations” of the particle), which in this case were primarily a simple contraction and expansion of the structure (see Methods and Fig. S13-18). The 3D diffraction intensities of such classes are shown in Fig S17. Subsequent iterative phasing of a class (Fig S17, class 1) produced the electron density map shown in Fig. 4 in red and compared with the crystal structure in blue. The resolution of this map is about 20 to 30 Å as determined from the phase-retrieval transfer function (PRTF) shown in Fig. 4 (d,e). The image shows the central hole and D6 symmetry as expected, and also reveals some internal features that agree with known densities calculated to low resolution (see Fig. S8, S25). There is no evidence in the map of a water layer around the particle, although residual densities from water and counterions may be present since the reconstructed maps closely match the simulated maps (Fig S25) with residual densities from MD simulations (see below) of PDB model with hydration layer and counterions, and the 2D phase-retrieved images also suggest the same (Fig. S36-37).

Although the obtained 3D classes differed in the scale of the reconstructed particle (Fig. S13-18), these were all more compact (Extended Data Fig. 5) than the crystal structure, especially along the short D6 symmetry axis (Fig S23-29), in contrast to the in-solution SAXS and cryoEM structures we obtained that were consistent with the PDB model of Ery (Fig. S1-S5, Fig. S40-42). This could not be explained by uncertainties in wavelength or detector distance, and instead reveals a compaction of the structure in the gas phase. Such a compaction is not unexpected[44] and depends on the protein fold, the solvent content, presence of solvent-filled cavities or protective layers. Generally, proteins with cavities undergo more compaction than those without[45–48].

### MD Simulations of Gas Phase Compaction

To test the hypothesis of gas-phase compaction, we carried out molecular dynamics (MD) simulations of the entire Ery particle with a minimal hydration layer of 0.5 nm, in vacuum, for up to 500 ns. Compared to the naked protein (Fig. S20), the rate of compaction was slower for Ery with minimal hydration layer. A reduction was seen in the radius of gyration of the structure, which fell initially by 2.5% in 1 ns and then more slowly and had plateaued by 250 ns (with 7.70 % compaction) (Fig. S21). The size of the central hole was also reduced, and of cavities between neighboring subunits at different radial locations. The bilayer 6-fold symmetry was preserved (Fig. S6). Although in the experiments Ery particles take tens to hundreds of milliseconds to travel from the electrospray nozzle to the interaction region, the MD simulations may be representative of the general trend. Simulations with different thicknesses of water layers show that the initial fast compaction takes longer with thicker water layers. While a hydration shell of 0.5 nm takes up to 1 ns for ~2.5% compaction, 1.25 to 1.5 nm thicknesses do not show compaction till 3 ns in the simulated time-window (Fig. S22, Fig S52), which may indicate that shock cooling could arrest compaction[49]. Using the MD simulations with different hydration shells, we performed an angular cross-correlation with the experimental 3D classes to estimate the hydration shell thickness in the experimental structures. This indicated 0.5 to 1 nm hydration layers, which are insufficient to avoid gas phase compaction (Fig. S46,47).

## Discussion

We have demonstrated three-dimensional diffractive imaging of an isolated single-protein system with hard X-ray laser pulses—a long-sought goal since the proposal of biomolecular XFEL SPI[6,33,50]. Despite the promise of XFEL SPI to provide time-resolved images of macromolecules without crystallisation nor cryogenic cooling, many technical challenges have stood in the way. Our demonstration here shows the feasibility of single-protein 3D imaging, which was achieved with improvements in beamline, sample injection, detection, and analysis, to produce diffraction patterns of sufficient SBR to be sorted and classified. From this analysis, 2D projection images of Ery were reconstructed at 15 Å resolution. This resolution was not maintained upon 3D analysis, where phasing of the 3D classes was only consistent to a resolution of about 20 to 30 Å. This result indicates a further challenge of aerosol XFEL SPI—particularly sample heterogeneity due to gas-phase compaction. While this could in principle be overcome by measuring orders of magnitude more diffraction patterns (requiring improvements in targeting molecules to the X-ray beam) to obtain high-resolution structures of the population (Fig S43) , next steps must mitigate against structural changes of the sample. This may be achieved through fine-tuning of droplet size coupled with minimizing trajectory time, or rapid freezing of droplets before entering vacuum, or avoiding aerosolisation entirely by delivering samples in a 100-nm thick flat liquid sheet[51]. Although the surrounding liquid adds to the scattering signal and reduces the contrast of the particle, the contribution from the surroundings could be minimised using advanced X-ray optics that focus the beam to close to the dimensions of the particles[52,53].

Our analysis had to contend with considerable shot-to-shot variations in XFEL pulse intensity. At present, this is less of a concern for SPI compared with incident fluence, pulse duration and hit rate. In the future, however, it could be substantially mitigated by recent developments at XFEL facilities in seeded X-ray pulses, which provide a stable wavelength and can be driven to saturation, thereby reducing pulse-to-pulse intensity fluctuations[54]. The pulse duration of 25 fs used here outruns the radiation-induced displacement of atoms, but simulations (Fig. S48-51) indicate there is still time for significant secondary ionisation to occur which reduces the integrated scattering strength of the molecule[55]. This can be addressed with shorter XFEL pulses without reducing pulse energy—that is, increasing the peak power of the pulses. Recently, pulses have been generated down to several femtoseconds with 1 TW peak power, which is 10 times greater than used here.

## Methods

### Beamline and experimental setup

The single-protein diffraction imaging experiment was carried out in the SPB/SFX instrument's upstream chamber with the nanofocus beam at 6 keV (~2.06Å). The x-ray pulses were delivered from the SASE1 undulator at the European XFEL (EuXFEL) with a pulse energy of 2.7 mJ and duration of 25 fs. The pulses were transported through the beamline with an estimated 50% transmission and focused by a Kirkpatrick-Baez mirror system to a 220 nm × 180 nm spot at the interaction region, as measured by ptychography, resulting in an estimated optimal peak intensity of approximately 30 mJ/μm$^2$ (3.1 × 10$^{13}$ photons/μm$^2$). During the measurements the focus was checked and re-optimized (see Extended Data Fig. 6, S10) for thermal drifts about every two hours, since the intensities of the strongest hits decayed over this time. The realignment used a YAG scintillator and high resolution microscope system capable of distinguishing spot sizes no less than 200 nm, so the peak intensity was usually lower than the best achievable.

Protein particles dispersed in a 25 mM ammonium acetate buffer, at a particle concentration of 10$^{13}$/ml, and stored at 4 °C, were aerosolized with $N_2$ and $CO_2$ gas using an electrospray ionization (ESI) source as described elsewhere.[32] The flow rates of $N_2$ and $CO_2$ were 1.9 and 0.25 liters per minute (lpm), respectively. The ESI source ejected the sample solution from a silica fused capillary with an inner diameter of 30 μm and a 30º conical tip. A voltage of 2 to 3 kV was applied to the sample to form a stable Taylor cone from which droplets of ~150 nm diameter were formed. The concentration of particles in solution was such that 90% of droplets do not contain any particles. Those drops that did therefore mostly held only one particle. The sample flow rate was controlled by adjusting the overpressure in the sample reservoir with a differential pressure regulator (Bronkhorst). Droplets emitted from the Taylor cone are charged. Here, they were neutralized by two VUV lamps (Hamamatsu) that photoionized the sheath gas to neutralize the droplets. Two skimmer boxes were used to remove excess carrier gas[32]. After the skimmers, He gas was introduced with a flow-rate of up to 250 psi, producing a gas mix that carried the particles through the aerodynamic lens and towards the X-ray focus (see Fig. 1). The focused particle beam had a diameter of about 100 μm and a speed of about 100 ms$^{-1}$. Gold nanoparticles (AuNPs) were aerosolized and injected to align the injection system to the X-ray beam, before injecting protein samples and collecting diffraction data. *Hummingbird*[56] was used for monitoring diffraction "hits" of particles, to optimise the delivery system and to reduce background scattering from residual gas. The software *Dragonfly*[42,43] was used for calculating 2D class averages during the beamtime.

Prior to the SPB/SFX experiment, preparatory measurements were performed at the Small Quantum Systems (SQS) instrument of the EuXFEL using soft-x-ray pulses of 1.2 keV photon energy with a micron focus beam and observed reduction in gas scattering background up to 90% with additional supply of He after the skimmers, which we employed at SPB/SFX as mentioned above (see supporting information for details and earlier observation of diffraction patterns from intact Ery at SQS with soft XFEL pulses S44-45). In contrast, we could achieve about only 50% reduction in gas-scattering background as a trade-off between hit-rate and background reduction due to nanofocus (Extended Data Fig 9).

**Sample preparation**

Ery was isolated and purified from the blood of earthworms (*Eisenia fetida*). Briefly, about 100 earthworms (Natursache, Germany) were washed thoroughly in DI water thrice to remove debris, blotted, and then washed with Tris buffer (50 mM Tris-HCl, 0.5 mM EDTA, 4mM $CaCl_2$, 10 mM $MgCl_2$ pH 7) twice, before homogenizing with added protease inhibitor (Pierce protease inhibitor tablets, Thermo Fisher). The homogenate was centrifuged at 12500 rcf at 4 °C for 30 min to remove the debris, and then it was filtered with a 0.22 µm sterile polyethersulfonate (PES) filter. The resulting Ery filtrate was centrifuged at 1000 rcf, the pellet was discarded and the resulting supernatant was filtered with a 0.22 µm filter again and was subsequently size-exclusion purified with Sephadex-300 Hi-prep column in Tris buffer (50 mM Tris-HCl, 0.5 mM EDTA, 4mM $CaCl_2$, 10 mM $MgCl_2$ pH 7) at RT. The purified fractions of intact Ery were further concentrated to reach concentrations of up to 13 mg/mL as stocks and stored at 4 °C. The concentration and oxidation state of Ery was determined by the UV-Vis spectrometer (Thermo Fisher nanodrop) (Fig S33). Before aerosol injection, 100 to 1000 nM Ery was buffer exchanged with 25 mM ammonium acetate at pH 7 using G25 spin-trap columns twice at 2000 rcf at 4 °C (leaving less than 1% of the residuals from the parent buffer). The conductivity of the final protein suspension was between 1900 to 2000 µS/cm. The quality of Ery was checked with DLS, ES-DMA-CPC, and TEM measurements (see below for details) just before sample injection at RT.

**Dynamic Light Scattering**

Dynamic light scattering (DLS) measurements were carried out with Wyatt nanostar DynaPro and Xtal Concepts GmbH DLS instruments. Briefly, Ery (100 - 1000 nM) was centrifuged at 1000 rpm and pellets were discarded and then filtered with 0.1 µm disposable filters. And the filtrate was analyzed (4 - 10 µL) placed in a dedicated quartz or disposable cuvette for the in-solution size distribution measurements (see Fig. S30).

**Mass Photometry Analysis**

Massphotometric measurements were made with a Refyn One massphotometer. Tris-buffer (pH 7) mentioned above was filtered with 0.22 µm filter and further filtered with 10 kDa spin filters before loading in the silicone gasket wells. Ery was centrifuged at 1000 rcf and any pellet discarded. Briefly, 2 µL of 10 nM Ery from the supernatant was introduced into the silicone gasket buffer well with preloaded 10 µL filtered Tris buffer, and subsequently images were acquired for 60 s. The obtained images were analyzed with Refyn software and the mass distribution was fitted with KDE (Fig. S32).

**Synchrotron SAXS characterization**

Small angle x-ray scattering measurements (Fig. S1-S4) were carried out at the P12 beamline at PETRA III, DESY, at a photon energy of 10.5 keV. Samples were flowed in standard quartz capillary tubes at a concentration of 3.6 mg/mL. 20 × 50 ms exposures for a total exposure time of 1 s were collected with standard batch collection mode. Individual frames were monitored for radiation damage, averaged, and corrected for instrument background. Buffer subtraction and concentration normalization were performed using ATSAS.[57] Raw experimental data were fit with a smooth curve using the denss.fit_data.py script.[58] Ab initio electron density reconstruction was performed directly from the experimental solution scattering data using DENSS (version 1.7.2).[59,60] Reconstructions were calculated on a 96 × 96 × 96 voxel grid while imposing six-fold dihedral symmetry (D6) enforced along the shortest principal axis of the molecule. A total of 20 independent reconstructions were aligned and averaged to generate the final electron density map using the denss.all.py script. Resolution was calculated using the built-in Fourier Shell Correlation (FSC) calculation which averages the FSCs of all twenty maps and estimates resolution where the averaged FSC falls below 0.5 (Fig S1-4).[60]

**Negative Stain EM**

Negative stain EM characterization was carried out with a Talos 120c and a JEOL 200 electron microscopes. Briefly, 3 µL of Ery was dropcasted onto a GloQube-plasma-treated (-ve/30s) Cu-300-mesh grid with 2-nm continuous carbon film for one minute. This was flash-washed twice with milli-q water droplets, blotted, and incubated for a second in 2% uranyl acetate and incubated with uranyl acetate again for 30 s, blotted with Whatman filterpaper, and then let dry at RT for 5 minutes before imaging in the electron microscope (Fig S39).

**Cryo-EM Analysis**

Grid preparation and vitrification: 4 µL of Ery (0.4 mg/ml) in 25 mM ammonium acetate (pH 7) was applied to glow-discharged (-ve/30 s) R1.2/R1.3 holey carbon grids with a 2-nm continuous carbon film. Grids were plunge-frozen using an FEI Vitrobot (Mark IV) operating at 4 °C and 100 % humidity, with a 60 s wait time, a 3.5 s blot time and a blot force of 0, followed by rapid plunging into liquid ethane.

Data collection: Frozen grids were transferred to a Titan Krios transmission electron microscope (Thermo Fisher Scientific) operated at 300 kV. Movies were acquired on a direct electron detector at a calibrated pixel size of 0.674 Å/pixel. The nominal defocus ranged from -1.0 to -3.0 µm, and the total electron dose was kept at approximately 1 $e^-$/ Å$^2$ per frame. Data collection was performed automatically using standard acquisition software and further processing was done using cryoSPARC (see Fig. S40-42). Resolution of the final 3D reconstruction was estimated with Fourier Shell Correlation with the threshold at 0.143.

**Differential Mobility Analysis (DMA)**

The size-distributions of electrospray-aerosolized Ery were measured with a differential mobility analyzer (DMA) with electrostatic classifier and a condensation particle counter (TSI). The aerosols were generated and neutralized as described above in the experimental setup (only with $CO_2$/$N_2$ carrier gases and no He) and delivered to the DMA-CPC through a conducting tube for size distribution sampling  (see Fig. S31).

**Aerosol size distribution estimation with optical light scattering**

Aerosol optical laser light scattering based estimation of size-distribution was carried out in a lab setup as described elsewhere.[61] For this, Ery molecules were electrospray-aerosolized and injected using an aerodynamic lenstack, and 532-nm laser light was used to record scattered intensities of Ery molecules. Based on the recorded intensities, with known polystyrene references as calibration as previously demonstrated, the size-distribution of Ery was determined (Fig. S34).

**XFEL-SPI Data Processing**

**Hit-finding**: In order to isolate detector frames that recorded X-ray diffraction from the sample (hits) from those containing only background due to gas scatter and detector noise (misses), we identified frames where the integral photon signal within a low angle region of the detector (the “hit-mask” as shown in Extended Data Fig. 7) was significantly above the background level. The hit-mask region was selected to provide a high expected SBR of the integral photon signal. Estimation of the background distribution was complicated by the high degree of variability of the He density within the vacuum chamber, leading to a corresponding variability in the number of gas scattering events over time. To address this issue, we employed a robust estimator for the standard deviation (σ) of the background signal within individual pulse trains (310 μs): 1.4826 (median - Q1) , where Q1 is the lower quartile of the photon counts. Hits were then identified as frames with photon counts 4σ above the median value within a pulse train. An example is shown in Extended Data Fig. 8, where a single hit is identified (See Fig S53 for the overall data processing pipeline).

**Background estimation**: An estimate for the background photon signal, *B*, (accounting for gas scattering, beamline background and false positive detector counts) associated with each hit was calculated by taking the mean per-pixel photon signal of all misses within an experiment run (3000 pulse trains). This was then multiplied by a factor that corrects for the pulse energy and the strength of the background within the pulse train where the hit was detected. This correction factor is given by the ratio of the per-frame pulse energy to the mean within the run multiplied by the ratio of the integrated photon signal of non-hits within the pulse train to the integrated mean background signal for the entire run.

**Signal-to-noise estimation**: The photon signals from each frame are assumed to be Poisson distributed. The SNR for each frame was calculated within the hit-mask region (defined above) as $S / \sqrt{K}$ where $S = K - B$ for an integrated photon signal $K$ and estimated background signal $B$. We restricted the SNR calculation to the low-angle region of the detector in order to obtain a statistically significant estimate for the sample scattering signal which ranges from about 15,000 photon counts for strong hits (from larger droplets or aggregated particles), to about 50 counts for weaker hits. Near the edges of the detector, corresponding to large diffraction angles, the estimated background scattering is often greater than the recorded photon signal, yielding an average value for the estimated scattering signal that is close to or less than zero. The signal to background ratio (SBR) is defined as $S/B$.

**Beam-drift and beam-centre correction**

Online and initial processing revealed significant beam drift during the data collection, resulting in beam-center shift affecting initial 2D classification (below). To correct for this, we implemented a maximum likelihood (ML) approach to refine the per-pattern beam center. Using the initial 2D EMC class averages, we calculated the log-likelihood ( $\sum P \log R$) of the diffraction patterns as a function of systematic detector shifts in the *x* and *y* (horizontal and vertical) directions. By maximizing the summed log-likelihood over all frames, we determined a global offset of -1.6 pixels in *x* and -2.3 pixels in *y*. Applying this corrected geometry significantly improved the 2D classification quality. A subsequent ML refinement pass on the improved classes yielded negligible further shift, confirming the convergence of the detector center (Fig. S10, 11).

**2D EMC classification**: Calculations were performed on the Maxwell computing cluster at DESY using a custom implementation of the EMC algorithm (github.com/andyofmelbourne/EMC2 commit b9632f2). 144 classes were initialised with random numbers drawn from a uniform distribution between 0 and 1, each on a 64 × 64 pixel grid of spatial frequencies (*q*) with an average binning ratio of 16 detector pixels to one class pixel (edge resolution of 3.8 nm or $q$ = 0.26 nm$^{-1}$). During the reconstruction each of the 50,000 frames were compared to each of the 144 classes for each in-plane rotation sample of 3300 orientations. A regularisation parameter was employed to flatten the probability distributions during early iterations,[29] this was raised exponentially from 0.001 (high flattening) to 1 (no flattening) for the first 100 iterations, and was then held constant for a further 50 iterations. Inversion symmetry about the origin was enforced for each class for all iterations. This approximates Friedel symmetry for pixels corresponding to low diffraction angles, but becomes less accurate for pixels corresponding to higher diffraction angles, where Ewald sphere curvature becomes significant.

For the purposes of phasing selected classes, a further 5 iterations were performed with the class pixel grid increased to 256 × 256 with the same pixel binning factor (edge resolution 14 Å) and the rotation sampling was also increased from 3300 to 9500.

Prior to classification, 500 of the stronger hits were manually classified as “good hits” and 500 as “bad”. The distribution of good and bad hits per class was then used to calculate the Degree-Of-Separation (DOS) per iteration and is given by the 1 - the normalised cross-correlation coefficient between these two distributions. The DOS started at 29% at the first iteration then increased to 98% by iteration 150, indicating that apparently good and bad diffraction frames were well separated during the classification process. The gini-coefficient, Q-factor and number of changes in orientation and class assignments were also monitored to ensure good convergence during the reconstruction. In supplementary movie 1, we show the classes as a function of iteration number. In addition we show a 2D plot which indicates the affinity of each frame to each class by projecting the probability simplex along its surface normal onto the two-dimensional plane. The probability simplex is here a 99-dimensional simplex whose vertices are the 100 standard unit vectors, one for each class, the position of a frame lies on the surface of this simplex and its coordinate along each axis is given by the sum of the probability-matrix for that frame over all orientations for the corresponding class.

**3D EMC reconstruction of diffraction volume of Ery**

After 2D classification a total of 26,908 diffraction patterns were selected for further processing. Reconstruction of the three-dimensional diffraction volume via the iterative EMC algorithm progressed in several stages. First the diffraction volume was recovered using the conventional approach.[43] This diffraction volume was defined on a 96 × 96 × 96 voxel grid with a binning factor 4 × 4 × 4 (64 detector pixels per voxel at low diffraction angles) and an edge resolution of 3.8 nm. The beta factor was increased exponentially for the first 30 iterations from 0.001 to 1, then held constant at 1 for a further 20 iterations. During these iterations 10,395 rotation samples were modelled to determine the sample's unknown orientation for each diffraction snap-shot. A further 5 iterations were performed on a larger grid of 256 × 256 × 256 with the same binning factor (edge resolution 1.4 nm) and with 79,191 rotation samples (see Extended Data Fig. 1 row 1). This reconstruction appears to exhibit poor signal-to-noise relative to that of the stronger hits included in the frame selection, see for example Extended Data Fig. 7. Distortions in the shape of the speckles also indicate that it suffers from heterogeneity in the dataset.

A second reconstruction was then performed with scale factor modelling. Here the set of orientations was extended to include scale factors: 0.9, 0.95, 1, 1.05 and 1.1. Here, each frame is compared to each of the 79,191 Ewald slices (corresponding to possible orientations of the molecule) through each of the 5 scaled diffraction volumes (corresponding to possible sizes of the molecule). The diffraction volume is then updated by inverting the recovered scale factors and orientations for each diffraction pattern before merging them into the initial reference frame of the molecule in the standard manner. Central sections through the recovered diffraction volume are shown in the second row of Extended Data Fig. 1. Here we can see an apparent improvement in the shape and symmetry of the central speckles.

In order to improve the SNR of the reconstruction at high resolutions, weaker diffraction frames were then excluded from the merge. This is beneficial for the SNR of the merge despite the fact that weaker diffraction patterns carry useful information about the sample, since the ratio of the sample diffraction signal to gas background at higher resolutions is well below that of the stronger hits. The distribution of photon counts as a function of distance from the detector center is shown in Extended Data Fig. 2.

For detector pixels that are dominated by high background, where $B \gg S$, we can approximate the SNR as $S/\sqrt{B}$. If we regard $S$ and $B$ as the cumulative signal acquired from a given frame selection, then adding an additional frame with a signal of $S_i$ and background of $B_i$ will only increase the SNR of the total merged signal when $S_i / B_i > S/(2B)$. That is, it is only beneficial to include the frame if its SBR is greater than half that of the total. Since this condition is invariant to scale in the signal and background independently, and the relationship between low and high angle signal strength in $S_i$ and $B_i$ will be approximately constant between frames, we can use the low angle estimates for the SBR to determine which frames to select in order to maximise the SNR at high resolution where it is hard to estimate the excess signal above the background. This criteria leads to a selection of 1,248 frames, out of a total of 26,908 (see Extended Data Fig. 3). The recovered diffraction volume is shown in the third row of Extended Data Fig 1, where the sample diffraction is more clearly resolved against the background level as expected.

In the fourth row of Extended Data Fig 1, the diffraction volume was merged with per voxel background-aware EMC modelling. Here, each frame d is modelled voxel by voxel (index i) as Poisson counts with mean F_dri = w_d·W_ri + b_d·B_i. Here W_ri is the 3D intensity model sliced at orientation r, w_d is the frame's relative fluence, B_i is a fixed detector background pattern and b_d is the frame's background level. The log-likelihood of orientation r is therefore log R_dr = Σ_i [K_di·log F_dri − F_dri], where K_di is the measured photon count. In the update step, the 3D intensity of each voxel, the fluence of each frame and the background of each pixel are re-estimated by maximising the expected log-likelihood, weighted by the orientation probabilities. Because the background sits inside the logarithm, these updates have no closed form, but each reduces to finding the root of a one-variable equation of the form $\Sigma_j a_j/(x + b_j) = c$, which is solved numerically. This step increases the fringe visibility of the merge.

This background aware algorithm removes the need to exclude low fluence frames from the merge, as each photon's contribution to a voxel is weighted by the likelihood that it originates from the particle rather than the background. In this way, weaker hits still contribute to the low angle speckles without adding noise to the high angle regions of the diffraction volume. Note that weak hits may still pose a problem if they carry too few photons to show which structural class they belong to in the image classification stage.

Finally in the fifth row of Extended Data Fig 1, we show the diffraction volume with all 26,908 diffraction patterns included, scale factor modelling, background aware refinement and with the point group symmetry of the molecule (D6) + Friedel symmetry (yielding D6h) applied. Radial profiles of the merged diffraction volumes are shown in Extended Data Fig 4, indicating that this volume contains the greatest signal to noise at the highest resolution. Although this captures the full measured signal above noise, the scale-factor model assumes compaction uniformly rescales a reference density, so structurally distinct states that do not follow this rule will be averaged together. To separate these states before phasing, we sort the data structurally, as described in 3D EMC Classification below.

**3D EMC Classification**

Similar to the 2D classification, 3D classification was carried out by classifying diffraction patterns into 3 distinct bins, denoted as Class 0, Class 1, and Class 2. All three classes were constrained to D6h point group symmetry. Unlike independent clustering methods, the EMC algorithm classifies frames probabilistically. This was carried out in four stages denoted as Stage 0, Stage 1, Stage 2, and Stage 3.

Class Continuity and Annealing: To prevent the three classes from collapsing into identical models or locking into local minima early in the reconstruction, two key regularizations were applied: 1. Class Continuity ($\sigma$): A Gaussian smoothing was applied along the class axis of the probability matrix. Early stages used $\sigma = 0.4$ (allowing ~4 % probability leakage between neighbouring classes), which was tightened to $\sigma = 0.2$ in later stages. This constrained the classification to yield a continuum of states rather than random fragmentation and is based on the prior assumption that frames in this set differ only in their degree of compaction. 2. Deterministic Annealing ($\beta$): The log-likelihood was scaled by an annealing factor ($\beta$), starting at 0.005 and increasing to 1 as frame assignments stabilized. This allowed the classes to initially hold broad distributions of frames before sharply sorting them at higher resolutions.

Likelihood Models and Class Divergence: The classification underwent a transition in its likelihood model. Stages 0 and 1 utilized a fluence-free marginalization ("W" mode), where per-frame fluence was integrated out. In stages 2 and 3, the algorithm switched to a Poisson likelihood with explicit background correction ('I' mode). This introduction of the background-aware likelihood heavily influenced the final class occupancy.

During the transition to the background-aware stage, Class 2 dropped from 27% to <1% of the frames, indicating that it did not represent a distinct state once the background-noise was properly modeled. The surviving classes, Class 0 and Class 1, exhibited a continuous drift in population. By the end of Stage 2, the frames split 65% : 34% (ignoring Class 2). The separation was driven by structural differences in the intensity models (Class 1 exhibited deeper minima between rings and higher-$q$ axial striping), rather than raw photon count, as the median photons per frame for both classes were statistically near-identical (1,414 vs. 1,363). Stage 3 models did not introduce fundamentally new class structures; instead, they extended the stage 2 ($160^3$ grid) class features to a larger voxel grid ($256^3$ grid, extending to 0.69 $nm^{-1}$). The high-$q$ extension revealed that beyond 0.4 $nm^{-1}$, the structural signal was dominated by a flat background excess (6-7 % of the background). Therefore, while stage 3 successfully provided the high-resolution expansion of the class models, the converged stage 2 models remained the most robust representation of the class occupancies (see Fig. S12-18, S.Table 3-4).

**Phase retrieval and resolution metrics**

For both the cases of 2D class averaged intensities and 3D merged intensities, iterative phase retrieval was carried out to obtain the real space images.

2D phase retrieval: All 144 2D class averages from the 2D EMC run were phased. The diffraction intensities were placed on a 342 x 342 grid of spatial frequencies (Fourier space) with a pixel size of 0.002 $nm^{-1}$ and a maximum at center edge of $q$ = 0.693 $nm^{-1}$. This accommodated the measured $q$-range, which corresponded to a range of spatial frequencies $q$ from 0.0247 $nm^{-1}$ (limited by central beamstip) to 0.670 $nm^{-1}$ (limited by detector center edge) spanning 405 Å to 14.9 Å full-period resolution. (The Fourier transform of this array has a real-space pixel spacing of 7.21 Å.) Pixels with exactly zero intensity were treated as unmeasured, while a hard low-pass constraint forced zero amplitude beyond a circle of a radius of 165 pixels (0.670 $nm^{-1}$). Each of the 144 classes was phased independently using the phasing.phase algorithm (from the 3D-phasing package, OpenCL/GPU). For each class, 100 random starts were launched, each beginning with a disc-shaped support (radius 30 nm) and uniformly random phases. The reconstruction ran in two stages using Difference Map (DM)[62] and Error Reduction (ER)[63,64] with shrinkwrap[65].

In stage 1 (real object): 4,000 iterations (20 cycles of 100 DM + 100 ER + 1 shrinkwrap). The object was constrained to be real, with the shrinkwrap $\sigma$ reducing from 3 to 1 pixel and a threshold of 0.25. In stage 2 (complex object), 6000 iterations (30 cycles of 100 DM + 100 ER + 1 shrinkwrap). The object was allowed to be complex, with $\sigma$ reducing from 1 to 0.1 pixels, a threshold of 0.20, and with a radial background correction. For each class, the 100 starts were aligned and a consistency filter was applied to keep the 30 mutually consistent starts. These 30 were realigned and averaged in the Fourier space. The phase retrieval transfer function (PRTF) was calculated as the magnitude of the averaged unit phasors ($\left|< e^{i\Phi} >\right|$) across the 30 starts. The resolution was determined as the point where the PRTF drops below the threshold of 1/e.[66] The resolution strongly correlated with the class population. Classes with $\geq 200$ frames reached a median resolution of 19 Å. About 38 strong classes had a resolution, limited by the forced low-pass filtering, of ~15 Å.

3D phase retrieval: Two complementary phase retrieval strategies were applied to the selected 3D merged intensities after 3D EMC classification. 3D phase retrieval was carried out using GPU-based Hybrid Input-Output (HIO)[64]/shrinkwrap framework.
High-*q* phasing with anisotropic taper: The class-1 merge (see S.Fig 17) was phased using the full *q* range (from the edge of detector hole to center edge of the detector) with an anisotropic taper of observed diffraction amplitudes to retain a signal out to the shell at a voxel-radius of 110 (0.60 $nm^{-1}$) within 15° of the 6-fold axis and the shell at voxel-radius of 60 (0.32 $nm^{-1}$) in-plane. A spherical incoherent background was fitted per shell beyond shell 50 to absorb the flat high-*q* excess (6-7% of the background). Measured amplitudes were imposed exactly on voxels inside the taper envelope. Each of 500 random starts ran five cycles of 80 HIO ($\beta$ = 0.9) and 5 ER with shrinkwrap (threshold 0.12 to 0.10 and blur $\sigma$ from 3 to 1.5 voxels), followed by 200 ER with the support fixed. Positivity and exact D6 symmetry were applied throughout. All 500 seeds converged (final errors 0.1161 to 0.1170). The 500-seed average is a two-layer hexagonal particle with an axial channel and a six-armed chiral pinwheel in each layer. The PRTF of the average smoothed over one speckle, crosses the 1/*e* threshold around 30 Å resolution in plane and 21 Å resolution along the 6-fold axis.

Low-*q* phasing with hard cutoff: We also performed a complementary phasing approach by imposing a hard cutoff at *q* = 0.26 $nm^{-1}$ (38.5 Å resolution), deliberately excluding the high-*q* region where the flat 6-7 % background excess and heterogeneity based modulation of high *q* speckles dominate. The support was a 16 nm sphere cut to a 21.6 nm slab along the six-fold axis, with the axis column removed. The algorithm used the same HIO/shrinkwrap framework but without positivity (real-object constraint only) and with  a radial background term. This low-*q* reconstruction recovered the internal features cleanly-the axial channel, the six-armed pinwheel, and the layer twist, but its PRTF never dropped below 0.83 out to the cutoff. The 79 converged seeds agreed at every measured *q*, and the only reason PRTF did not decay is that no data beyond 38.5 Å was supplied. The 38.5 Å cut-off is not a physical limit but a deliberate choice to exclude high-*q* contamination while retaining enough data to constrain phases.

Both methods independently recovered an axial channel. In the low-*q* run a narrow vacuum-like hole was enforced by removing the central axis from the support mask (motivated by the observation of transparent-hole-like density in the phase-retrieved 2D class-averages of face-on hits). In the high-*q* phasing, which used a spherical support with shrinkwrap and imposed no explicit central constraint, a zero-density channel was still recovered. The agreement between the unconstrained 3D result and the 2D projection confirms that the channel is a genuine structural feature, and the weak density within it is consistent with residual solvent rather than being empty.

**Fluence Estimation with EMC Scaling**

The EMC model only fixes the product of $w_d$ (the relative fluence of frame d) with W (the model's diffraction volume), so the fluences $w_d$ are known up to one global scale factor. To set this scale, we calculated the diffraction volume of the atomic model (PDB ID: 2gtl) for an incident fluence of 1 photon/μm², including the classical electron radius, pixel solid angle and polarisation, on the same 3D grid as the EMC merge. After aligning the two volumes, we found the scale factor s by least-squares fitting $s \cdot I_{sim}$ to the EMC intensities. We restricted the fit to low scattering angles, where the signal depends mainly on the molecule's total mass and is least sensitive to structural differences. Because $w_d \cdot W_{ri}$ gives expected photons per pixel, the absolute fluence of each frame is then $\Phi_d = s \cdot w_d$ photons/μm². This calibration is only as good as the match between the atomic model and the measured molecule, especially its mass and compaction state. The simulated ($I_{sim}$) and merged (W) volumes, time dependent fluence signal and fluence distributions are shown in Extended Data Fig. 6.

**MD simulations on the hydration-shell and vacuum effects on Ery**

Simulations of the whole Ery molecule were initiated using the PDB structure 2GTL. The PDB file was first prepared by PDBFixer to model missing loops based on the known sequences. Oxygen was coordinated to the heme groups by replacing carbon monoxide. The CHARMM27 forcefield was chosen for its parameterization of heme residue.[67] Neutralization was required since the system was predicted to have -1620 e charge based on side chain pKas, in contrast to the neutral particles in the experiment. For the bare protein and the protein with water-shell thinner than 2 nm, neutralization was performed by adding protons based on solvent accessible surface area of residues. For the other thicker water shell (to find sweet-spot hydration shell to avoid gas-phase compaction), which could accommodate counterions, neutralization was obtained by first hydrating with TIP3P water model and then randomly replacing 1620 water molecules with sodium ions. All simulation models were placed in a cubic box with an edge length of 60.7384 nm, defined by firstly centering the protein and imposing a distance of 15 nm from the surface of the protein to the box boundary. Linear Constraint Solver (LINCS) algorithm was used to constraint bond lengths[68].

A steepest-descent algorithm was used to find a local energy minimum during energy minimization (EM). 100 ps of 300 K temperature equilibration (NVT) via Velocity rescaling thermostat[69] was conducted to heat the system. Particle mesh Ewald (PME) was used for calculating long-range interactions. Van der Waals interactions were modeled using a Lennard-Jones cutoff of 1.2 nm with a force-switch modifier applied between 1.0 and 1.2 nm, smoothly decreasing the forces to zero at the cutoff distance.

A production of molecular dynamics of 3 ns with 0.002 ps of time step was employed on all models to observe initial compaction. The 0.5 nm hydration shell underwent 500 ns simulation to enable closer comparison with the experimental conditions. All simulations were run with GROMACS v.2024.4[70] on DESY Maxwell HPC cluster.

**Radiation damage simulations**

To better understand how ultrafast radiation damage might affect the observed data, simulations were performed to generate synthetic damage-affected 2D scattering patterns. A representative Ery target was constructed from the deposited 2GTL PDB structure by adding missing non-H atoms and then H atoms in *Coot* and *phenix.ready_set*, respectively. The structure was then solvated with a 1 nm hydration shell using GROMACS 4.5.4[71]. Hybrid plasma-MD simulations of the target during illumination were performed with the non-Maxwellian plasma code *AC4DC*[72] in conjunction with the hybrid plasma/MD framework of *MolDStruct*[71,73], as described in (S.K.P, C.C., Cardoch S., A.V.M, N.A.Z, manuscript in preparation). The X-ray pulse was modeled with a spatially uniform fluence of $8.3 \times 10^{12}$ photons/um$^2$ over a 25 fs FDHM Gaussian temporal profile, at about half the highest fluence observed in the experiment. Snapshots of the structure at different times during the pulse are given in Fig. S48 and as a movie in Fig. S49. Atomic displacements over the duration of the pulse are visualized in Fig. S50.

Damage-affected diffraction patterns were computed using the time-dependent atom electron configurations and positions from the dynamical simulations. Fig. S51 shows <I>/σ(I) for 500 × 500 pixel screens configured for edge resolutions of 30 Å and 4 Å. Damage starts to affect the measure at a resolution of about 4 nm, but only creates a significant change to the speckle contrast at 1 nm and beyond, and is unlikely to cause resolution loss in our measurements. The calculations indicated a reduction in pulse-integrated scattering strength of the particle by 30%, due to ionization and assuming that the electrons leave the particle.

**Hydration-shell quantification with correlation analysis**

The size of the protein hydration shell was estimated using angular cross-correlation analysis of the experimental and simulated diffraction images.[74] For a given set of $M$ diffraction images considered in the polar coordinate system $(q, \varphi)$ defined in the detector plane, the average difference-pattern angular cross-correlation function (CCF) $C^{diff}(q, q', \Delta) = \langle\langle I_i^{diff}(q, \varphi) I_i^{diff}(q, \varphi + \Delta)\rangle_\varphi\rangle_i$ was determined, where $0 \leq \Delta, \varphi < 2\pi$ are angular coordinates, $\langle \bullet \rangle_\varphi$ denotes averaging over the azimuthal angle $\varphi$, and $\langle \bullet \rangle_i$ defines averaging over $M/2$ difference diffraction patterns, $I_i^{diff}(q, \varphi) = I_j(q, \varphi) - I_k(q, \varphi)$, each constructed from a unique pair $(j, k)$ of images. Subsequently, the normalized Fourier components (FCs) of the CCF, $C_n(q, q') = \frac{1}{4\pi}\int_0^{2\pi} C^{diff}(q, q', \Delta) e^{-in\Delta} d\Delta$ , were calculated for even orders $n = 2, 4, 6, 8, 10, 12$. The small-angle x-ray scattering (SAXS) profile for the same set of frames was determined as, $I^{SAXS}(q) = \langle\langle I_j(q, \varphi)\rangle_\varphi\rangle_j$ , where $\langle \bullet \rangle_j$ denotes averaging over $M$ individual frames. The FCs $C_n(q, q')$ and SAXS intensities $I^{SAXS}(q)$ were individually determined for a selected set of $M = 522$ experimental high-intensity single-particle hits, as well as for simulated diffraction datasets corresponding to several Ery structures obtained from MD simulations. These included a “dry” protein (PDB entry 2GTL), a “wet” protein (with biological water included in the model), and also structures with hydration shells of different thickness (0.5, 0.75, 1.0, 1.5, and 2.0 nm). In each model case, $M = 2000$ two-dimensional coherent diffraction patterns were simulated in forward-scattering geometry under conditions similar to the experimental setup: photon energy of 6 keV, sample-detector distance of 715 mm, square detector of size 102.4 mm x 102.4 mm, and pixel dimensions of 400 $\mu$m x 400 $\mu$m. All SAXS profiles were normalized by the corresponding scaling factors $A = \langle I^{SAXS}(q)\rangle_q$, where averaging $\langle \bullet \rangle_q$ was performed over the range $0.18\, nm^{-1} \leq q \leq 2.0\, nm^{-1}$ , and the FCs $C_n(q, q')$ were normalized by the respective values of $A^2$. Finally, the quantities $\langle Log(|C_n(q, q')|)\rangle_{q,q'}$ and $\sum_{n=2,4,6} 1/\left|\langle Log(|C_n(q, q')|)\rangle_{q,q'}\right|$ were evaluated for all datasets, where averaging was performed over the range $0.18\, nm^{-1} \leq q, q' \leq 2.0\, nm^{-1}$ (see Fig. S46). The estimated hydrations-shell thickness in the experimentally studied Ery aerosols lies in the range of 0.5-1.0 nm.

**Author Contributions**

PLX conceived this project. PLX and HNC wrote the proposals, obtained and co-ordinated the beamtimes 3004 (02-06 Nov 2023) and 7927 (27-30 Sep 2024) together with the collaborators at SQS and SPB/SFX instruments of EuXFEL. SD, MM, YO, RBo, SU, TB, ADF, BS, MP, AC, ID, and CD operated SQS-NQS instrument and the DSSC detector. JB, RB, FK, TP, TS, CK, RL, MV, and FT operated the SPB/SFX nanofocus instrument and the AGIPD detector. LG, ES, OT, JE, AJM, CL, WZ, MB, OY, EC, NDL, SBa, HNC and PLX participated in the online data collection and analysis. AJM led the off-line data-analysis together with PLX, HNC with contributions from FRNCM, AB, and OY. JB, SRZ, LVH, AKS, RAK, RA, KJ, SL, JH, PLX, JK, and MBFM operated the injection set up. XG, EGM, CC, and EDS carried out MD simulations of vacuum effects on Ery with different hydration-shells. SKP, CC, NZ, and AVM carried out radiation damage simulations. RPK and TBB carried out hydration-shell quantification with correlation analysis. PLX purified and characterized Ery. PLX and TDG carried out synchrotron SAXS simulation and data analysis. CSc, ER, and HH supported in the XBI lab EuXFEL. PLX, ER, CoCa, and CSe performed negative-stain EM and cryoEM measurements and PLX carried out the cryoEM-SPI data-analysis. JL, PLX and AKS performed the aerosol optical scattering based sizing and analysis. PLX, RS, MA, DAB, RDK, SB, SO and APM contributed to different samples or simulations and participated in the designing and execution of the experiment. PLX, AJM, and HNC wrote the manuscript with input from all the co-authors.

**Acknowledgements**

We acknowledge European XFEL in Schenefeld, Germany, for provision of X-ray free-electron laser beamtimes at SQS and SPB/SFX under proposal numbers 3004 and 7927, respectively, and would like to thank the staff of European XFEL and DESY for their assistance. PLX thanks Joachim Herz Stiftung for a fellowship. We acknowledge financial support by Deutsches Elektronen-Synchrotron DESY, a member of the Helmholtz Association (HGF), also for the provision of experimental facilities and for the use of the Maxwell computational resources operated at DESY. This work was supported by the Cluster of Excellence "CUI: Advanced Imaging of Matter" of the Deutsche Forschungsgemeinschaft (DFG, EXC 2056, project ID 390715994). PLX and HNC acknowledge support from the Human Frontiers Science Program (RGP0010/2017) and European Research Council—Frontiers in Attosecond X-ray Science: Imaging and Spectroscopy (AXSIS) (ERC-2013-SyG 609920). AJM acknowledges an early career fellowship of the Australian Research Council (ARC). TDG acknowledges funding from National Institutes of Health, National Institute of General Medical Sciences award R35GM158053. SKP, NZ, and AVM thanks the Australian Research Council. NZ acknowledges funding from the Australian Research Council Future Fellowship FT220100405. We thank SPC facility, EMBL for assistance in SAXS (at P12 PETRA, DESY) and mass-photometry measurements and CSSB and XBI-EuXFEL for access to electron microscopes. We acknowledge David Baker, Neil King, Erin C. Yang, Shunzhi Wang, Marcos C. Miranda, Hao Shen, Alexis Courbet, and Nathan M. Ennist from the Institute for Protein Design for their participation in and contributions to the beamtimes 3004 and 7927.

**Data Availability**

The original experimental data can be accessed at https://doi.org/10.22003/XFEL.EU-DATA-003004-00 and https://doi.org/10.22003/XFEL.EU-DATA-007927-00 from the EuXFEL. Processed data and darks used for analysis and reconstruction presented in this article can be accessed at the Coherent X-ray Imaging Databank (CXIDB) under the ID 248.

**Code Availability**

The codes used for analysis can be accessed at https://github.com/FXIhub/xfel7927 and https://github.com/FXIhub/xfel3004

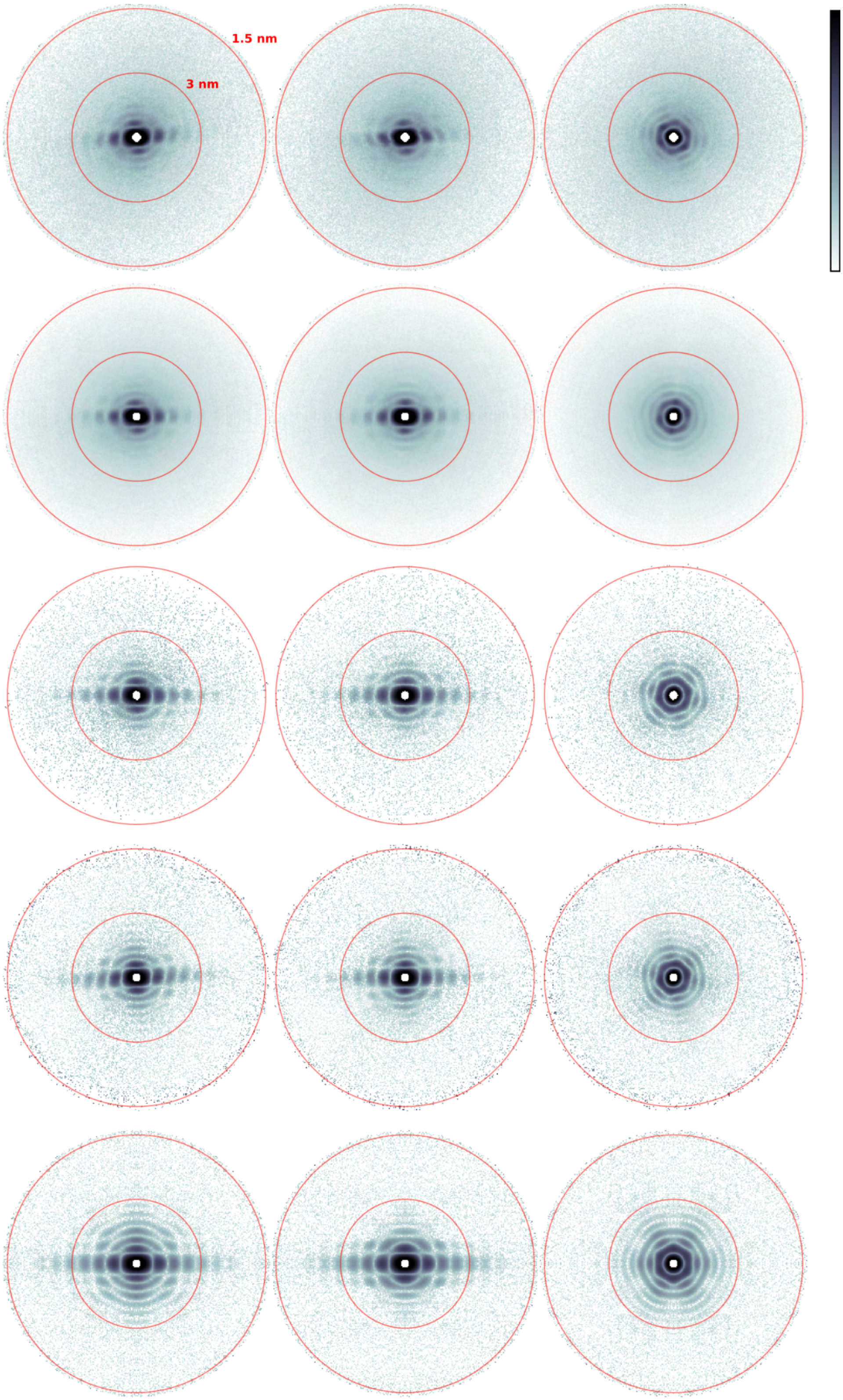

**Extended Data Fig. 1 |** Central sections through the reconstructed diffraction volumes. **Row 1**: initial reconstruction. **Row 2**: as in **1** with scale factor modelling. **Row 3**: as in **2** with SBR based frame selection. **Row 4**: with scale factor modelling and background aware EMC. **Row 5**: as in **4** with the point group symmetry of the molecule (D6) enforced. Logarithmic color scale.

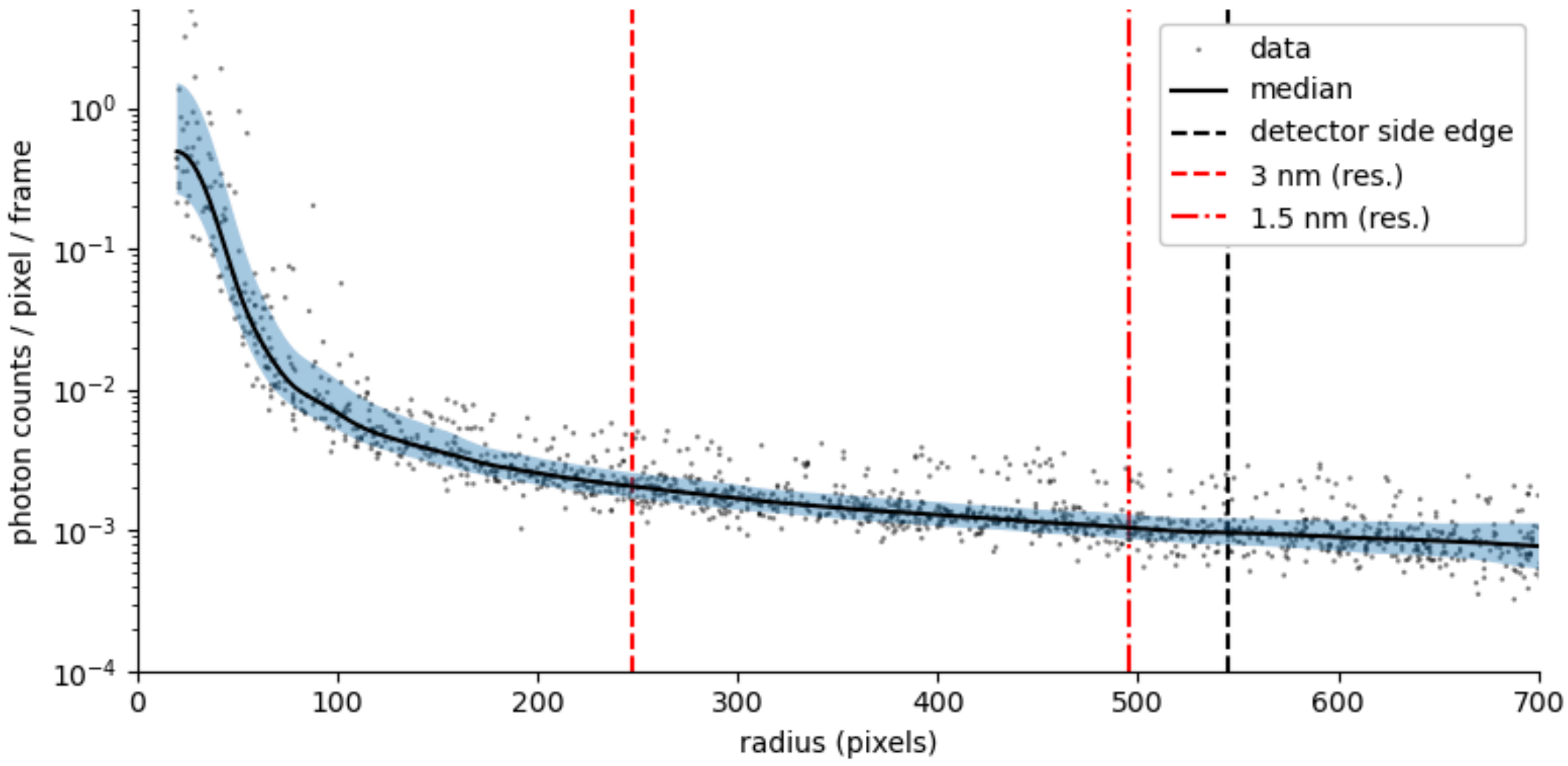

**Extended Data Fig. 2 |** Radial profile of the mean photon counts per pixel per frame within an annulus (centered on the optical axis, 10 pixels thick) for the selected frame list. The median value over all frames is shown in solid black. Lower 16'th to upper 84'th percentiles are shaded in blue. A random selection of individual data points (the mean pixel value within a radial window for a single frame) are shown as black dots.

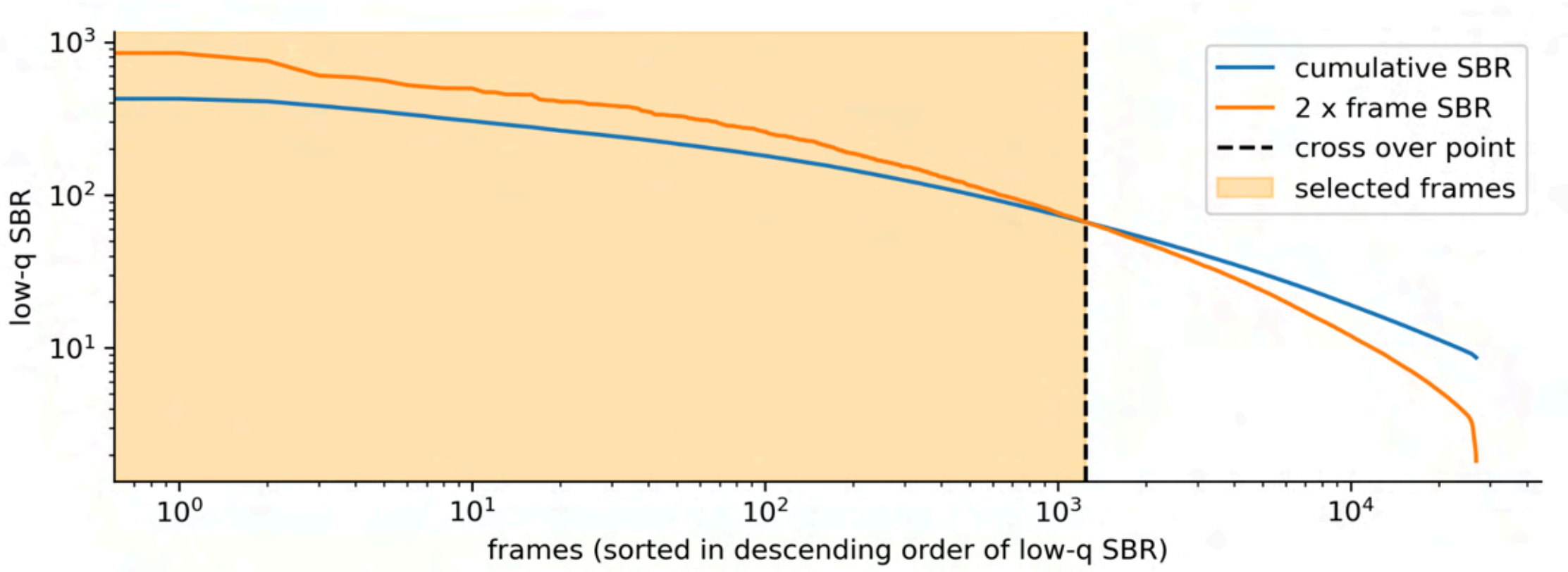

**Extended Data Fig. 3 |** Twice the Signal-to-Background Ratio (SBR) for individual frames, evaluated in a low-q high-signal region of the detector, are shown in orange. The cumulative SBR, for all frames to the left of a given frame index, are shown in blue. Frame indices along the abscissa were sorted by SBR in descending order. The 1,248 frames indices selected for further analysis (out of a total of 26,908) are shaded beige.

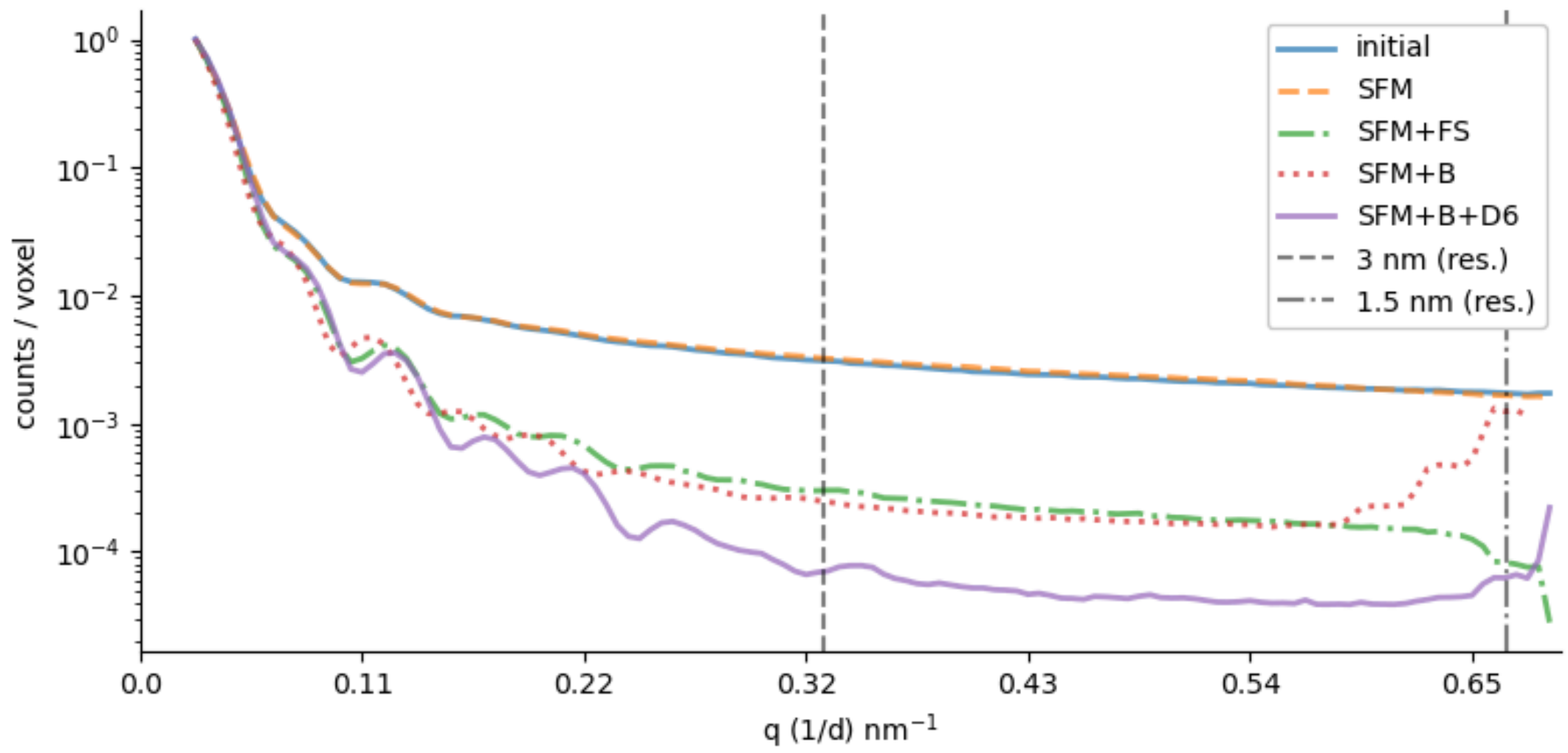

**Extended Data Fig. 4 |** Radial profiles of the 3D diffraction volumes shown in Extended Data Fig. 1. The labels in the plot legend correspond to Scale Factor Modeling (SFM), SBR based Frame Selection (FS), Background aware EMC (B) and D6 point group symmetry (D6). Labels are displayed in order of row number in Extended Data Fig. 1.

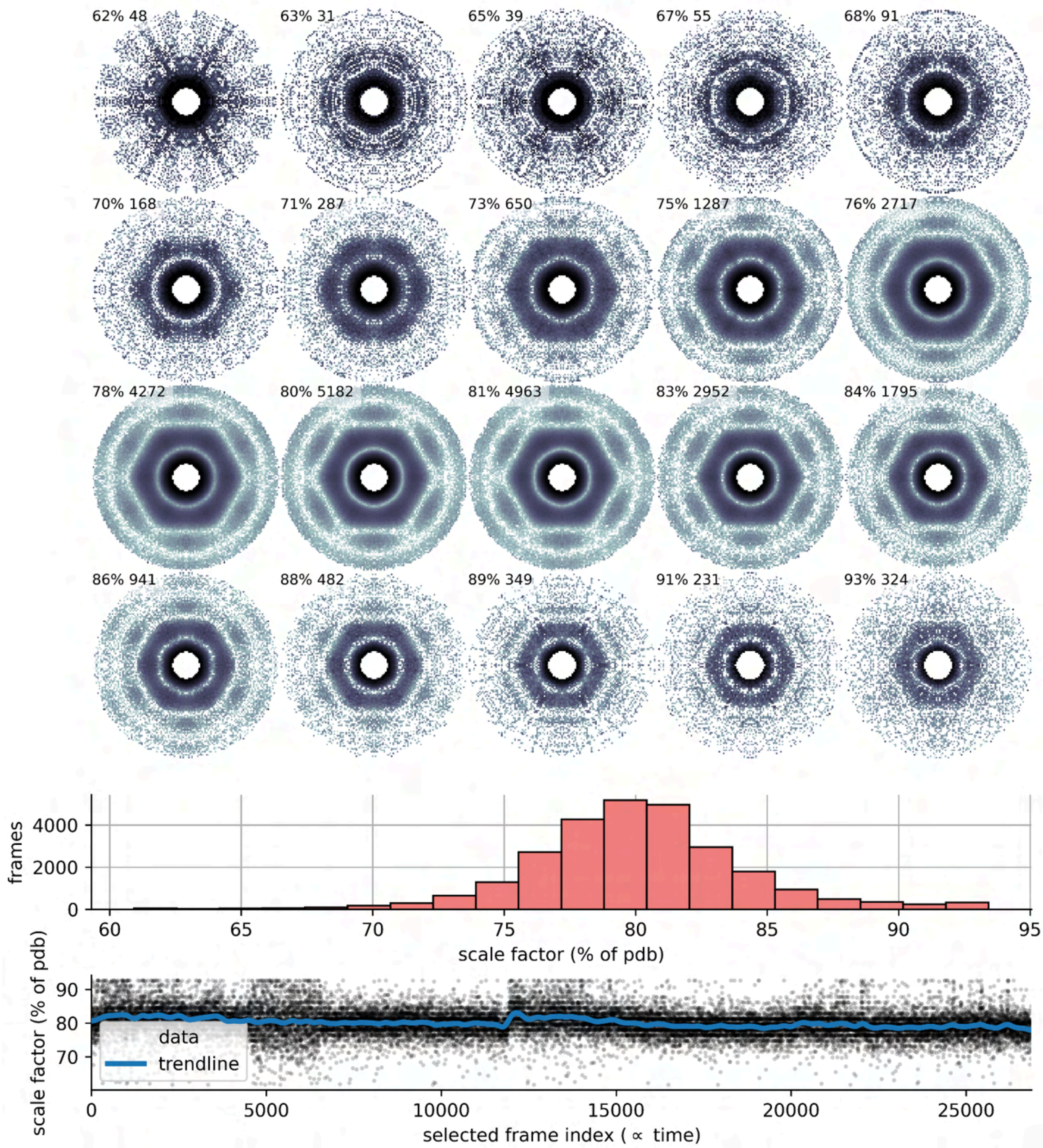

**Extended Data Fig. 5 |** Reconstructed scale factors via EMC. Central sections through the diffraction volumes of the merged frames corresponding to each scale factor in the *x*,*y* plane (top). The scale factor and number of assigned frames are shown in the top left of each image (respectively). Histogram showing the number of diffraction frames assigned to each scale factor bin (middle). The expectation value of the recovered scale factors for each frame as a function of frame index, which is monotonically increasing with respect to frame acquisition time (bottom).

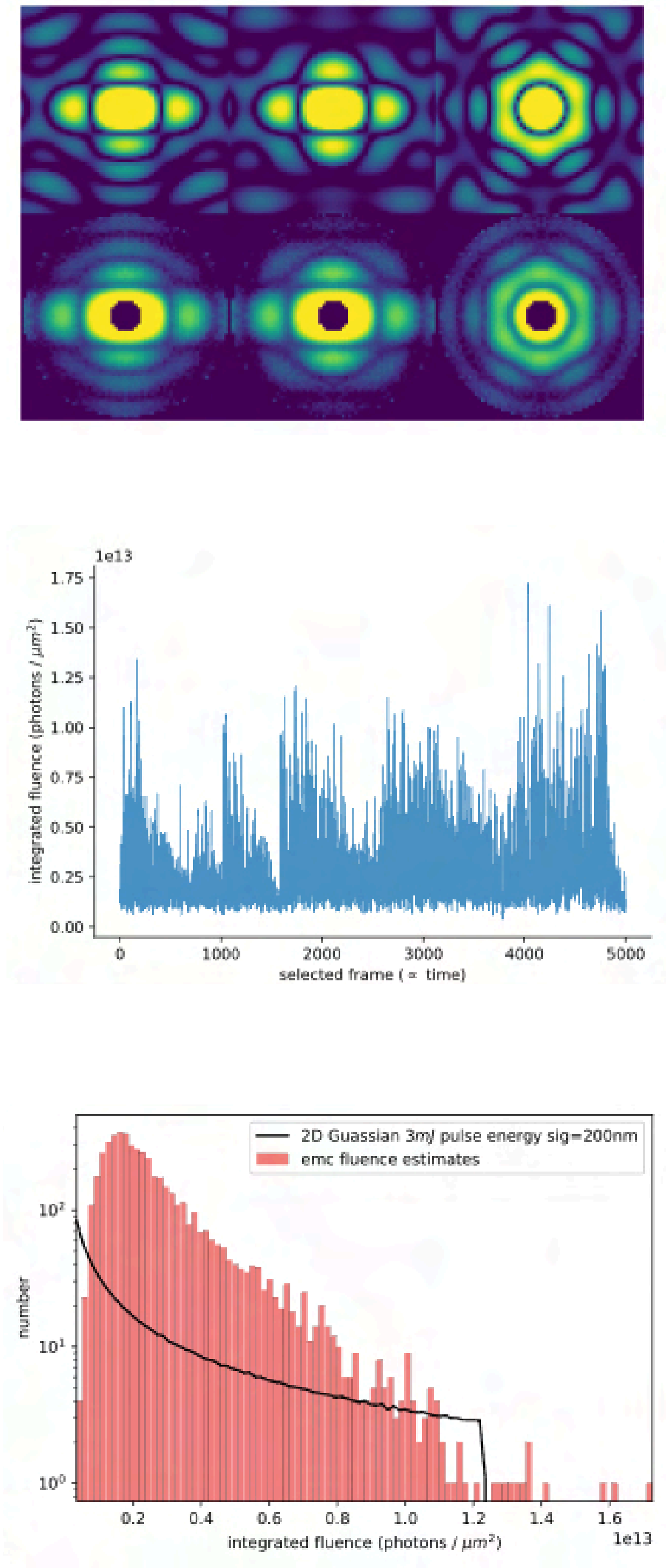

**Extended Data Fig. 6 |** Fluence estimation of the SPB/SFX nanofocus with experimental 3D merged intensity of Ery. With the known electron density of Ery and cold (zero-dose) atomic scattering factors, the fluence is estimated by comparing the observed diffraction intensities to the calculated. **Top:** Orthogonal slices of intensities simulated from PDB ID 2GTL (top) and obtained from the 3D merge of XFEL data (bottom). **Middle:** Intensities of experimental frames included in the 3D-merge sorted as a function of data collection time (intensity drops due to focal spot drift, and right after focal-spot adjustment the intensity increases before degrading again as the beam drifted). **Bottom:** Histogram of pulse energy estimates, shown a highest of $1.75 \times 10^{13}$ ph/μm$^2$.

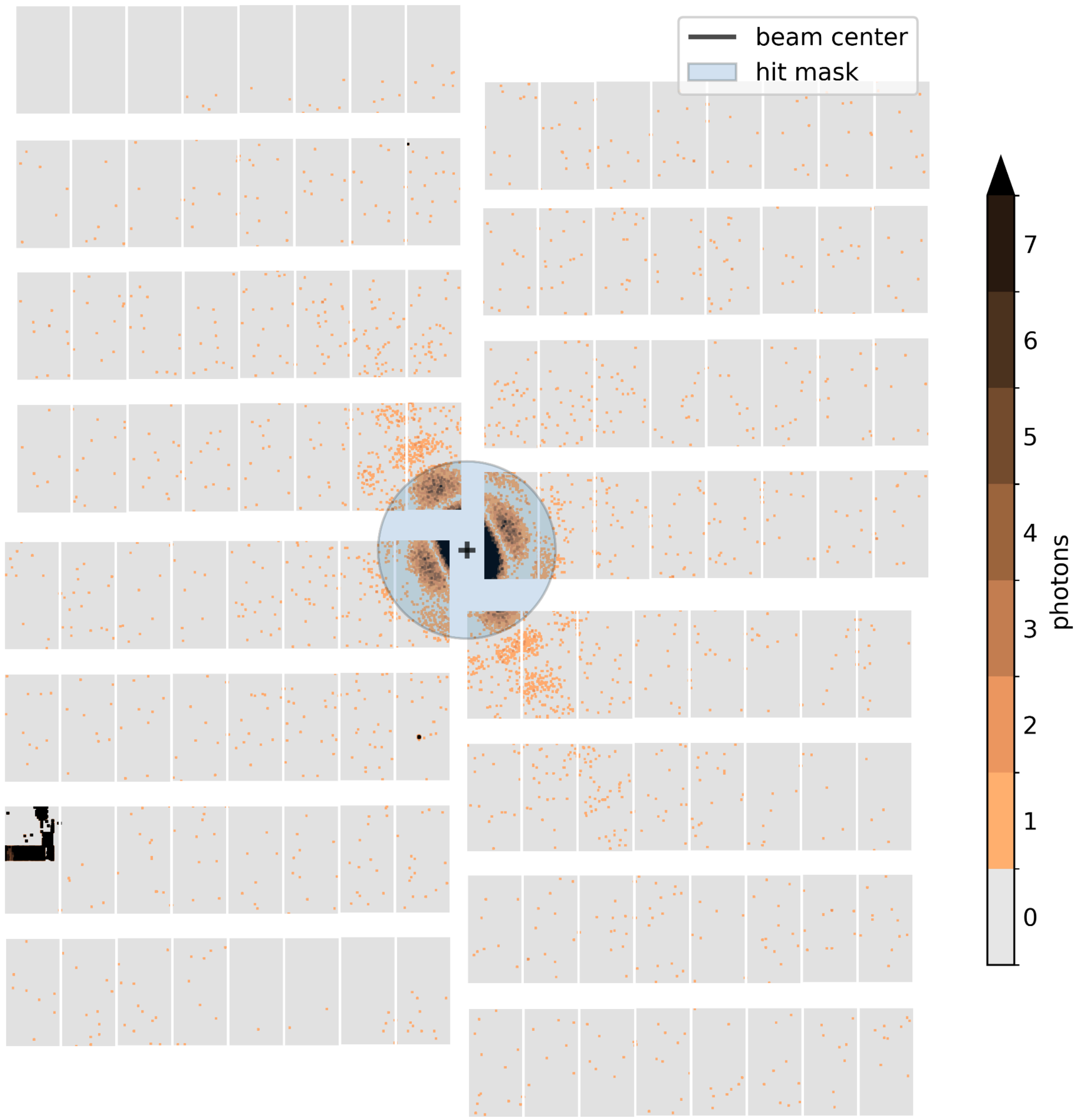

**Extended Data Fig. 7** | Photon signal recorded from a single x-ray pulse by the AGIPD detector. The central region of the detector which was used to calculate the hit-score is shaded in blue.

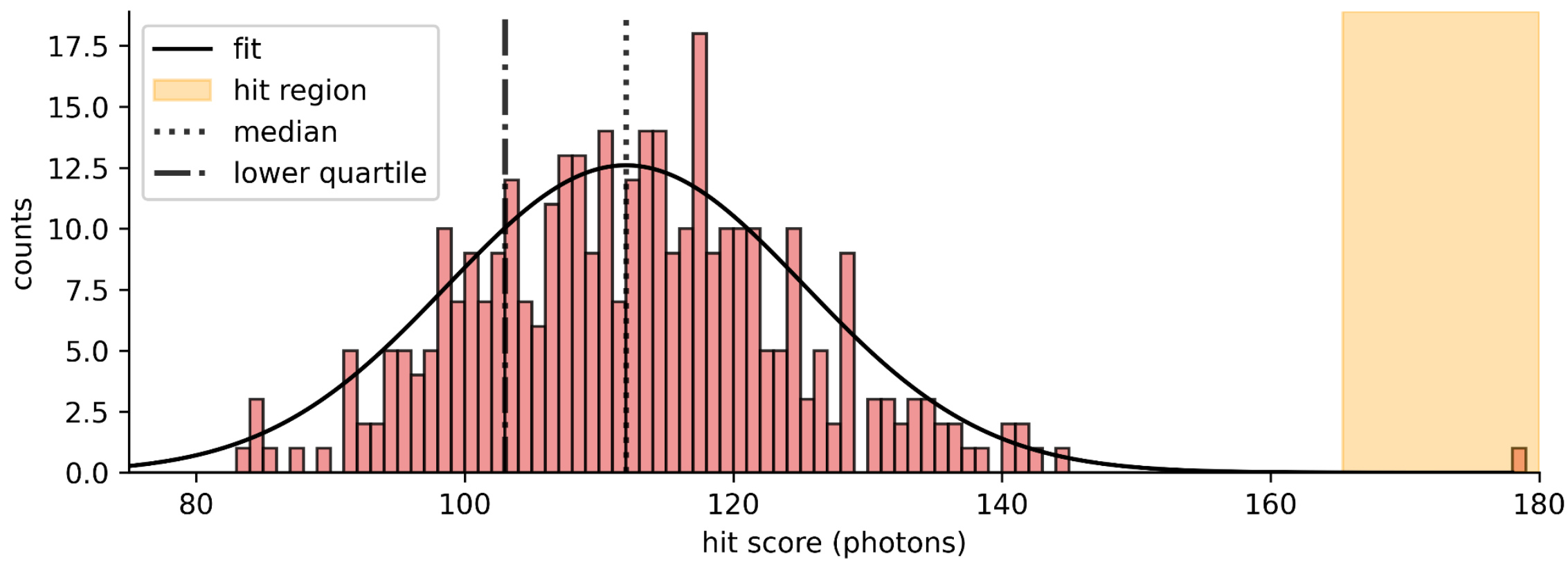

**Extended Data Fig. 8 |** Distribution of hit-scores within a single pulse train. The median (dots) and lower quartile (dash-dot) values used to determine the underlying distribution of the background (line) are also shown.

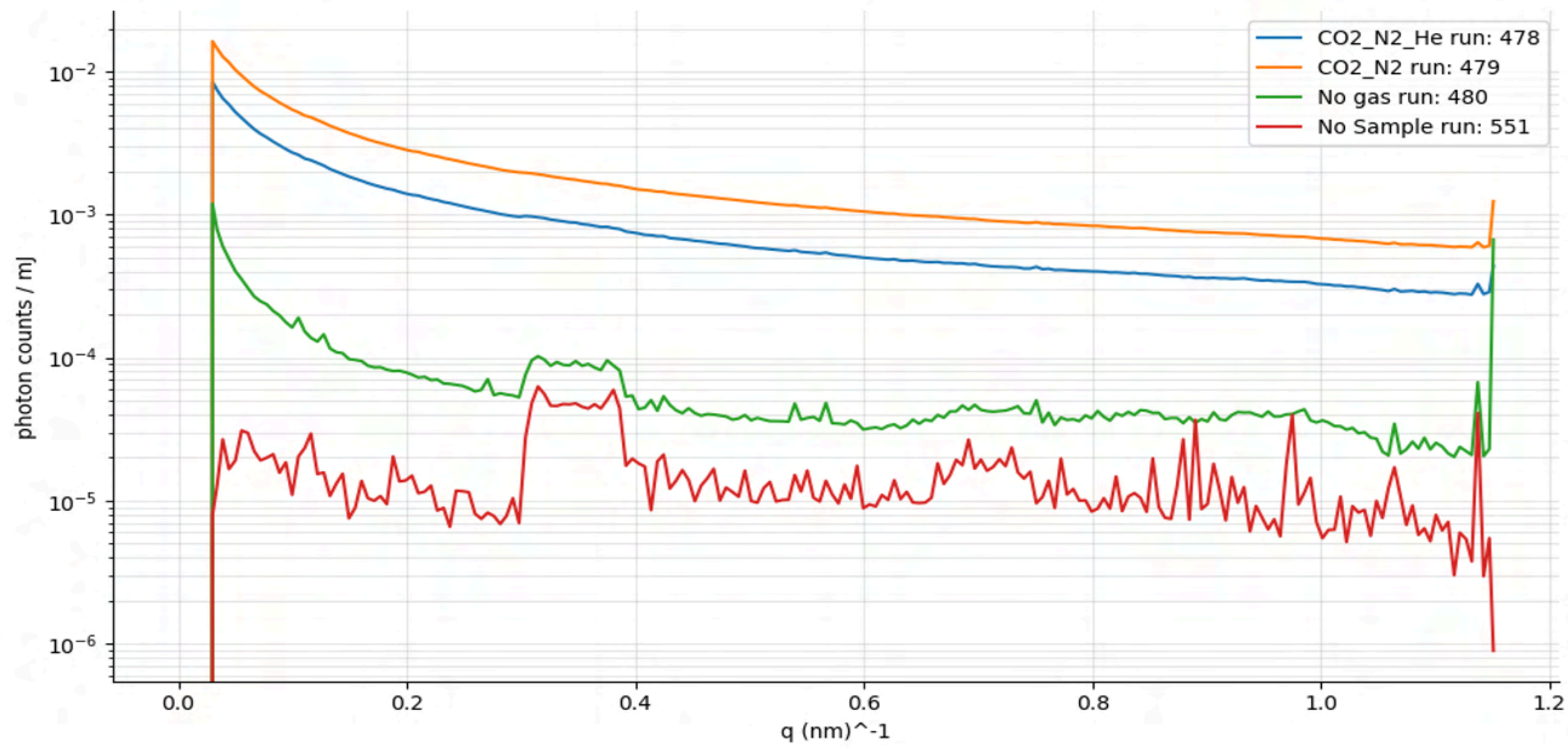

**Extended Data Fig. 9 |** Gas background scattering signal at SPB/SFX. He-injection led to background reduction by 50%. While reduction is possible up to 90% (Fig. S10), in order to keep up the meagre hit-rate, we limited the additional He-supply to achieve 50 % background reduction.

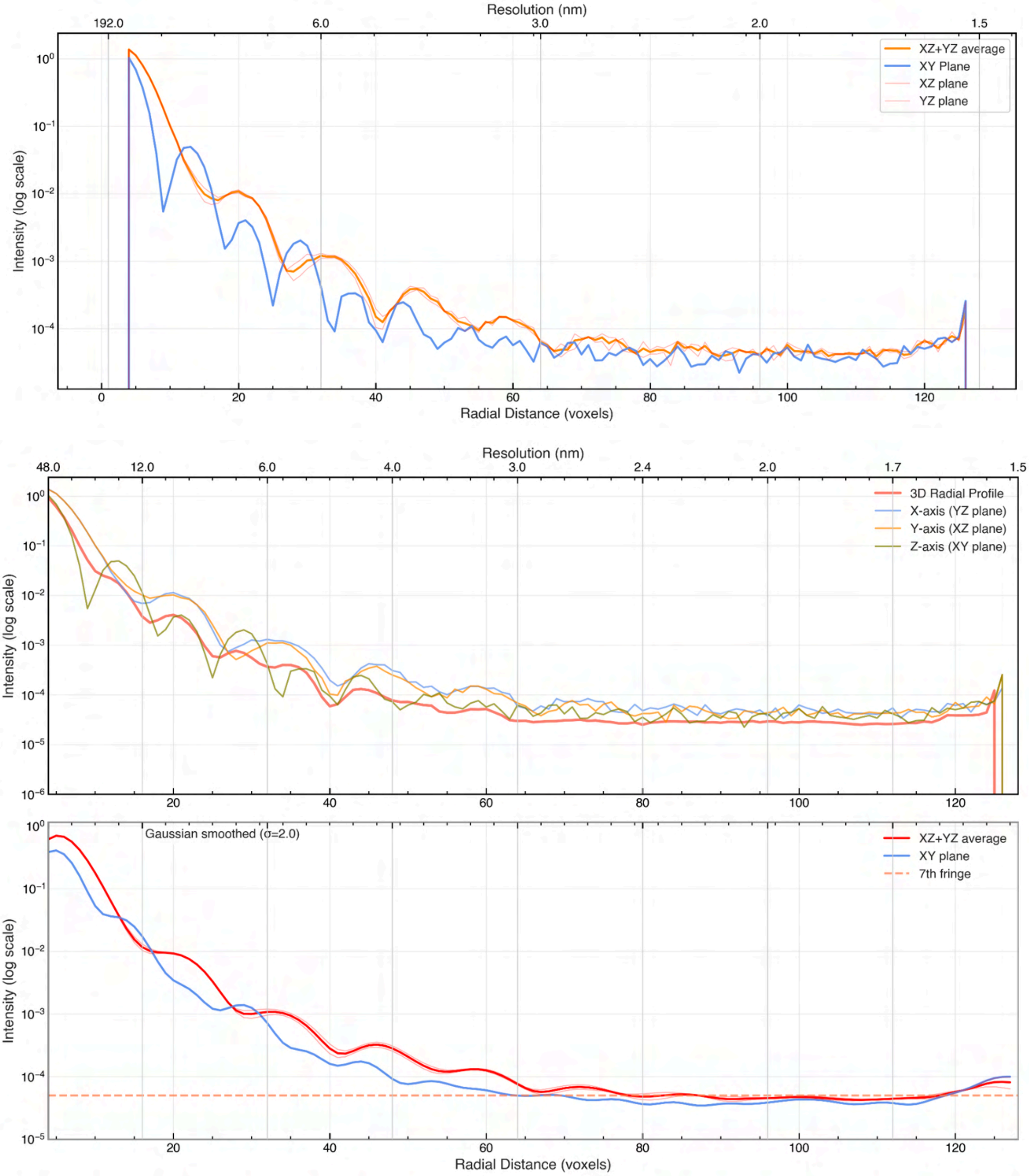

**Extended Data Fig. 10 |** Axial radial profiles of the reconstructed 3D intensity merge of Ery with XFEL-SPI.

Supplementary Information

# Femtosecond Three-Dimensional Imaging of Single-Protein with Hard X-ray Laser

Paul Lourdu Xavier,[1,2,3*] Andrew J Morgan,[1,6] Johan Bielecki,[2] Amit K. Samanta,[1] Simon Dold,[2] Chufeng Li,[1] Xinyue Gao,[15] Spencer K. Passmore,[22] Ruslan P. Kurta,[2] Tim B. Berberich,[2] Emiliano De Santis,[17] Wenhui Zhang,[1] Juncheng E,[2] Faisal H. M. Koua,[2] Romain Letrun,[2] Luca Gelisio,[2] Oleksii Turkot,[2] Egor Sobolev,[2] Lukas V. Haas,[1,4] Jannik Lübke,[1,4] Jingxuan He,[1,4] Madeline B. F. Memovich,[1] Kevin Janson,[1] Stefanie Lenzen,[1,4] Safi Rafie-Zinedine,[2] Mansi Butola,[1] Oleksandr Yefanov,[1] Fabian Trost,[2] Chan Kim,[2] Tomas Popelar,[2] Tokushi Sato,[2] Mohammad Vakilli,[2] Yevheniy Ovcharenko,[2] Rebecca Boll,[2] Sergey Usenko,[2] Thomas M. Baumann,[2] Alberto De Fanis,[2] Björn Senftleben,[2] Matteo Porro,[2,21] Andrea Castoldi,[23,24] Ibrahym Dourki,[2] Cyril Danilevski,[2] Ekaterina Round,[2] Christina Schmidt,[2] Huijong Han,[2] Roberto Alvarez,[7] Ervin S. H. Chia,[8] Cornelia Cazey,[25] Carolin Seuring,[25] Ruojie Sha,[9] Maia Azubel,[10] David A. Bushnell,[10] Sébastien Boutet,[11] Sergey Ovchinnikov,[12,13] Adrian P. Mancuso,[2] Michael Meyer,[2] Richard Bean,[2] Anton Barty,[1] Filipe R.N.C. Maia,[14,18] Richard A. Kirian,[7] Thomas D. Grant,[20] N. Duane Loh,[8] Carl Caleman,[1,16] Erik G. Marklund,[15] Nadia Zatsepin,[22] Andrew V. Martin,[19] Saša Bajt,[1,4] Roger D. Kornberg,[10] Jochen Küpper,[1,4,5] and Henry N. Chapman[1,4,5*]

[1]Center for Free-Electron Laser Science CFEL, Deutsches Elektronen Synchrotron DESY, Notkestrasse 85, 22607 Hamburg, Germany
[2]European XFEL, Holzkoppel 4, 22869 Schenefeld, Germany
[3] Max-Planck Institute for the Structure and Dynamics of Matter, Luruper Chausee, 22761 Hamburg, Germany
[4]The Hamburg Centre for Ultrafast Imaging, Universität Hamburg, Luruper Chaussee 149, 22761 Hamburg, Germany
[5]Department of Physics, Universität Hamburg, Luruper Chaussee 149, 22761, Hamburg, Germany
[6] University of Melbourne, Victoria 3010, Australia
[7]Arizona State University, Tempe, Arizona 85287, USA
[8]National University of Singapore, Singapore 117551, Singapore
[9]New York University, New York City, New York 10003, USA
[10]Stanford School of Medicine, Stanford University, CA 94305, USA
[11]Linac Coherent Light Source, SLAC National Accelerator Laboratory, Menlo Park, CA 94025, USA
[12]FAS, Harvard University, Cambridge, MA 02138, USA
[13]Massachusetts Institute of Technology, Cambridge, MA 02139, USA
[14] Laboratory of Molecular Biophysics, Department of Cell and Molecular Biology, Uppsala University, Husargatan 3 (Box 596), SE-75124, Uppsala, Sweden
[15]Department of Chemistry for Life Sciences, Uppsala University, Box 576, SE-75123, Uppsala, Sweden.
[16]Department of Physics and Astronomy, Uppsala University, Box 516, SE-751 05 Uppsala, Sweden
[17]Department of Physics, University of Rome Tor Vergata and INFN, Via della Ricerca Scientifica 1, 00133, Rome, Italy
[18]NERSC, Lawrence Berkeley National Laboratory, Berkeley, CA 94720, USA
[19]Royal Melbourne Institute of Technology, Melbourne VIC 3000, Australia
[20]Department of Structural Biology, Jacobs School of Medicine and Biomedical Sciences, University at Buffalo, Buffalo, NY USA
[21]Department of Molecular Sciences and Nanosystems, Ca' Foscari University of Venice, 30172 Venezia, Italy
[22]Optical Sciences Centre, Swinburne University of Technology, Hawthorn VIC 3122, Australia
[23]Dipartimento di Elettronica, Informazione e Bioingegneria (DEIB), Politecnico di Milano, Piazza Leonardo da Vinci 32, Milano, 20133, Italy
[24]INFN, Sezione di Milano, Via Celoria 16, Milano, 20133, Italy
[25]Center for Structural Systems Biology (CSSB), Notkestrasse 85, 22607 Hamburg, Germany

*lourdu.xavier.paulraj@desy.de, henry.chapman@desy.de

# *Supplementary notes*

## Comparative spatial analysis of the XFEL-SPI reconstructed density map with the PDB model

**Thresholding and pre-processing of maps**

The XFEL-SPI reconstructed density map and the reference density map generated from PDB 2GTL were analyzed in parallel using an identical pipeline. The reconstructed map was used in its native voxel size 7.21 Å. The reference map was generated from the 2GTL atomic coordinates at 29 Å resolution using the *molmap* command in ChimeraX 1.10.1, with a voxel size of 9.67 Å. Voxel coordinates were converted to lengths using their respective voxel sizes so that all the spatial metrics are reported in Å or $Å^2$.

Binary masks were created by thresholding each map. For the reconstruction, a threshold of 0.02 was applied (in raw density units), while for the reference density map, a threshold of 0.08 was applied. These values were selected to capture the full molecular envelope while excluding low-density noise. After thresholding, the reconstruction contained 18,928 voxels while the reference contained 7,811 voxels. All subsequent analyses—density-weighted PCA, angular symmetry analysis, Voronoi tessellation, radial density profiling, Mahalanobis distance, and distance metrics—were performed on voxels inside these masks, in physical units.

**Density-Weighted Principal Component Analysis and Shape Comparison**

We performed a density-weighted PCA (dw-PCA) on the XFEL-SPI reconstruction—a second-moment analysis in which we use density values as statistical weights in the covariance matrix. Eigendecomposition revealed three principal components PC1, PC2, and PC3 explaining 41.7%, 41.7%, and 16.6% of the spatial variance, respectively ($\lambda_1$ = 3268.2 $Å^2$, $\lambda_2$ = 3264.7 $Å^2$, $\lambda_3$ = 1301.3 $Å^2$). The near-degeneracy of PC1 and PC2 indicates in-plane isotropy, consistent with the six-fold rotational symmetry. The distinct third eigenvalue (~40 % of $\lambda_1$) indicates a flattened, disk-like geometry with the short-axis oriented along z. In case of the reference PDB 2GTL, the dw-PCA yielded eigenvalues ($\lambda_1$ = 4677.4 $Å^2$, $\lambda_2$ = 4675.8 $Å^2$, $\lambda_3$ = 2272.6 $Å^2$), corresponding to explained-variance ratios for PC1, PC2 and PC3 of 40.2%, 40.2% and 19.5%, respectively. The same near-degeneracy of PC1 and PC2 is observed, confirming in-plane isotropy in the reference structure as well, and the distinct third eigenvalue (~49% of $\lambda_1$) again indicates a flattened, disk-like geometry with the short-axis along z, consistent with D6-symmetric architecture expected for Ery and with the angular analysis below. Both the reconstructed map and the reference agree on the patterns of eigenvalue degeneracy and the quantitative shape classification (Fig. S12 and S13 for the PC1-PC2 projections). The absolute eigenvalues of the reference are slightly larger across all three axes, with the reconstructed map spanning approximately 85% of the reference along its principal linear dimensions, consistent with the expected gas-phase compaction.

| Parameter | XFEL-SPI reconstruction | 2GTL PDB reference |
|---|---|---|
| $\lambda_1$ (in-plane, $Å^2$) | 3,268 | 4,677 |
| $\lambda_2$ (in-plane, $Å^2$) | 3,265 | 4,678 |
| $\lambda_3$ (axial, $Å^2$) | 1,301 | 2,273 |
| $\lambda_3/\lambda_1$ | 0.398 | 0.486 |
| Voxels above threshold | 18,928 | 7,811 |
| Voxel size (Å) | 7.21 | 9.67 |
| Physical volume ($Å^3$) | 7.09 x $10^6$ | 7.06 x $10^6$ |

**S.Table 1 |** Comparison of the dw-PCA parameters between XFEL-SPI reconstructed map and the PDB model.

The reconstruction is compacted relative to the reference along all principal axes, with the largest compaction along the axial direction: $\sqrt{\lambda_1}$ and $\sqrt{\lambda_2}$ indicate linear compaction in-plane of about 16%, while $\sqrt{\lambda_3}$ indicates compaction of 24% along the axial direction—overall indicating anisotropic compaction with gas-phase compaction more pronounced along the six-fold symmetry axis.

**Voronoi Tessellation Analysis**
To characterize the local packing geometry of the projected density, we performed a two-dimensional Voronoi tessellation on the unique XY projections of the voxels above the density threshold. Duplicate XY coordinates were removed using numpy.unique, reducing 18,928 voxels to 1,026 unique points for the reconstruction (threshold 0.02) and 7,811 voxels to 655 unique points for the 2GTL reference (threshold 0.08). A small random displacement (0.001 x voxel size, fixed seed 42) was applied before tessellation to avoid degenerate cells, and the reconstruction point set was subsampled to 8,000 points using the same seeded random state for reproducibility. Cell areas were computed from the shoelace formula after sorting vertices counter-clockwise around the centroid. Cells with non-finite or artifact areas were excluded. The coordination number of each point was defined as the number of unique Voronoi neighbours.

The mean coordination number was 6.0 for both the reconstruction and the 2GTL reference map, and the coordination distributions in both the maps peaked at 5-6 neighbours (42.6 % in reconstruction and 48.5% in PDB). This is the signature of locally hexagonal packing, consistent with the known D6 architecture of Ery in which 12 globin dodecamers assemble around a central linker core. The preservation of mean coordination 6.0 indicates that the local hexagonal ordering of the reference is retained through the gas-phase measurements. Small deviations from the reference are observed in the tails of the coordination distribution (the reconstruction shows a slightly broader range of coordination numbers, 3–15 vs. 3–14 ) (see Fig. S12 and S13) consistent with the local distortions induced by axial compaction and likely by residual non-volatile material retained after desolvation.

| **Metric** | **XFEL-SPI reconstruction** | **2GTL PDB reference** |
|---|---|---|
| Voxels above threshold | 18,928 | 7,811 |
| Unique XY points | 1,026 | 655 |
| Cells analyzed | 7,930 | 7,781 |
| Mean cell area (Å$^2$) | 8.4 | 14.6 |
| Mean coordination | 6.0 | 6.0 |
| Modal coordination | 5 (23.7%) | 5 (26.5%) |
| Coordination range | 3–15 | 3–14 |
| Fraction with 5-6 neighbours | 42.6 % | 48.5% |
| Fraction with coordination = 6 | 18.9% | 22.0% |

**S.Table 2 |** Comparative Voronoi tessellation analysis between XFEL-SPI reconstructed map and the PDB model.

**Mahalanobis distance distribution**

To characterize the radial mass distribution in the shape's own metric, we computed the density-weighted Mahalanobis distance[1] $D_M = \sqrt{(r - c)^T \Sigma^{-1} (r - c)}$ for every voxel above the density threshold,

Where r is the position of the voxel, **c** is the density-weighted center of mass and **Σ** is the density-weighted covariance matrix from the PCA. The distribution of $D_M$ across voxels acts as a shape-corrected radial mass profile, as the inverse covariance matrix $\Sigma^{-1}$ whitens the particle's anisotropy; voxels at the same $D_M$ sit at the same shape-relative distance from the center regardless of which spatial axis they lie along.

In the Mahalanobis distance histogram, the 2GTL reference exhibits a bimodal distribution (mean $D_M$ = 1.69, max = 2.47 ), characterized by a main peak near $D_M \approx 1.8$ and a distinct low-$D_M$ shoulder between 0.5 and 1.0. The two modes likely correspond to the two structural components of Ery: the central linker core (low $D_M$) and the surrounding hexagonal globin ring (main peak). In contrast the XFEL-SPI reconstruction shows a unimodal distribution (mean $D_M$ = 1.85, max = 2.75), peaked near $D_M \approx 2.1$ with no low-$D_M$ population. The absence of the low-$D_M$ mode might indicate that the distinct central density present in PDB 2GTL is absent in reconstructed density, even as the overall particle remains compact. This is corroborated by MD simulations, where the linker units' finger-like protrusions at the center shrivel up and move inward toward the hexagonal ring, which is consistent with the cavity ratio increasing from 0.40 (reference) to 0.83 (reconstruction).

Plotting local density against $D_M$ has provided further spatial insights on the structural compaction. In the 2GTL reference, the highest density is located near the outer periphery ($D_M \approx 1.8$), consistent with the tightly packed hexagonal globin ring, and the density decreases smoothly toward the center with a low-$D_M$ shoulder between 0.5 and 1.0, reflecting the extended, continuous linker core. The XFEL-SPI reconstruction exhibits a markedly different distribution. The highest-density voxels reach much further inward toward lower $D_M$, forming a dense, compact core, while the outer periphery ($D_M \approx 2.5$) becomes notably sparser. Furthermore the peak absolute density in the reconstruction (0.40) is significantly higher than in the reference (0.30).

Taken together, these observations indicate a two-sided gas-phase compaction. The sparse periphery reflects outer globin ring contracting inward, while the depleted center and the emergence of denser intermediate core indicate that the extended linker units have retracted. This is corroborated by MD simulations, where the linker units' finger-like protrusions at the center shrivel up and move inward toward the hexagonal ring.

**Euclidean and Manhattan distances**

To characterize the overall spatial extent and internal density distribution of the reconstructed particle, we computed both the Euclidean and Manhattan (city-block) distances between all thresholded voxels.

The Euclidean distance between any two voxels i and j is defined as:

$$d_{ij} = \left\|r_i - r_j\right\|2 = \sqrt{(x_i - x_j)^2 + (y_i - y_j)^2 + (z_i - z_j)^2}$$

The maximum Euclidean distance ($d_{max}$) identifies the farthest two points in the particle. For 2GTL reference, the $d_{max}$ was 300.3 Å, whereas the XFEL-SPI reconstruction yielded a smaller maximum extent of 272.6 Å.

While Euclidean distance measures the straight line separation, the Manhattan distance ($d^{(1)}$) measures the distance along a grid, effectively summing the absolute difference along each co-ordinate axis.

$$d_{ij}^{(1)} = \sum_{k=1}^{3} \left|r_{i,k} - r_{j,k}\right|$$

The maximum Manhattan distance across all pairs was 447.0 Å for the reconstruction and 502.7 Å for the reference PDB. And when evaluating the Manhattan distance specifically between the two points that were farthest apart in Euclidean space, the distances were 439.8 Å for reconstruction and 454.3 Å for the reference.

The ratio of these two metrics ($d^{(1)}/d$) provides a measure of path of tortuosity. The ratio of Manhattan-to-Euclidean distance between this farthest pair was ≈ 1.61 for the reconstruction and ≈ 1.51 for the reference.

Taken together with other spatial metrics, this ratio indicates that the internal trajectory across the particle deviates significantly from a straight line. This is consistent with a ring-like or porous structure where mass is distributed across the hexagonal body rather than in a solid-core. Importantly, the preservation of a high tortuosity ratio (1.61 vs 1.51) in the reconstruction demonstrates that the overall topology and ring-like architecture of the molecule are conserved, even as the overall envelope compacts.

**Radius of gyration**

To provide an outlier-insensitive measure of the particle's overall size, radius of gyration ($R_g$) was calculated as follows.

$$R_g = \sqrt{\frac{1}{N}\sum_{i=1}^{N}\left\|r_i - \bar{r}\right\|^2}$$

where $\bar{r}$ is the unweighted geometric center of the voxels. The $R_g$ was 107.8 Å for the 2GTL reference and 97.2 Å for the XFEL-SPI reconstruction. The consistent reduction in maximum extent ($d_{max}$) and $R_g$ in the reconstruction (by ≈ 10% in both metrics) aligns with the compaction observed in the PCA and Mahalanobis analyses, indicating a globally smaller envelope in the gas-phase. However, because the maximum Euclidean distance is dominated by the in-plane extent (which compresses less than the axial direction), the reduction in maximum linear extent is less pronounced than the 24% reduction observed along the axial principal axis (as observed in dw-PCA).

See Fig S27-28 collage plots for dw-PCA, Voronoi tessellation, Mahalanobis, Euclidean, Manhattan, and radius of gyration analysis between PDB model and XFEL-SPI reconstruction. FigS23-25 shows slices through the density maps of PDB model (S23), XFEL-SPI reconstruction (S24), and 500-ns gas-phase compacted (S25) models and S26 shows the data-distribution of reconstructed map and how threshold levels used as in ChimeraX.

**Angular distribution and rotational symmetry analysis**
To characterize the global rotational symmetry of the particle in the projection plane, we analyzed the angular distribution of the threshold density. Voxels above the density threshold were projected onto the XY plane. To prevent multiple Z-slices from artificially weighting a single in-plane coordinate, duplicate XY positions were removed using numpy.unique.

The azimuthal angle $\theta_i$ of each unique projected point was computed relative to the density-weighted center of mass in the XY plane ($C_{xy}$):

$$\theta_i = atan2(y_i - c_y, x_i - c_x)$$

The angles were binned into a histogram at 1° resolution to visualize the distribution. To quantitatively assess the symmetry without being biased local shot noise in individual histogram bins, we computed the rotational power spectrum *P(k)* of the angular distribution for harmonic orders k = 1 through 20:

$$P(k) = \left| \frac{1}{N} \sum_{j=1}^{N} e^{ik\theta_j} \right|^2 . N$$

where N is the total number of unique projected points. A particle with perfect m-fold rotational symmetry will exhibit a dominant peak in *P(k)* at harmonic order k=m.

**Radial density profile**

To characterize the internal density distribution and the central cavity, the mean voxel density was computed as a function of distance from the density-weighted 3D center of mass. The profile was binned to 60 radial shells. The cavity ratio was defined as the ratio of the mean density in the interior region to the mean density in the surrounding rim (the 40th to the 70th percentile of the radial range).

The rotational power spectrum of XFEL-SPI reconstruction exhibits a clear, dominant peak at harmonic order k = 6, with the power at order k = 3 fully suppressed.

The rotational power spectrum confirms a dominant six-fold symmetry in both XFEL-SPI reconstruction and the 2GTL reference, with harmonic order 6 dominating and order 3 fully suppressed. Concurrently the radial density profile reveals that the deep central cavity of the solution-state crystal structure (ratio 0.40) is largely filled or compacted in the gas-phase (ratio 0.83). Together, these analyses demonstrate that the global hexagonal architecture of Ery is preserved during the gas-phase compaction, even as the central pore collapses (as seen in MD simulations too).

See Fig S29 for angular distribution and radial density profiles.

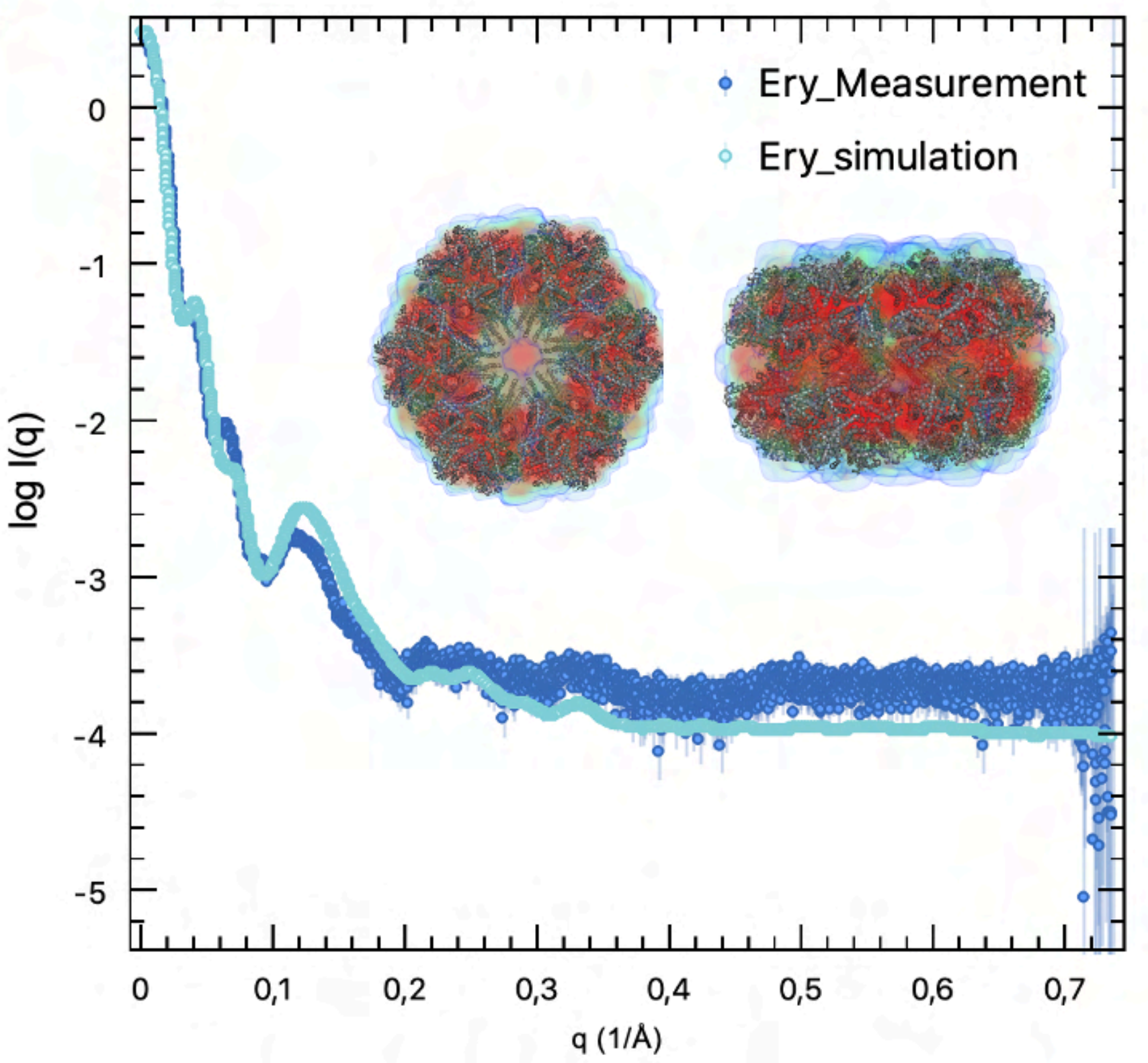

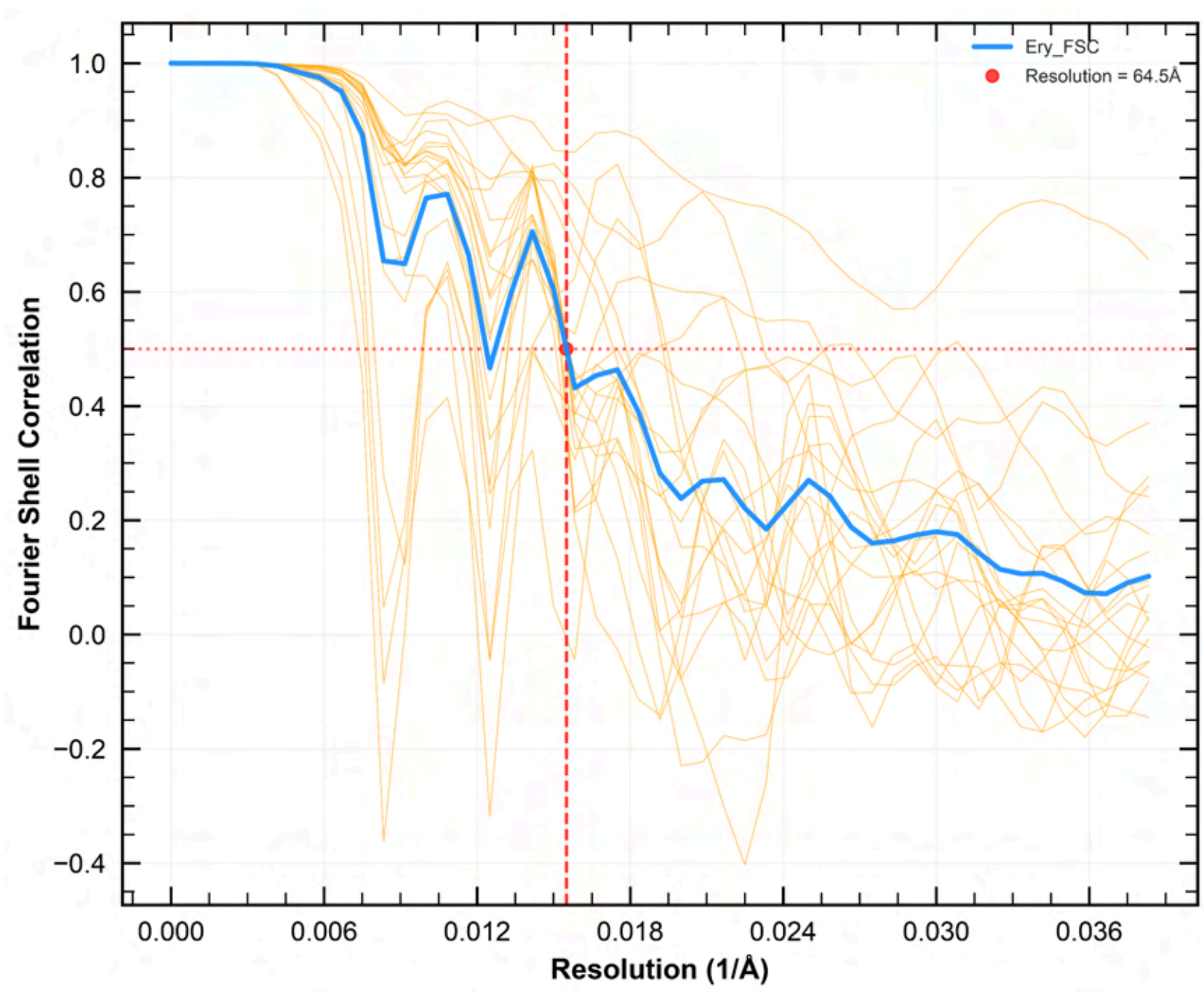

**Fig. S1 | Top:** Synchrotron SAXS pattern of Ery matching simulation at 10.5 keV. Inset: DENSS based 3D reconstruction of Ery at two different orientations (transparent envelope) from the SAXS pattern superimposed with the PDB 2GTL (cartoon). Excellent agreement between the reconstructed envelope and the atomic model validates the high structural fidelity and conformational integrity of the sample. **Bottom:** Fourier-Shell Correlation (FSC) curve, with the resolution estimated at the FSC = 0.5 threshold, indicates the resolution is ~64.5 Å.

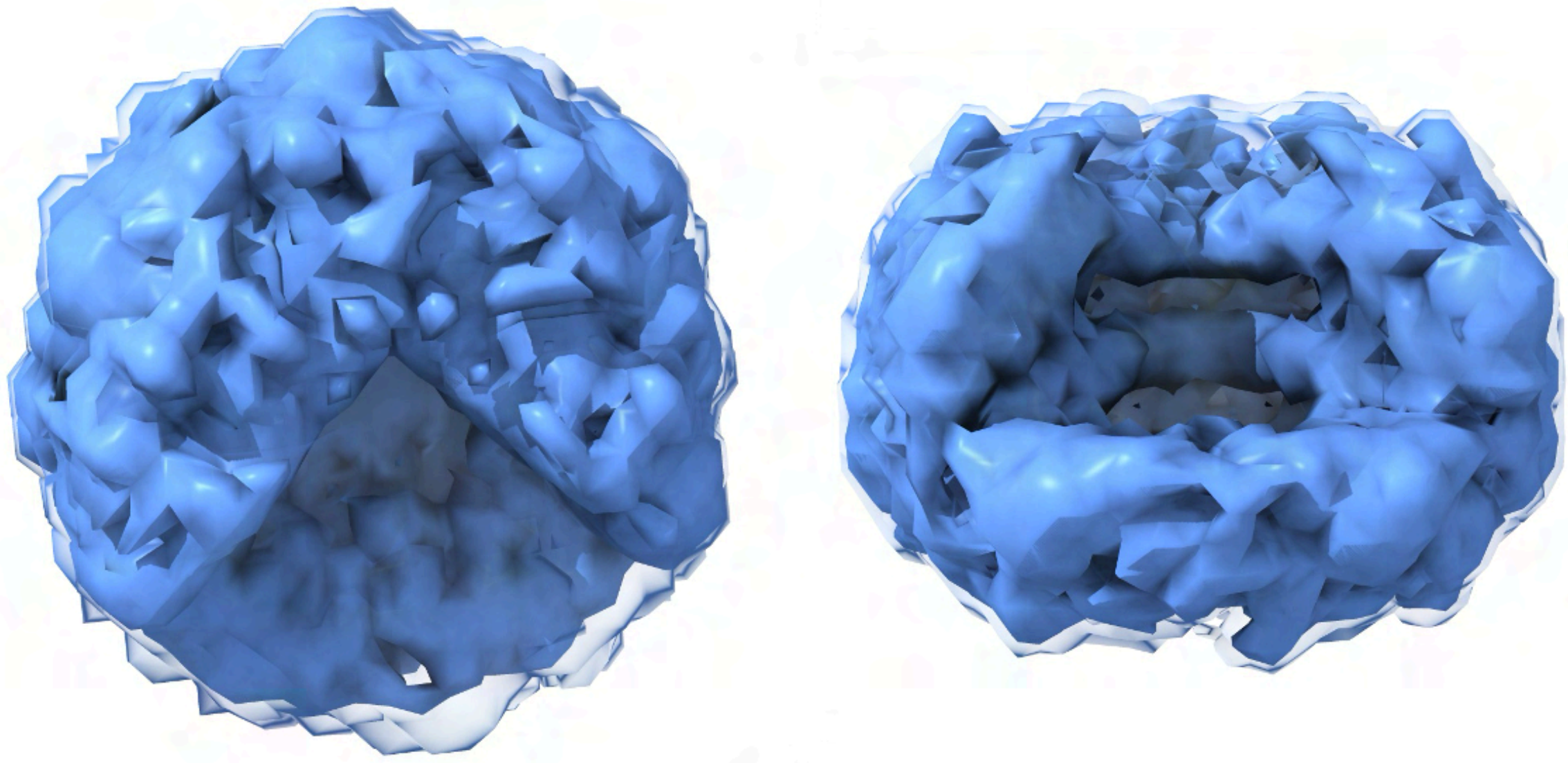

**Fig. S2 |** Octant cut through the SAXS density map of Ery (inset in the plot above) at 65Å resolution obtained through iterative phasing  from in-solution synchrotron SAXS at 10.5 keV at two different thresholds of isosurface density view. The transparent outer envelope (cornflower blue transparent, 1 $\sigma$) and the inner octant cut showing internal cavities (cornflower blue opaque, 3.5 $\sigma$).

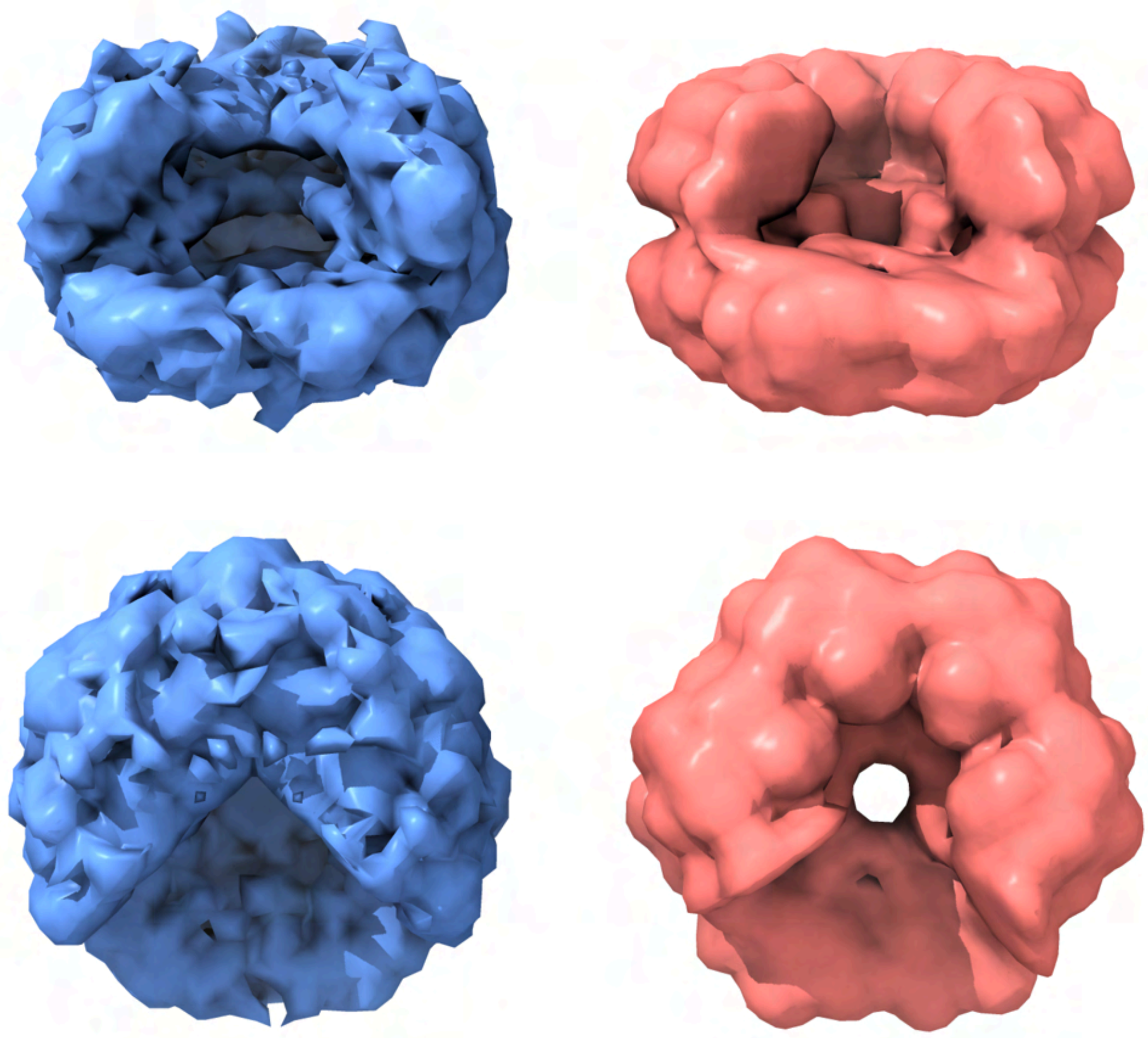

**Fig. S3 |** Comparison of octant cuts of density maps from synchrotron SAXS at 10.5 keV  (cornflower blue)  and simulated-from-PDB (salmon) in two different orientations. Except for the central hole, the in-solution-SAXS-derived density map through iterative phasing shows reasonable similarity with the PDB-derived simulated low-resolution density map indicating the sample quality.

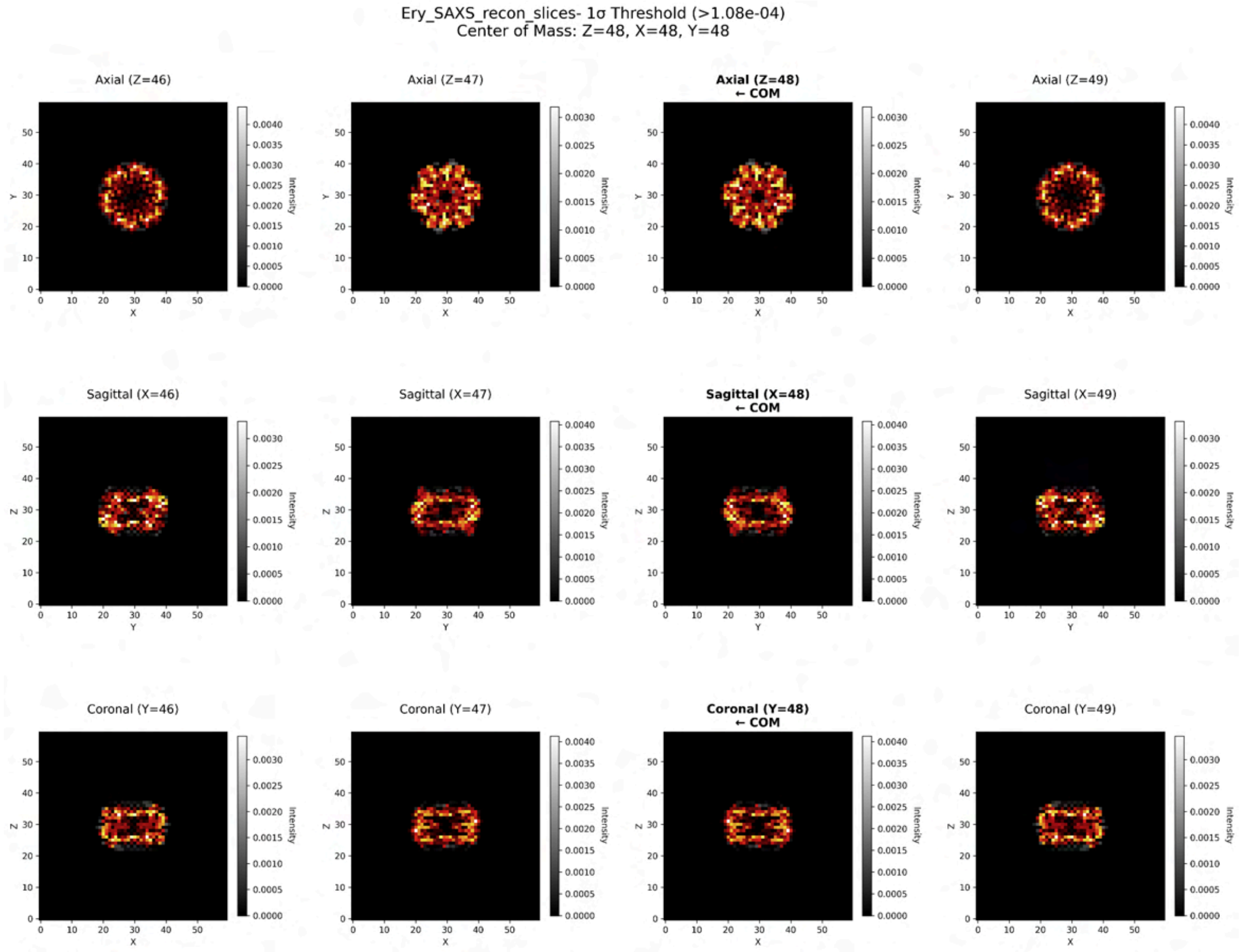

**Fig. S4 |** Slices of the reconstructed map from synchrotron SAXS at 10.5 keV

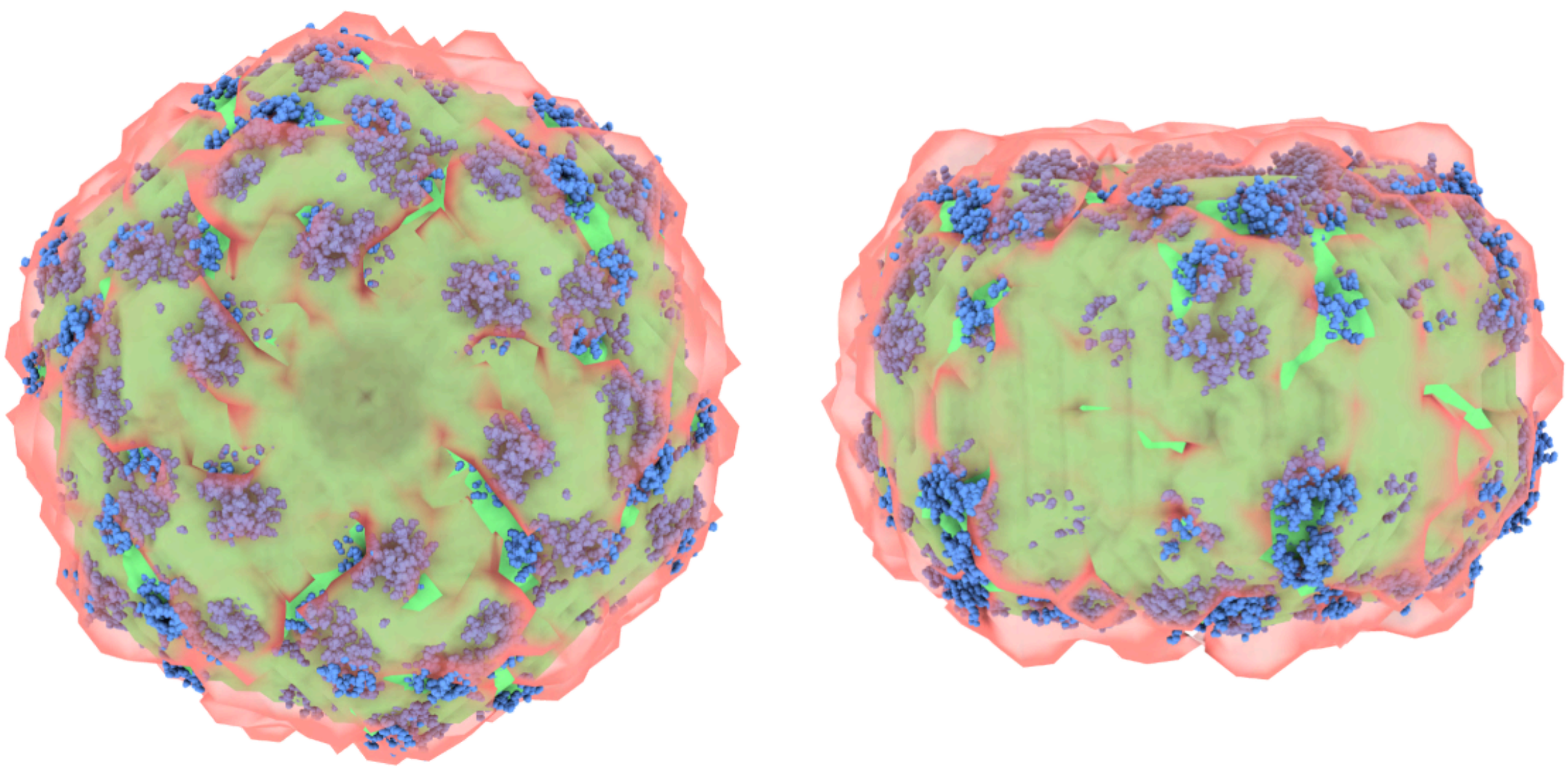

**Fig. S5 | In-solution vs aerosol density maps with x-rays:** Comparison of molecular envelope of the density maps obtained from synchrotron SAXS (transparent salmon) and XFEL-SPI reconstruction (forest-green) with PDB model (cornflowerblue) as the ground truth in two-different-orientations. Map alignment and fits were carried out automatically in ChimeraX 1.10.1

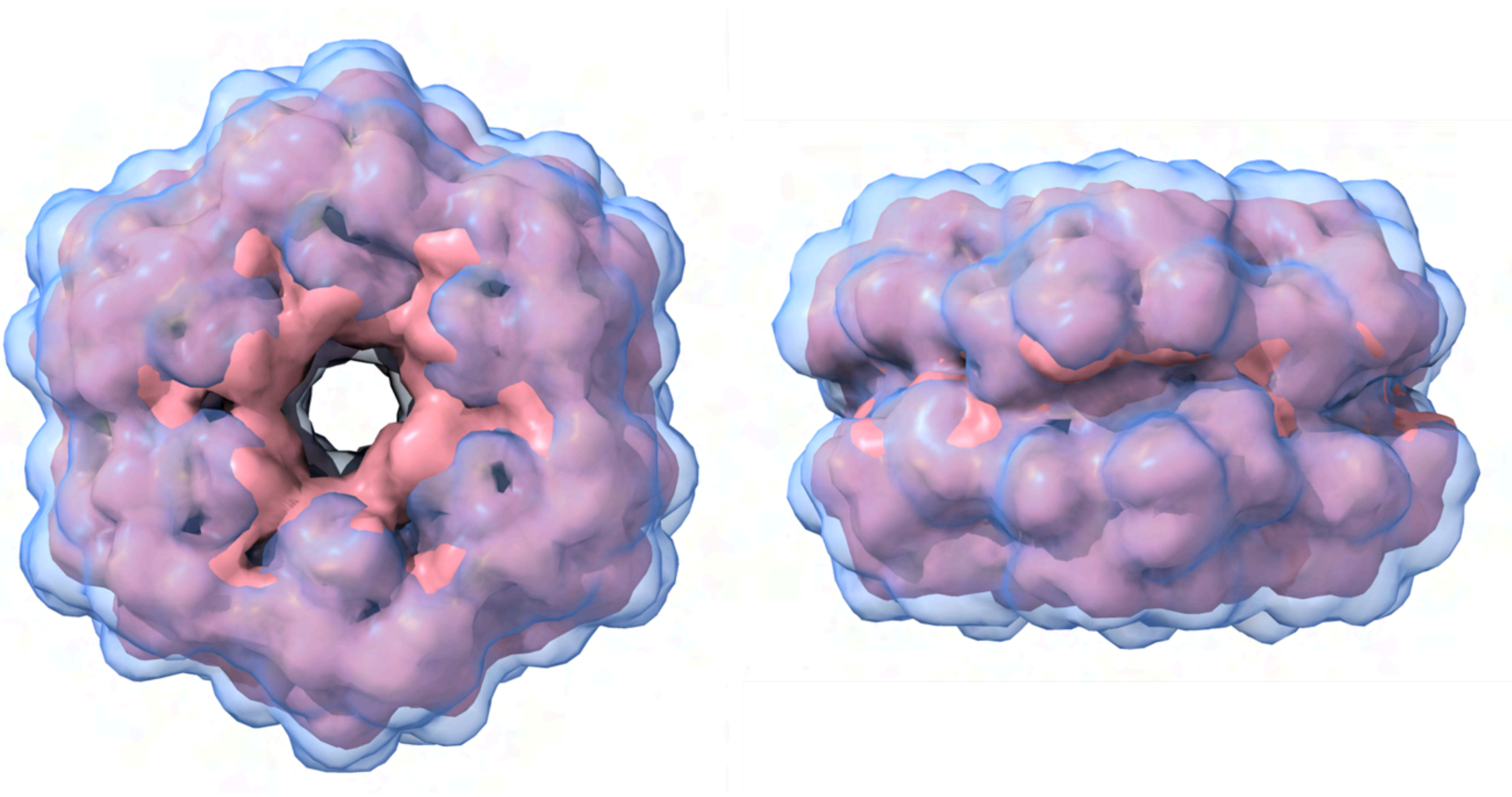

**Fig. S6 |** Molecular envelope comparison between PBD and MD-simulated compacted model at 20Å resolution: Autofit by ChimeraX 1.10.1. Isosurface density comparison shows the compaction: Ery (ground truth with 2GTL PDB model, cornflowerblue transparent) and 0.5 nm water-shell after 400 ns in vacuum (salmon opaque, molecular map made from the 400ns-compacted Ery PDB). The residual water layer present after 400 ns is not shown here. Corresponding Rg plot in S8

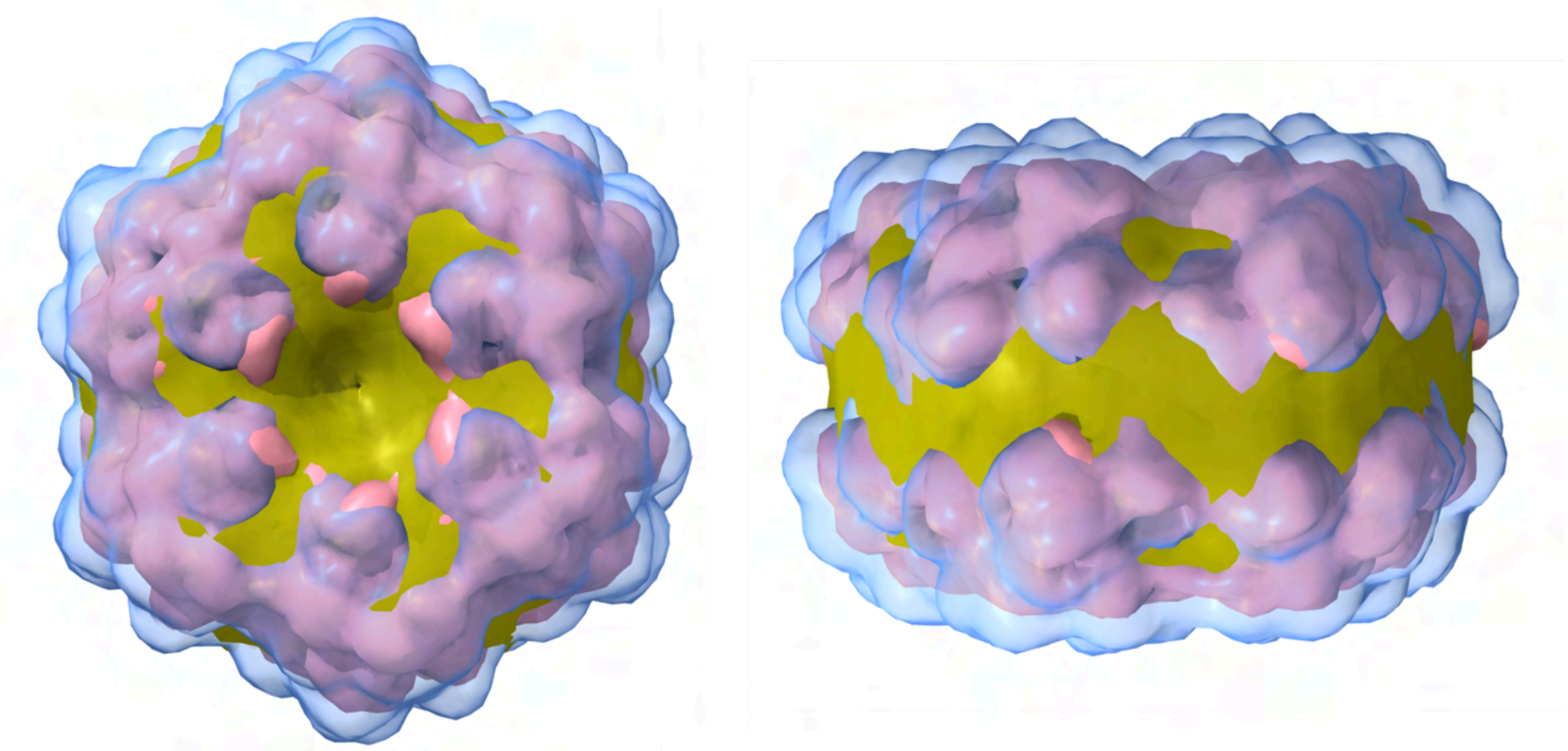

**Fig. S7 |** Isosurface density comparison between the ground truth pdb model, gas-phase compacted model at 400 ns, and the XFEL-SPI reconstructed model, at rmslevel 0.5 on the same grid. Ground truth with 2GTL PDB model (cornflowerblue transparent) and with 0.5 nm water-shell after 400 ns in vacuum (salmon opaque) (both 20Å resolution molecular maps). Reconstructed density from XFEL-SPI (olive-green).

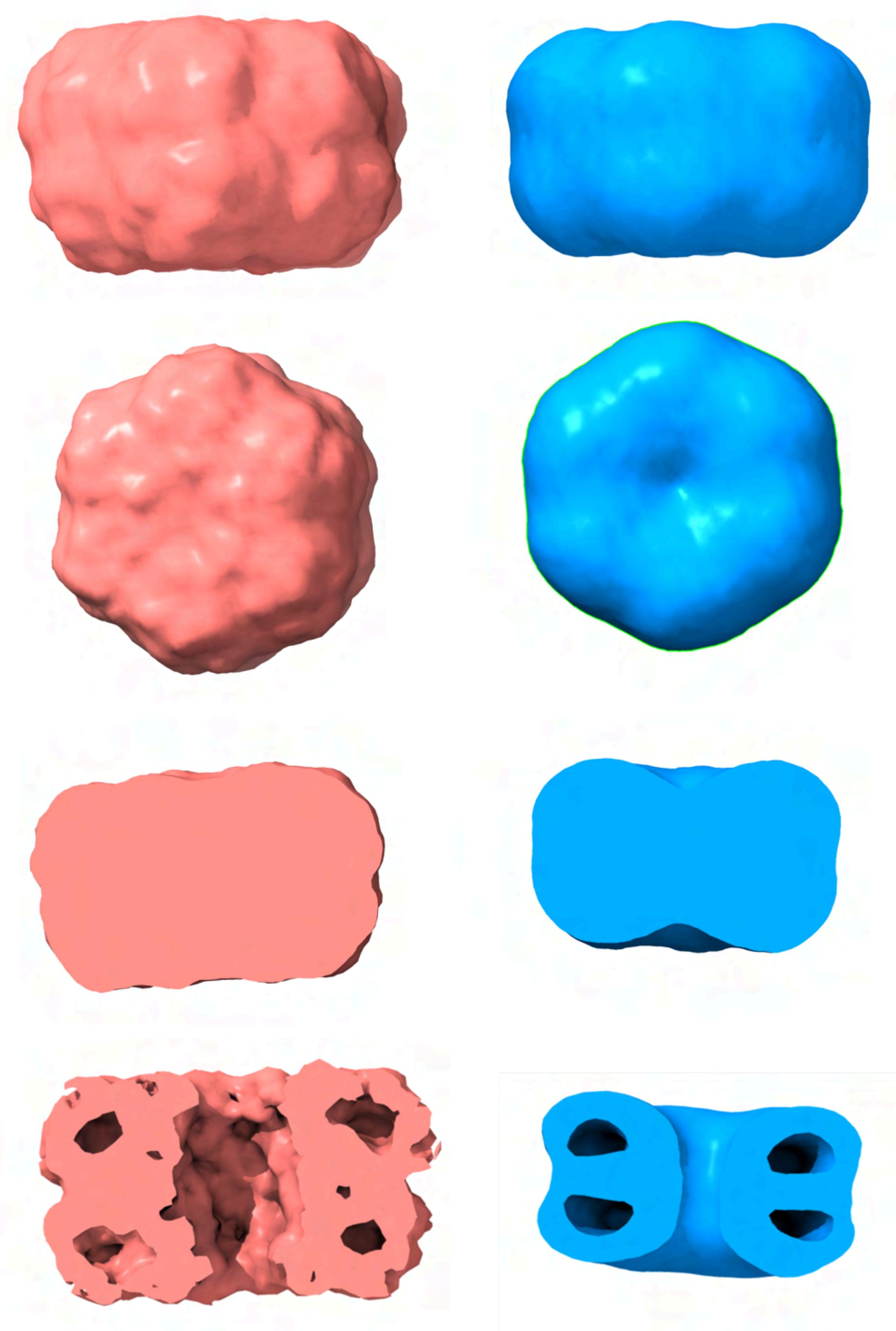

**Fig. S8 | Side-by-side comparison** of molmap of gas-phase-compacted Ery (initial condition with 0.5 nm water layer) at 20 Å with residual solvent (salmon) and the XFEL-SPI reconstruction (blue) on the same grid (top three rows at rmsLevel 1) and bottom row showing cross section at absolute contour level ~0.34 for recon (blue) and 0.37 for molmap. The reconstruction appears further compacted than the MD simulations, more along the six-fold symmetry axis. **Note:** the representative molmaps with residual solvent are from a single structure, not accounting for shot-to-shot heterogeneity encountered in experiments and the ensemble reconstruction.

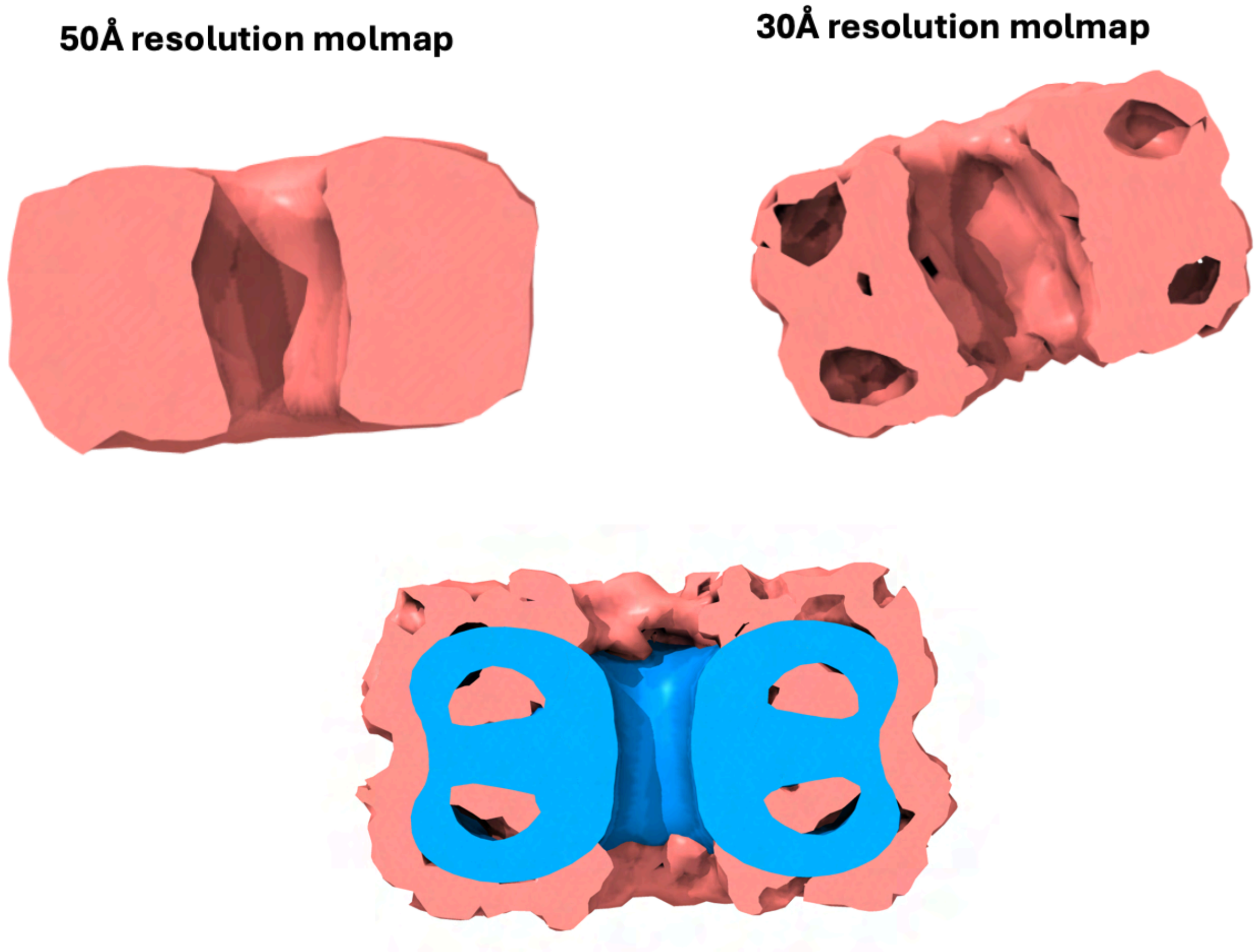

**Fig. S9 | Cross-section comparison validating the resolution estimated from the PRTF. Top row:** Molmaps were generated from the gas-phase compacted Ery model (500 ns) with residual solvent layer, at 50 Å (shown at the maximum absolute threshold before the isosurface density disappears) and 30 Å (shown at a lower absolute threshold). Cross-sectional views show that ~30 Å, cavities between subunits (radial), become visible in addition to the central hole. The appearance of inter-subunit cavities in the reconstruction therefore indicates that the effective resolution is equal to or better than 30 Å. **Bottom row:** Overlaid cross-sections of 500 ns molmap at 20 Å with residual solvent (salmon) and blue (XFEL-SPI recon) which shows more compaction relatively but with cavities (reduced in size) at expected positions (isosurface density visualized in Chimera X 1.10.1 at relatively closer absolute threshold (0.37 for molmap and 0.34 for recon) in the same grid and voxel size).

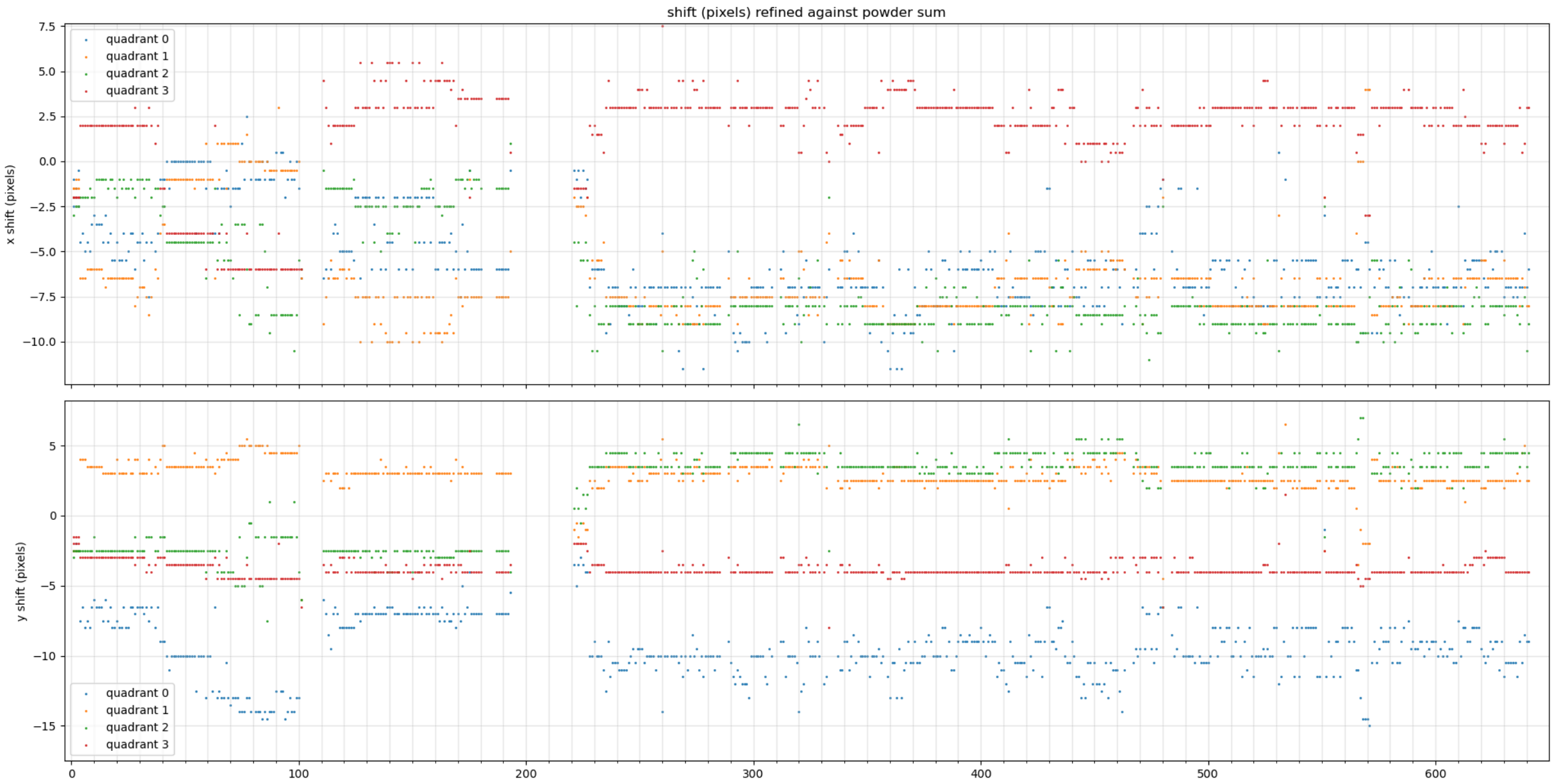

**Fig. S10** | Shift of pixels refined against powder sum for all the runs (had different samples other than Ery) during the 7927 beamtime at SPB/SFX with the AGIPD detector.

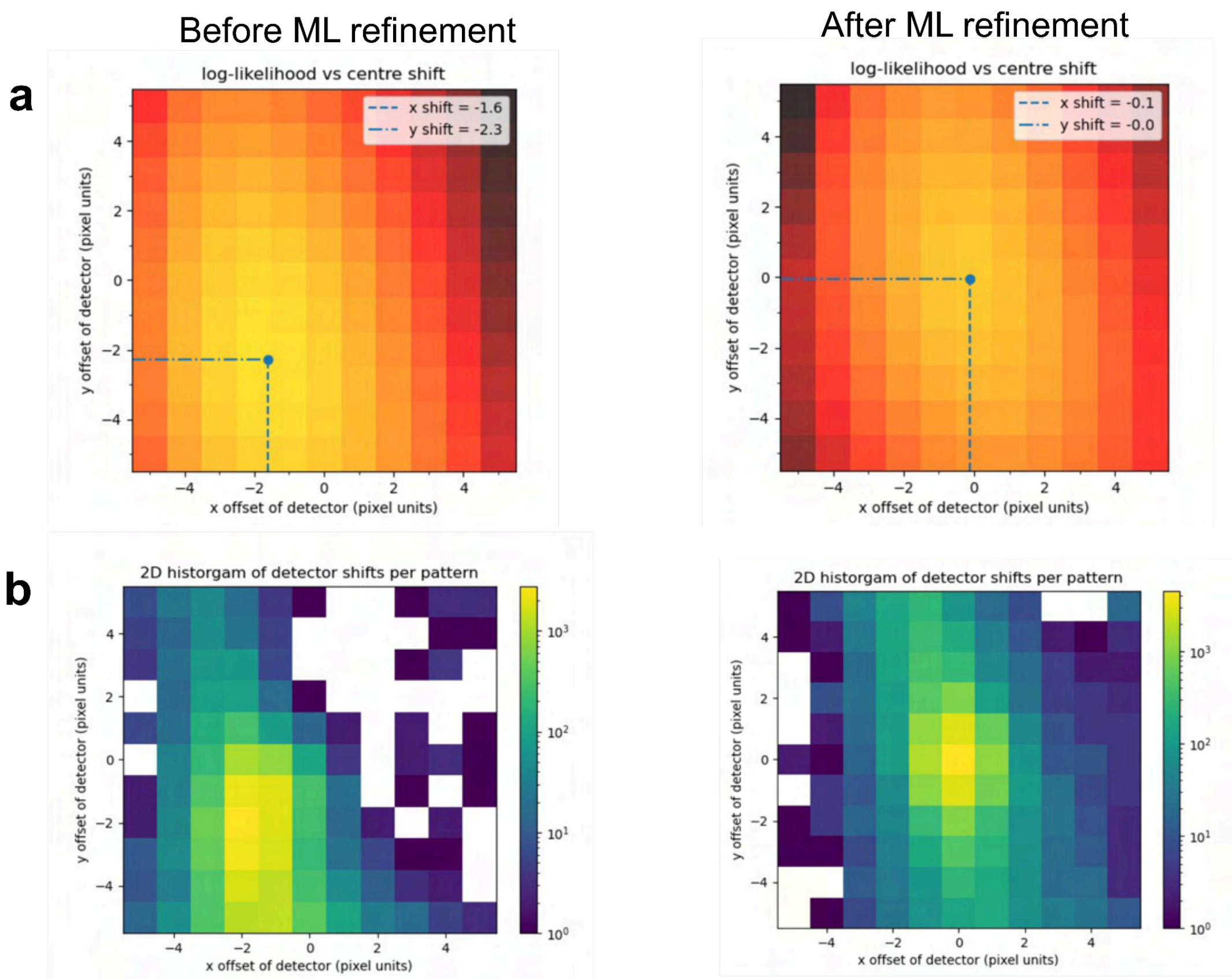

**Fig. S11 |** Maximum-Likelihood (ML) based beam-center refinement to overcome beam drift. Top: a: log-likelihood vs central shift (before (left) and after ML (right) refinement). Bottom: 2D histogram of detector shifts per pattern (before (left) and after (right) ML refinement)

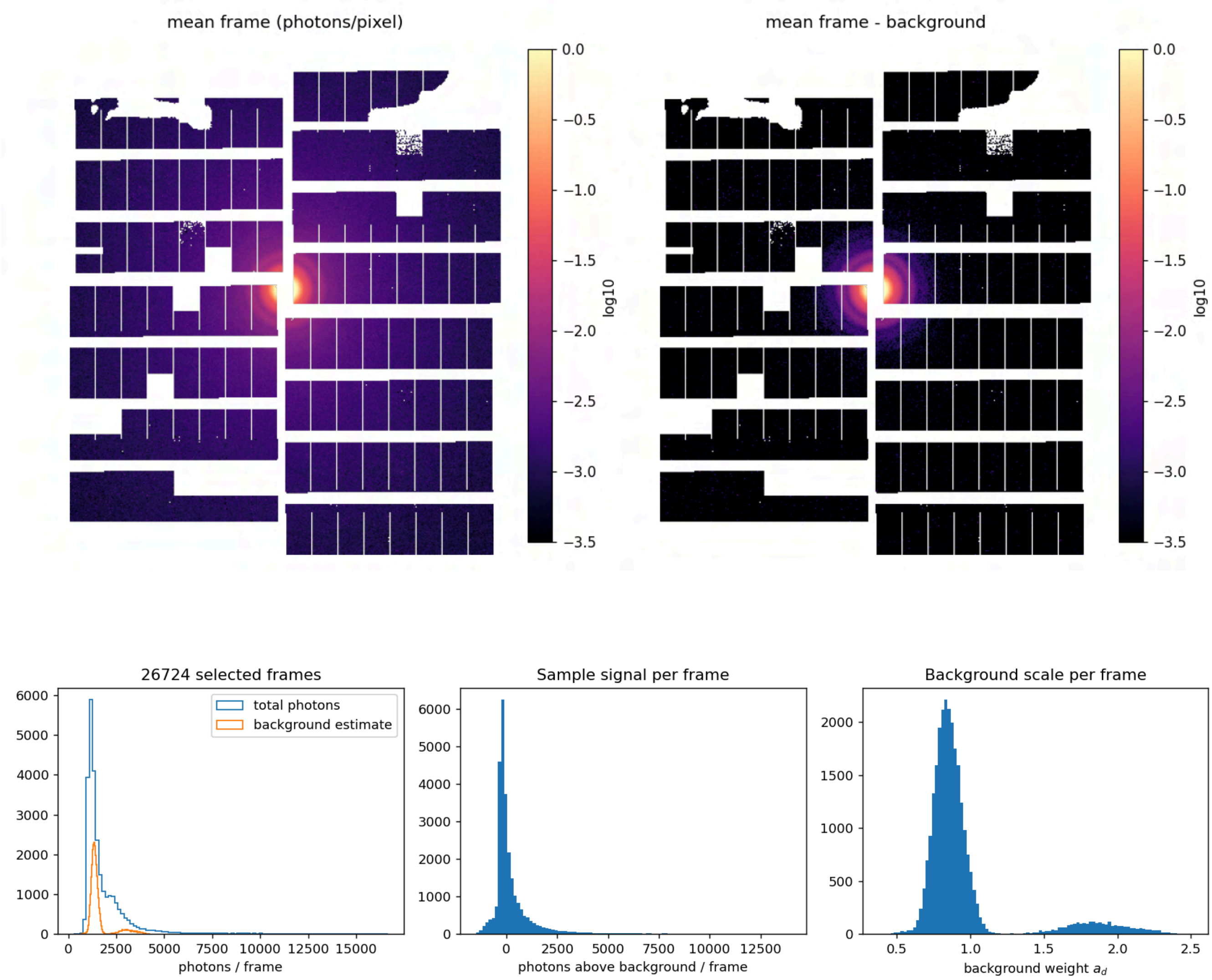

**Fig. S12 |** Per-frame photon statistics. **Top:** The median frame has 1,380 photons, against a background estimate of 1,385 photons. Most hits are therefore weak, and a long tail reaches 5,000–15,000 photons. Mean frame (right) and mean frame - background (left). **Bottom:** Histogram of photons/frame of selected 26724 frames that went into the 3D classification. Total photons and background estimate per frame (left), sample signal per frame above the background (middle), and bimodal distribution of background from different sets of Ery runs.

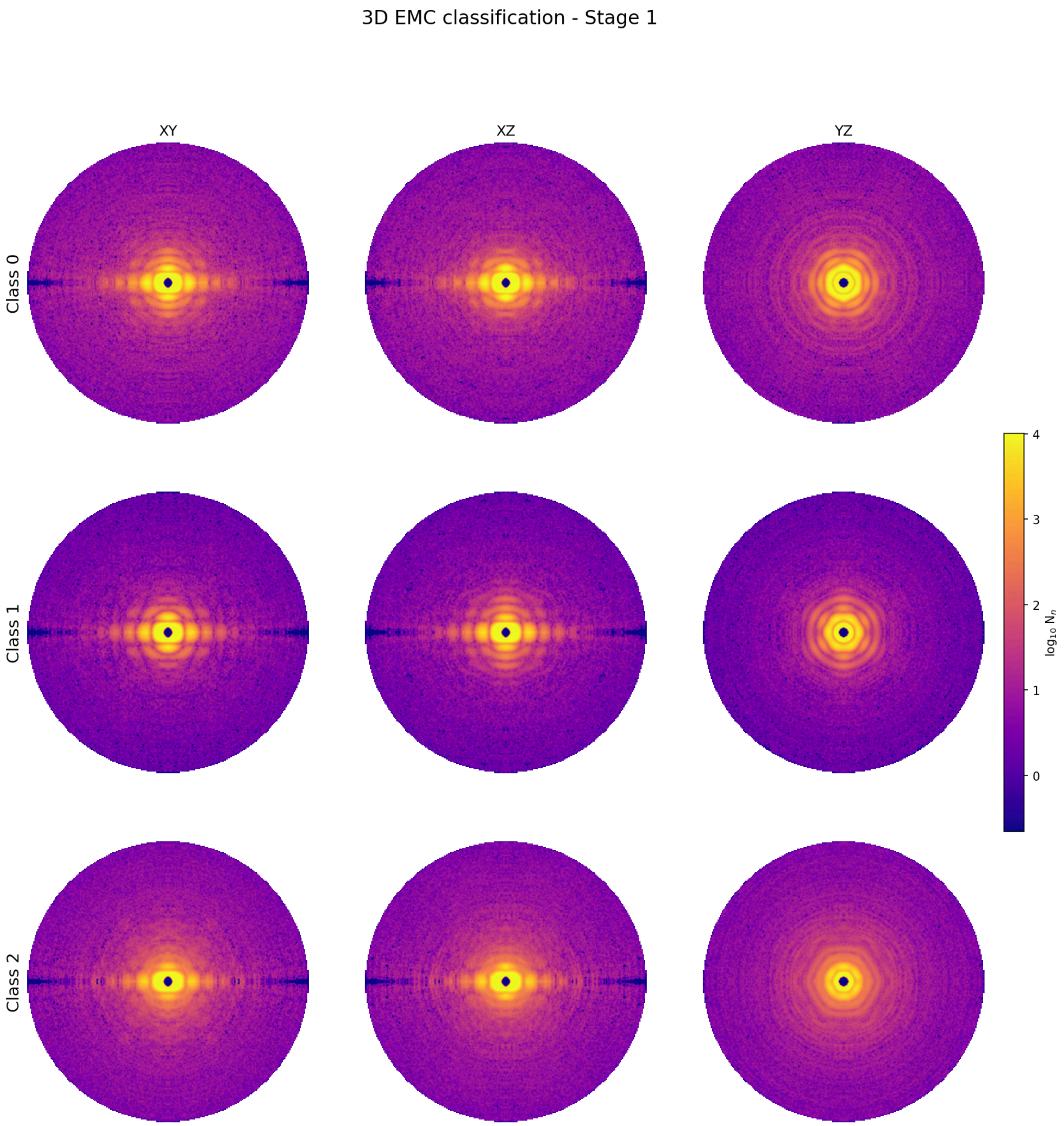

**Fig. S13 |** 3D EMC classification resulting in 3 distinct classes in Stage 1 before introducing background-aware likelihood in Stage 2-3. Three orthoplanes XY, XZ, YZ of each class are displayed. **Top:** Class 0 **Middle:** Class 1, and **Bottom:** Class 2.

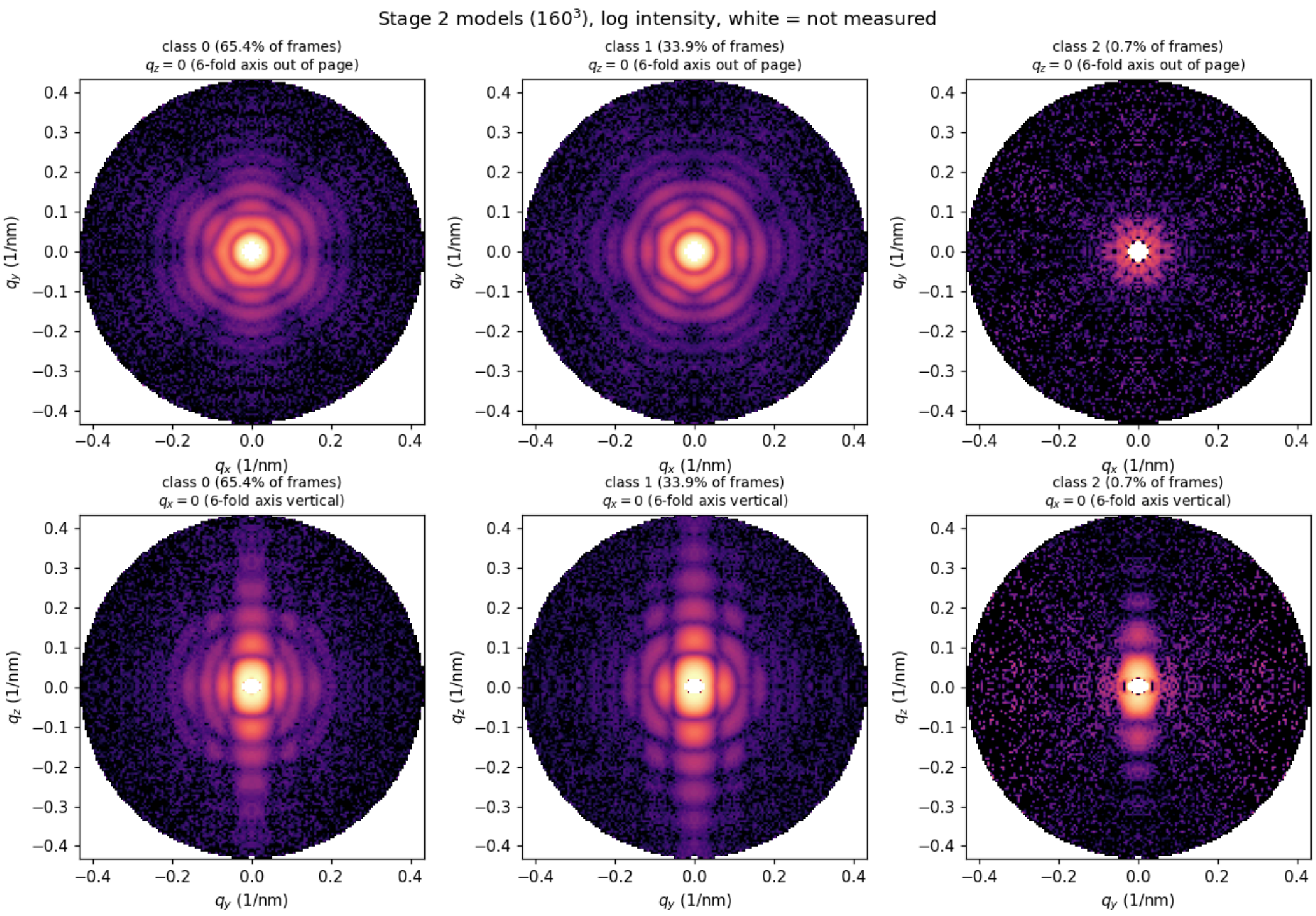

**Fig. S14 |** Stage 2 level of the EMC 3D classification with $160^3$ voxel grid after the introduction of background-aware likelihood. Class 0 (left), Class 1 (middle), and Class 2 (right). Note the visibly different central speckle size and fringe spacing between class 1 and 2 indicating size difference.

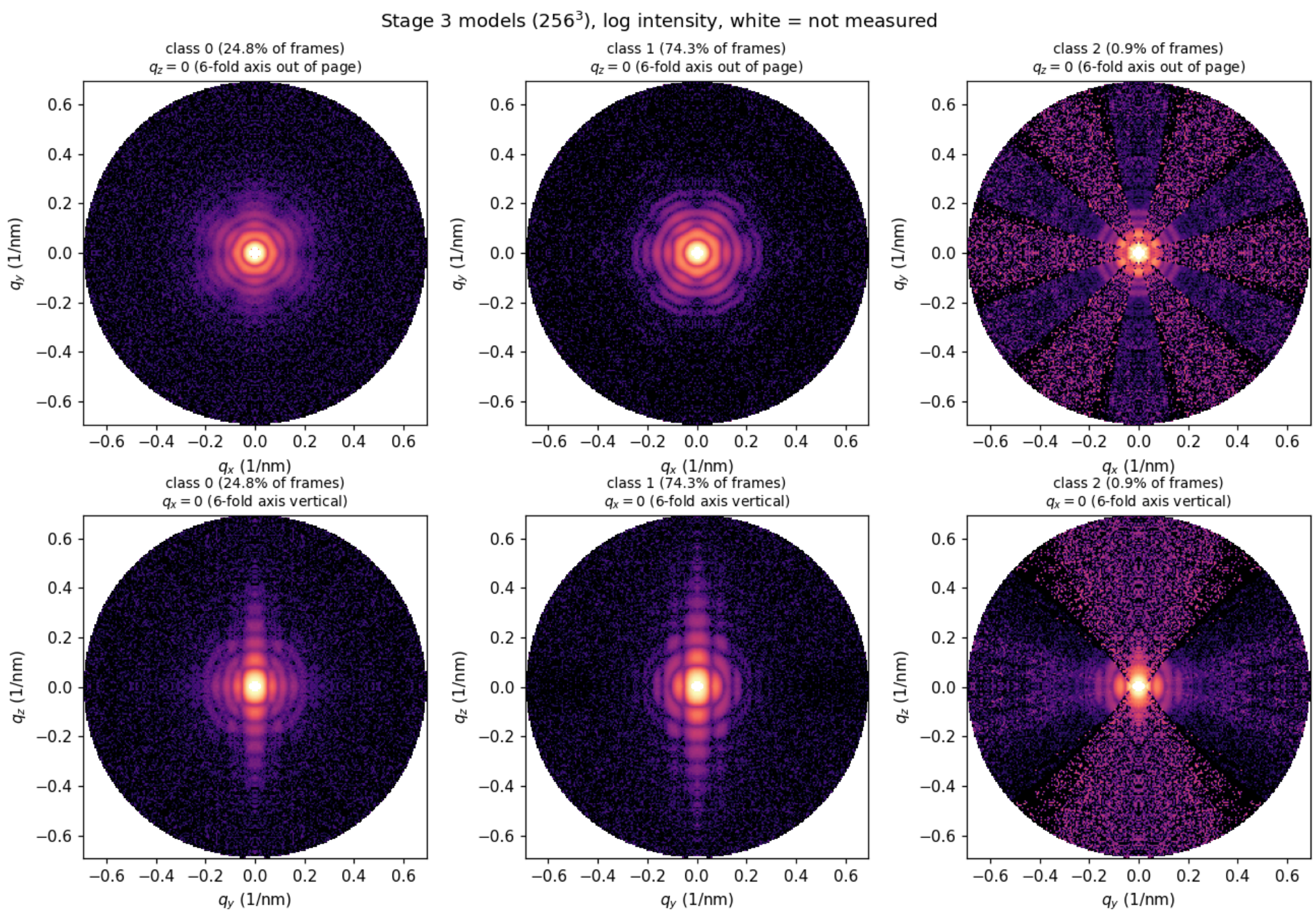

**Fig. S15 |** Stage 3 level of the EMC 3D classification expanded to the $256^3$ voxel grid (0.69 $nm^{-1}$). Class 0 (left), Class 1 (middle), and Class 2 (right). Note the visibly different central speckle size and fringe spacing between class 1 and 2 indicating size difference.

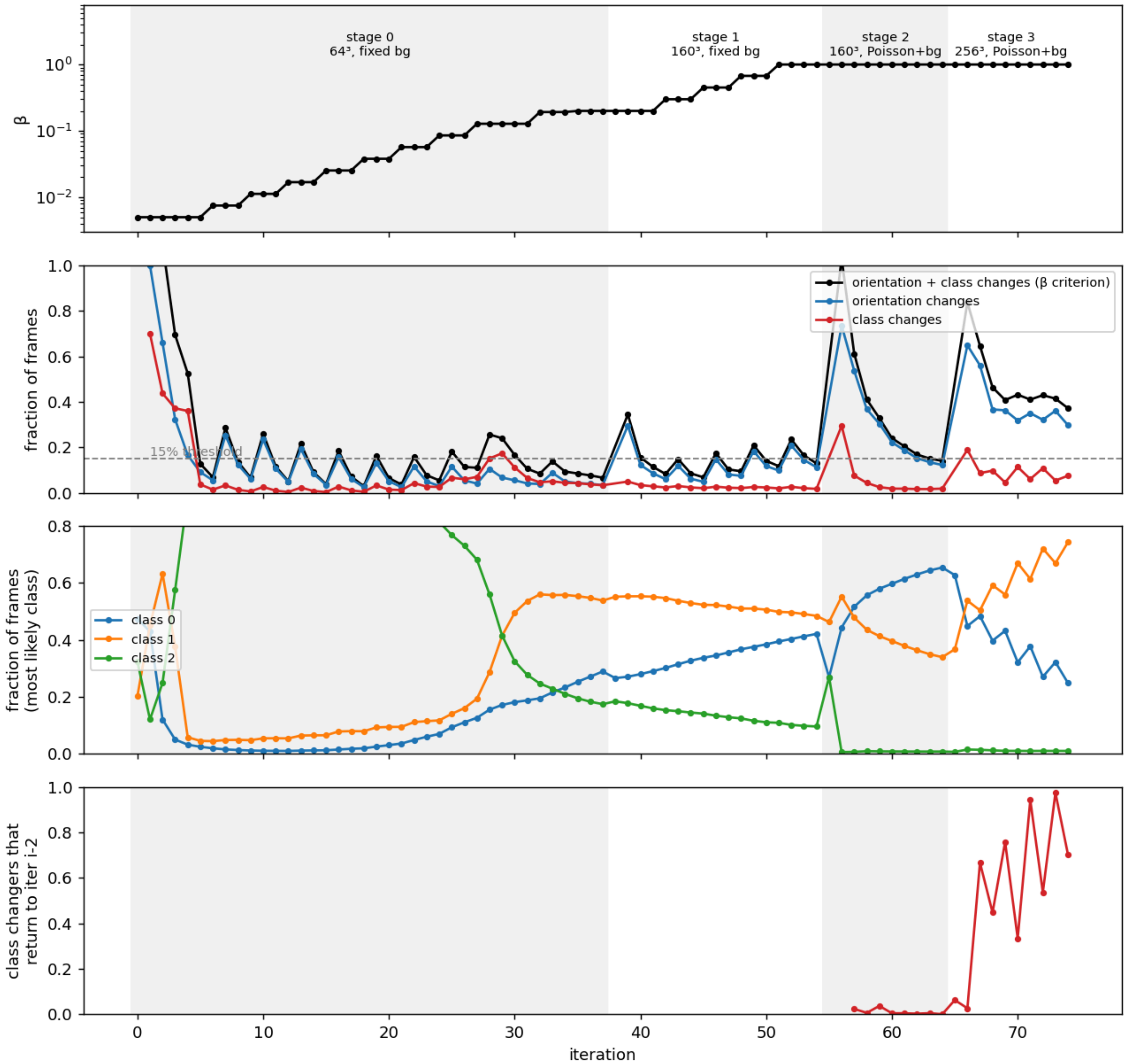

**Fig. S16 |** Monitoring convergence during the 3D classification. Top row: Stage 0-2 converged based on the $\beta$ criterion, however in stage-3 frames were still changing orientations as seen in the second row from top.

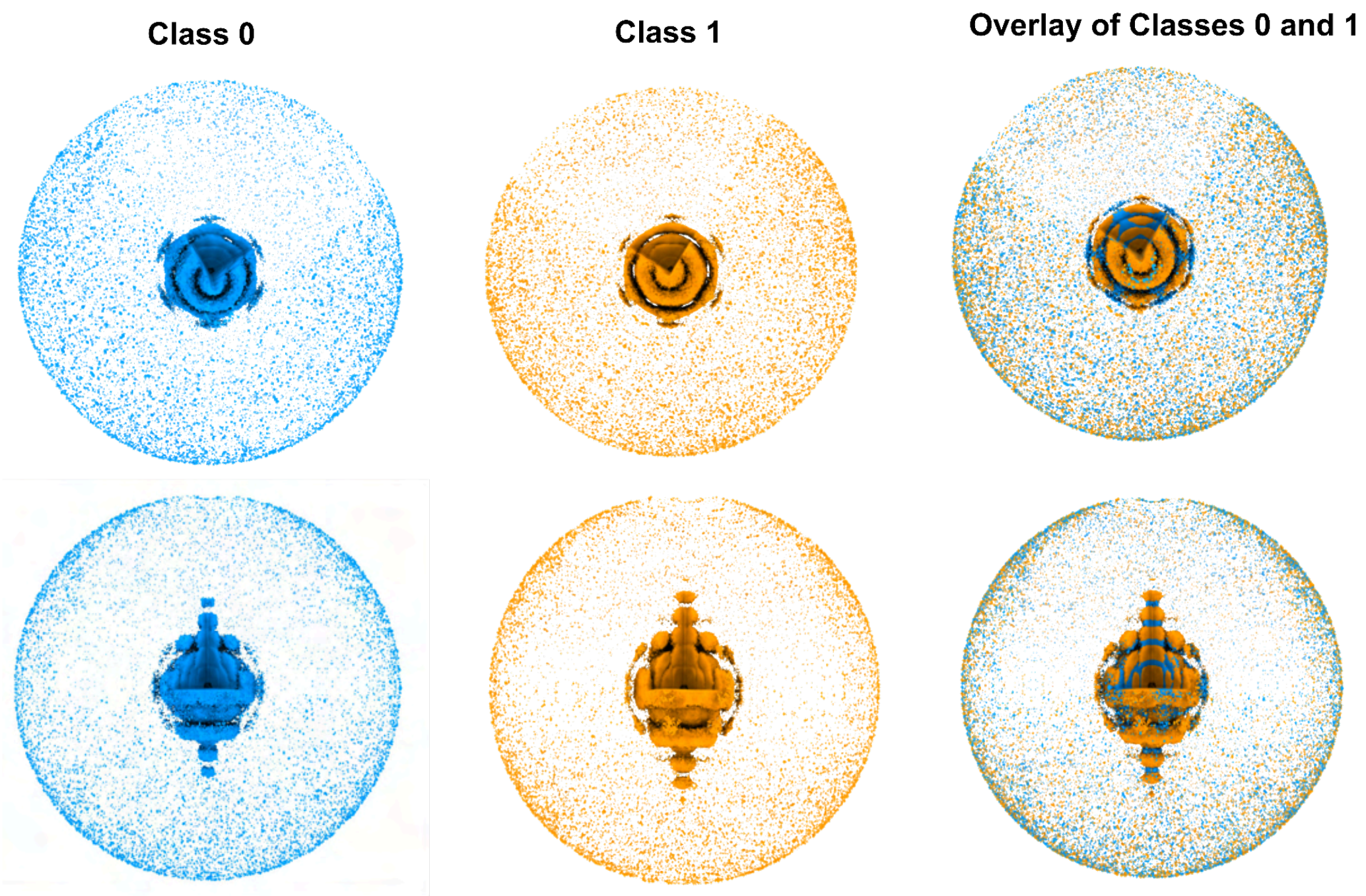

**Fig. S17 |** 3D EMC classification of Ery results after Stage 3: 3D merged intensities of the two classes Class 0 (blue) and 1 (orange) after final Stage 3 classification with background-aware likelihood. Note the fringe differences between two classes. See below for the radial profile ( Fig. S18) below which reflects the differences clearly. Maps are visualized in the same grid and isosurface density levels.

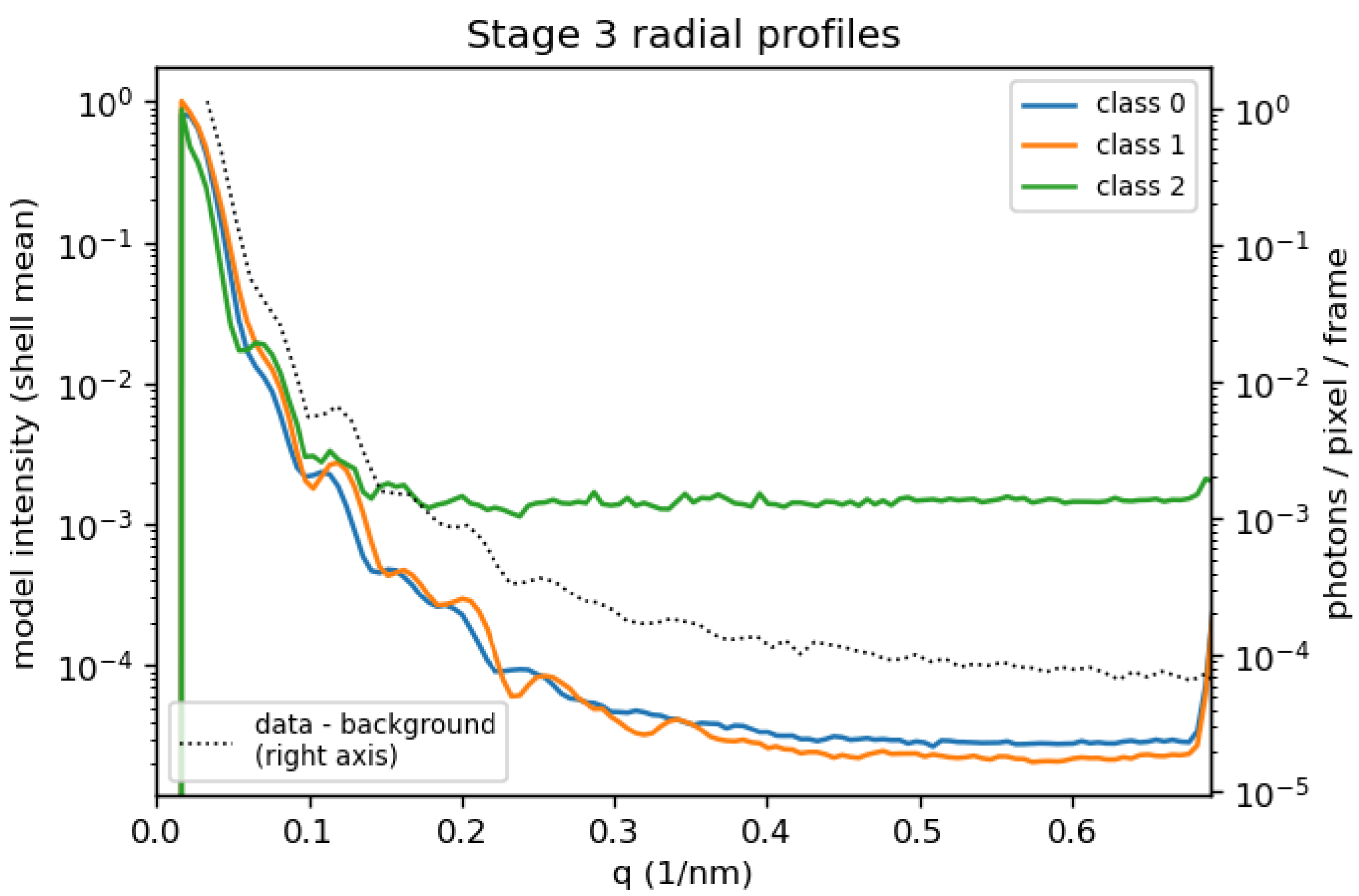

**Fig. S18 |** Radial profiles of final 3D maps after the Stage 3 level EMC 3D classification.

| Stage | Grid | q_max ($nm^{-1}$) | Order (step) | Orientations | $\beta$ | $\sigma$ | Background, likelihood | P_thresh | Iterations/ max |
|---|---|---|---|---|---|---|---|---|---|
| 0 | $64^3$ | 0.17 | 8 (7.1°) | 4,515 | 0.005 → 0.2 | 0.4 | fixed, fluence_free, 'W' | 0 | 38/60 |
| 1 | $160^3$ | 0.43 | 12 (4.8°) | 14,7664 | 0.2 → 1 | 0.2 | fixed, fluence_free, 'W' | 0 | 17/30 |
| 2 | $160^3$ | 0.43 | 12 (4.8°) | 14,764 | 1 | 0.2 | Poisson + background, "I" | $10^{-3}$ | 10/20 (converged) |
| 3 | $256^3$ | 0.69 | 15 (3.8°) | 28,910 | 1 | 0.2 | Poisson + background, "I" | $10^{-3}$ | 10/10 (not converged) |

**S.Table 3 |** Parameters used in different stages of 3D EMC classification. $\sigma$ is the continuity parameter (see methods).

| End Stage | Iterations | Class 0 most likely | Class 0 weighted | Class 1 most likely | Class 1 weighted | Class 2 most likely | Class 2 weighted |
|---|---|---|---|---|---|---|---|
| Stage 1 | 54 | 11, 244 | 11,393 | 12,942 | 12,723 | 2,538 | 2,608 |
| Stage 2 | 64 | 17, 485 | 17,531 | 9,057 | 9,017 | 182 | 176 |
| Stage 3 | 74 | 6,620 | 6,662 | 19,867 | 19,829 | 237 | 233 |

**S.Table 4 |** Results of the different stages of the 3D EMC classification.

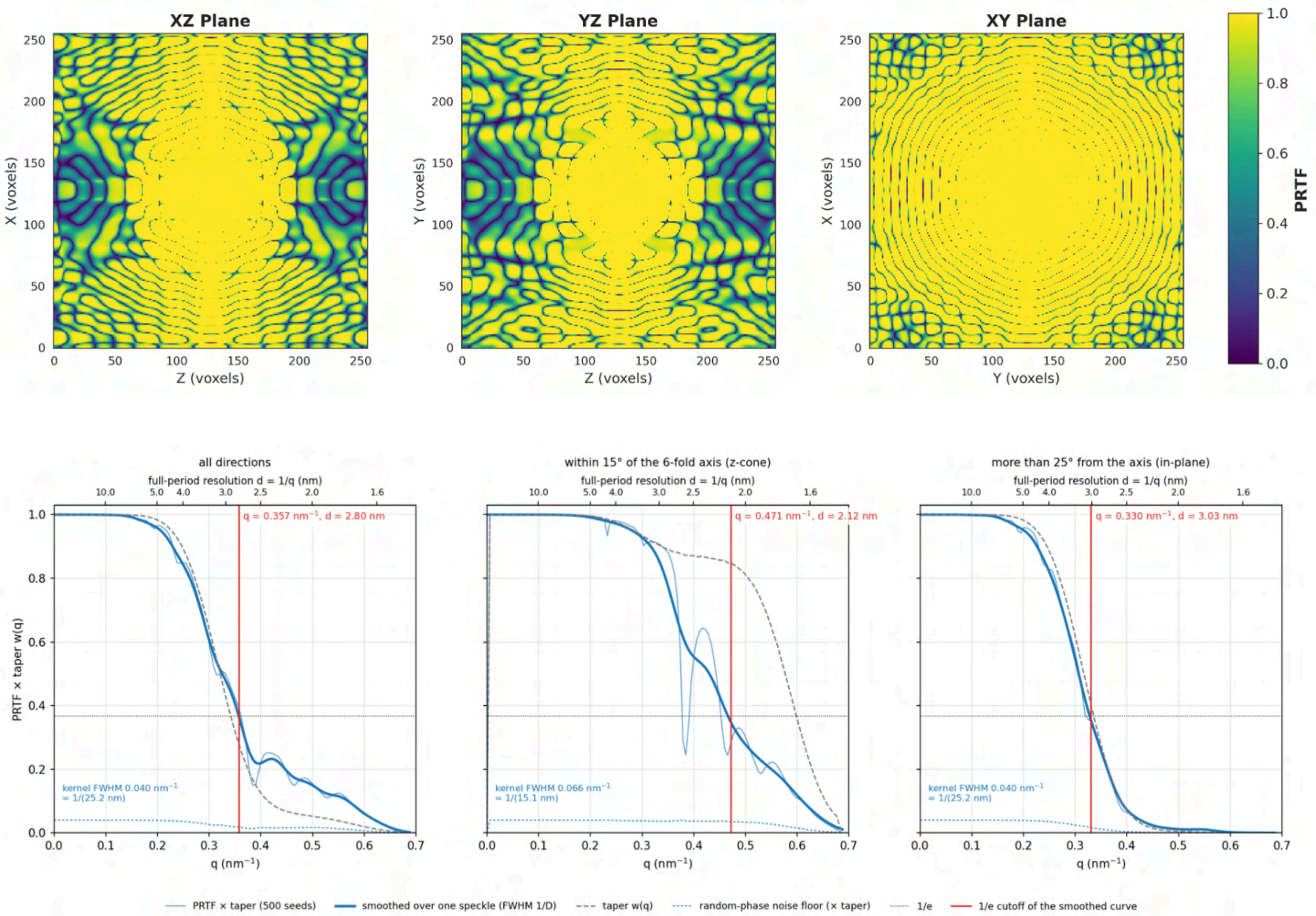

**Fig. S19 | Top:** Orthogonal planes of PRTF: XZ, YZ and XY planes showing anisotropy. **Bottom:** Individual PRTF plots of all-directions, z-cone, and in-plane, with taper w(q) (dashed line).

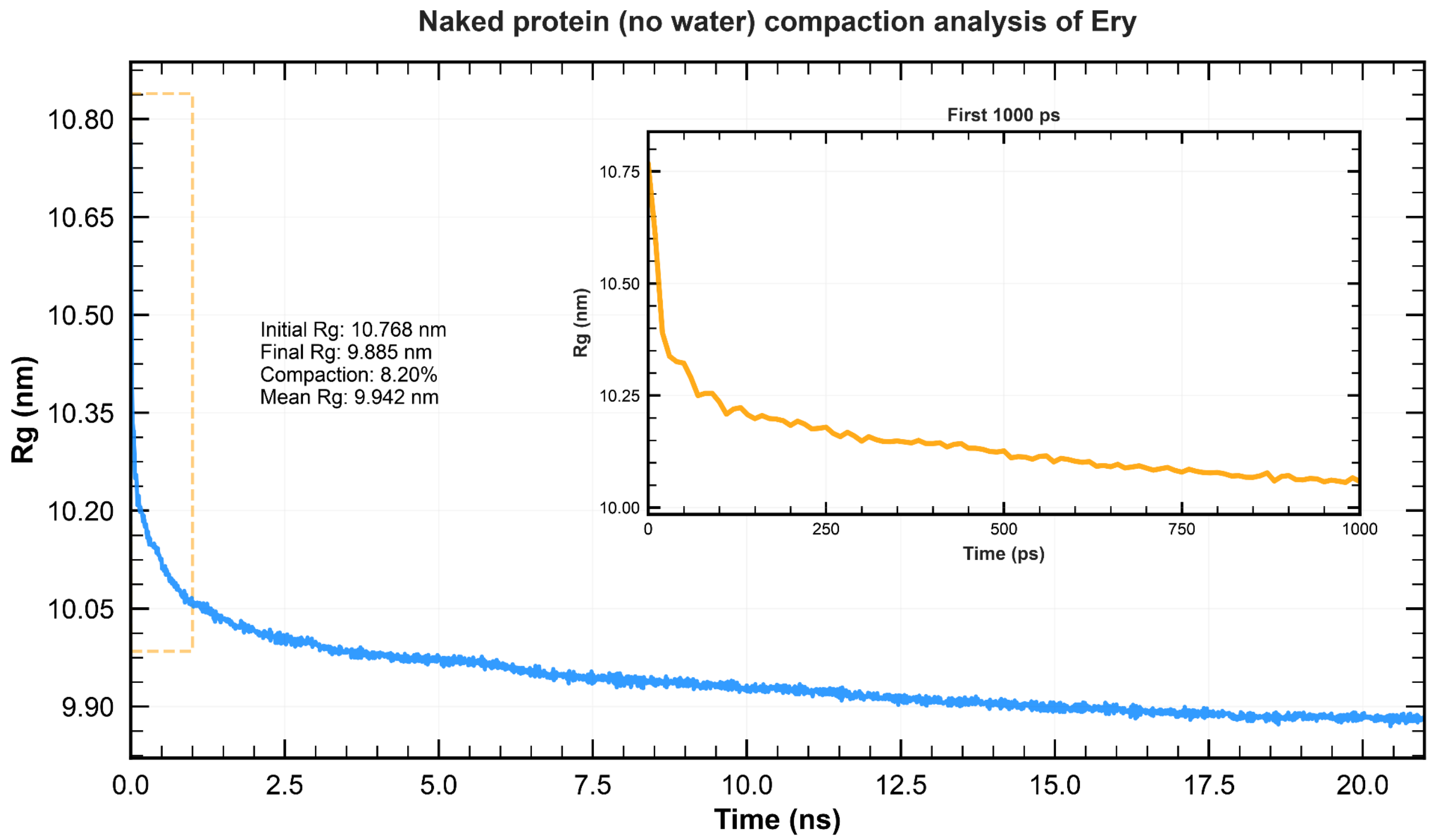

**Fig. S20 |** MD simulation compaction analysis of naked Ery protein (no water added) in vacuum, showing rapid fall in Rg in the first few hundred picoseconds (roughly about 6% reduction in Rg in the first ns).

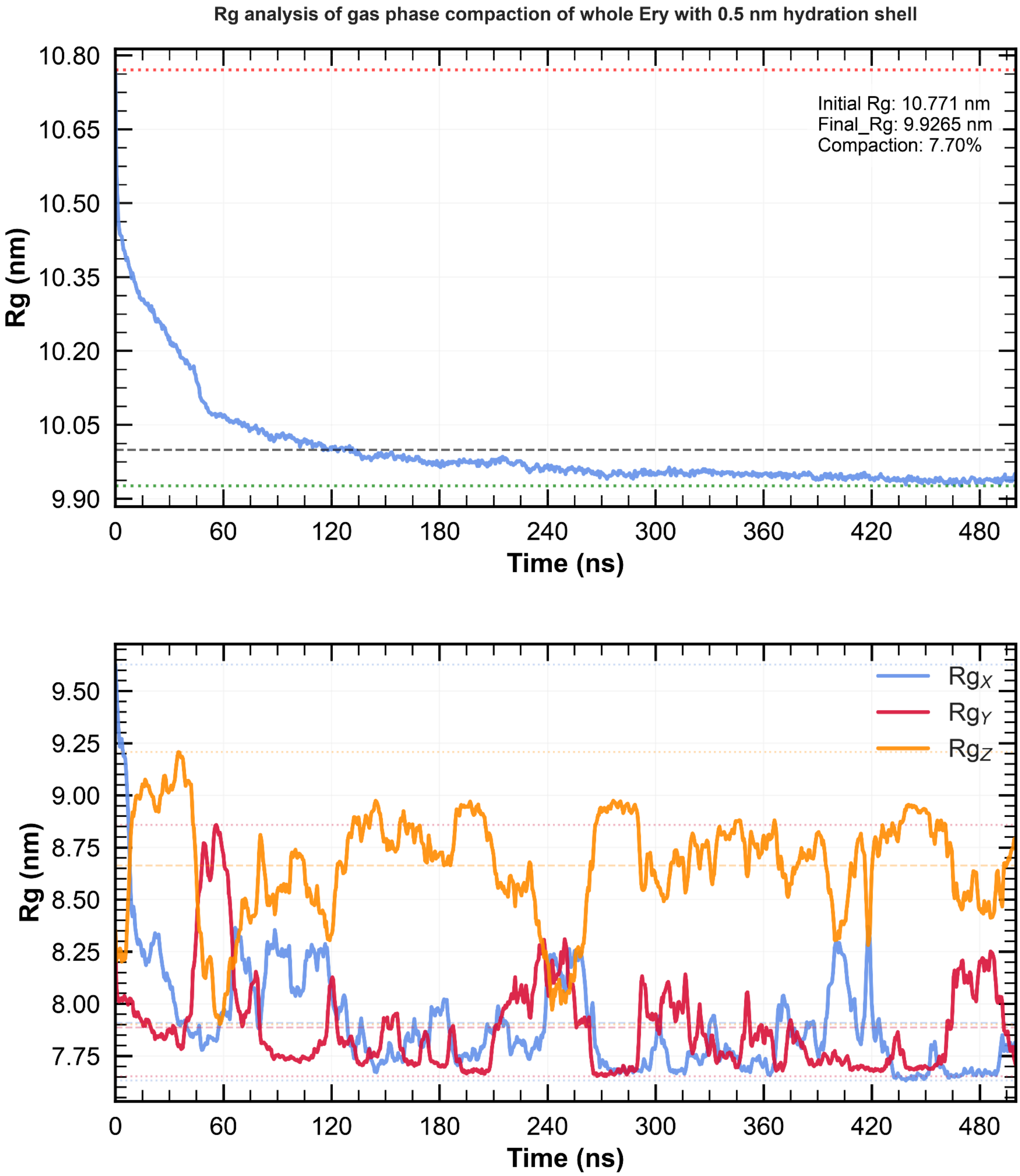

**Fig. S21 |** Rg analysis of gas phase compaction from MD simulations of whole Ery molecule with 0.5 nm water-shell up to 500 ns.

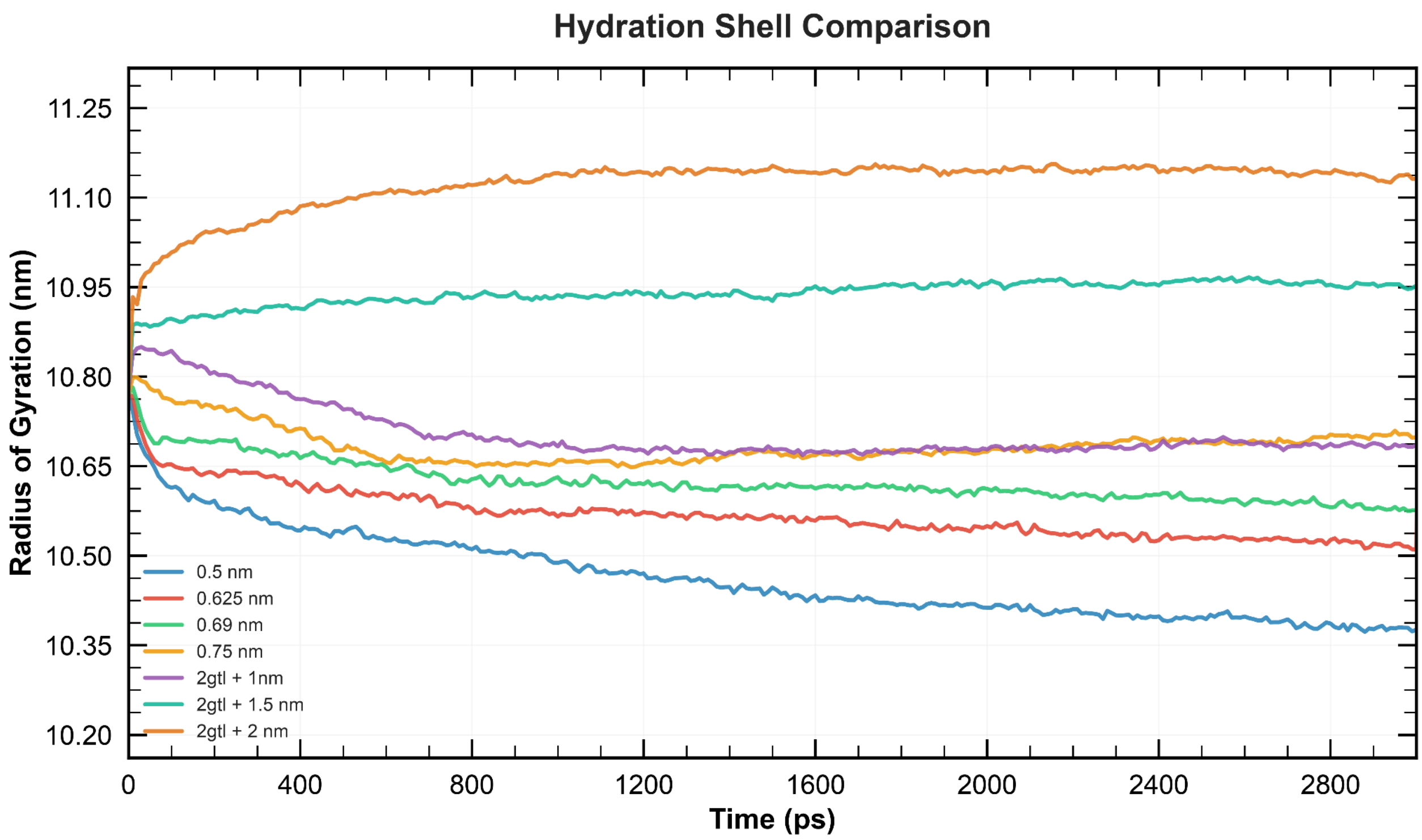

**Fig. S22 |** MD simulations with varying water-shell to find the sweet-spot hydration shell to avoid gas-phase compaction. The simulations were carried out up to 3 ns to observe the initial trends on the effect of varying hydration shell on compaction.

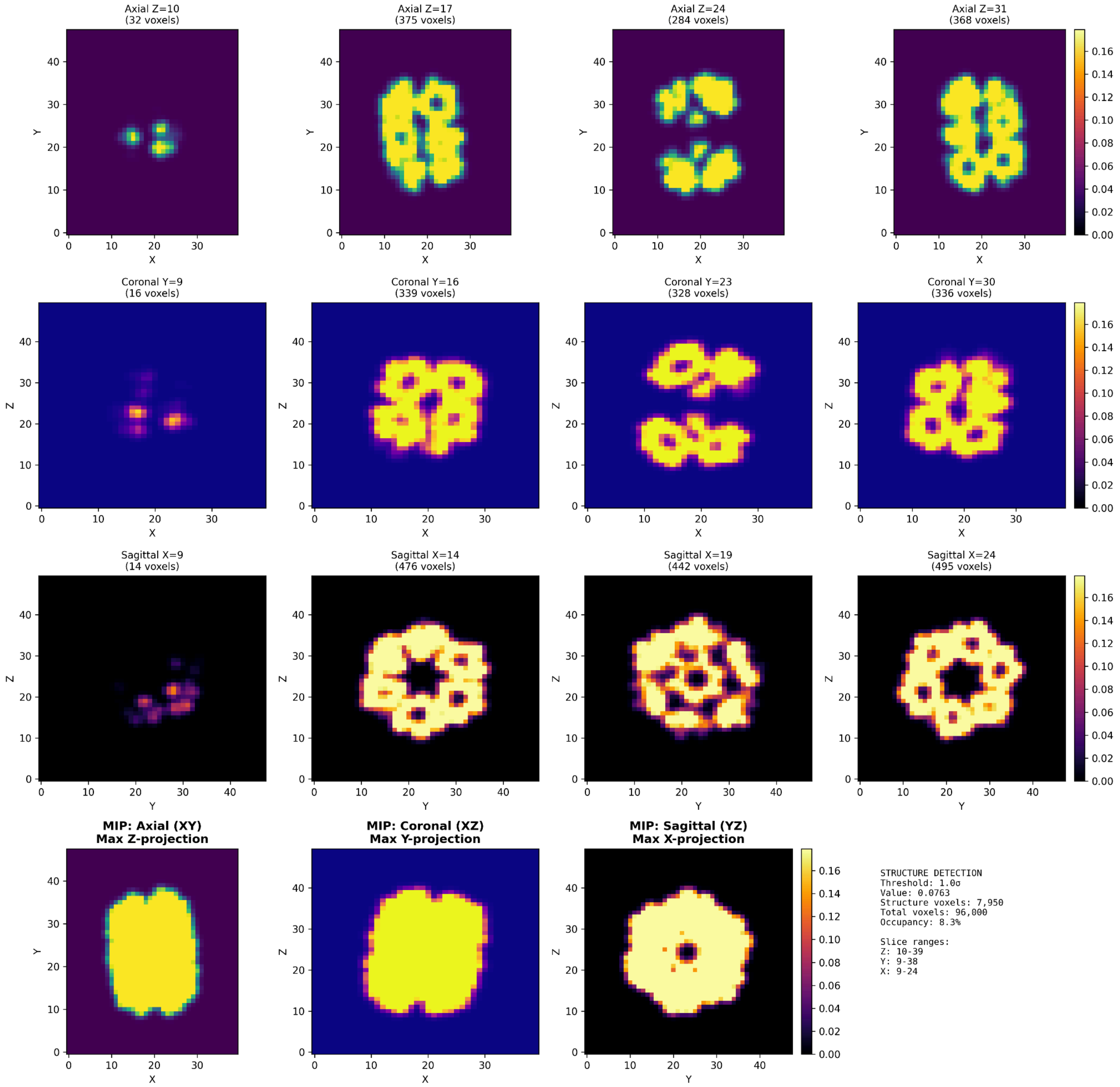

**Fig. S23 | Slices of the simulated density of map from PDB 2GTL at 29A resolution for comparison** at axial, coronal, and sagittal planes, at first, second, and third rows, respectively. The fourth row shows max Z, Y, X projections for comparison with the reconstructed structure below in Fig. in S6.

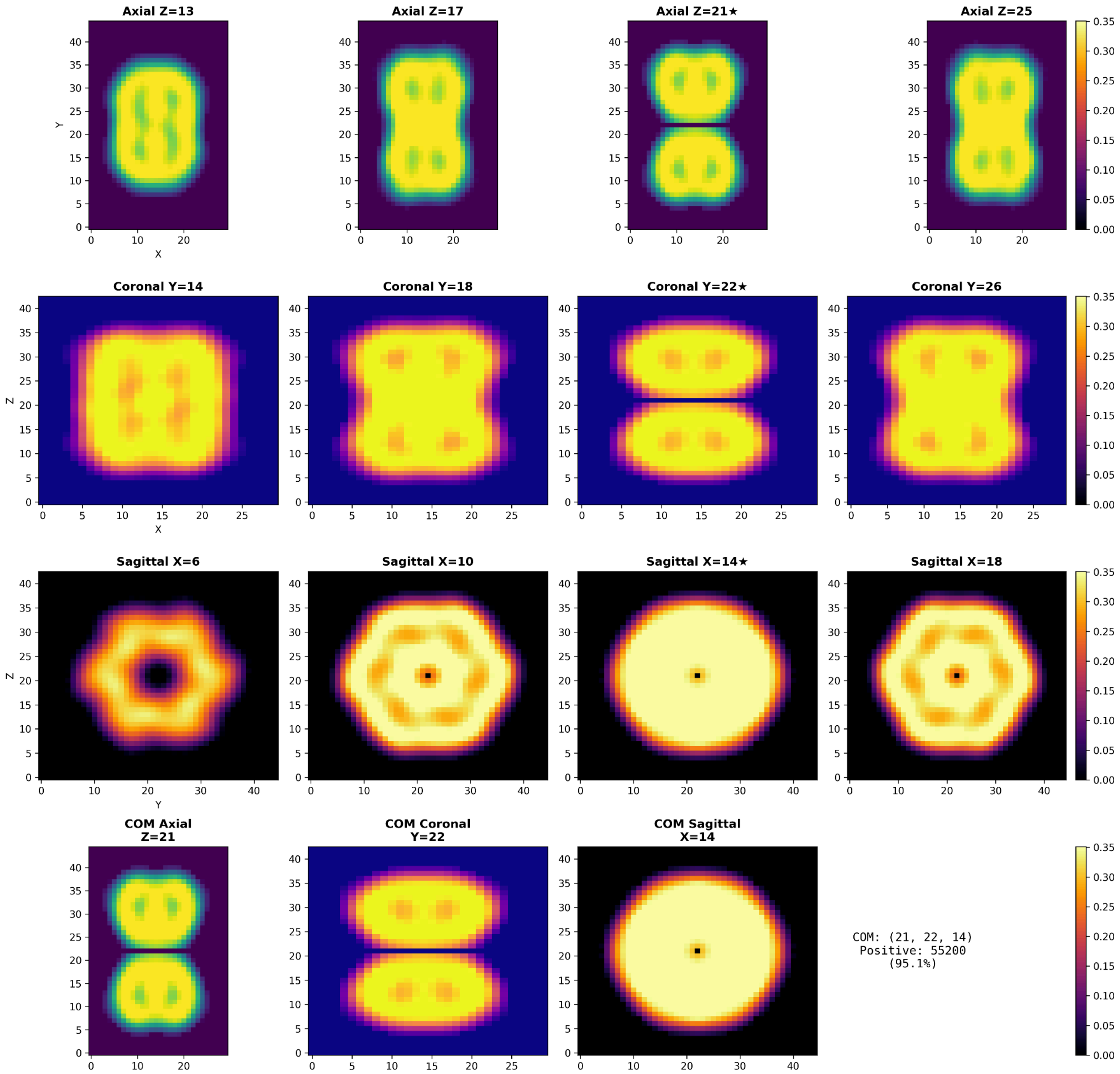

**Fig. S24 |** Slices through the reconstructed XFEL-SPI 3D structure of aerosolized Ery at 6 keV. See Fig. S6 for comparison

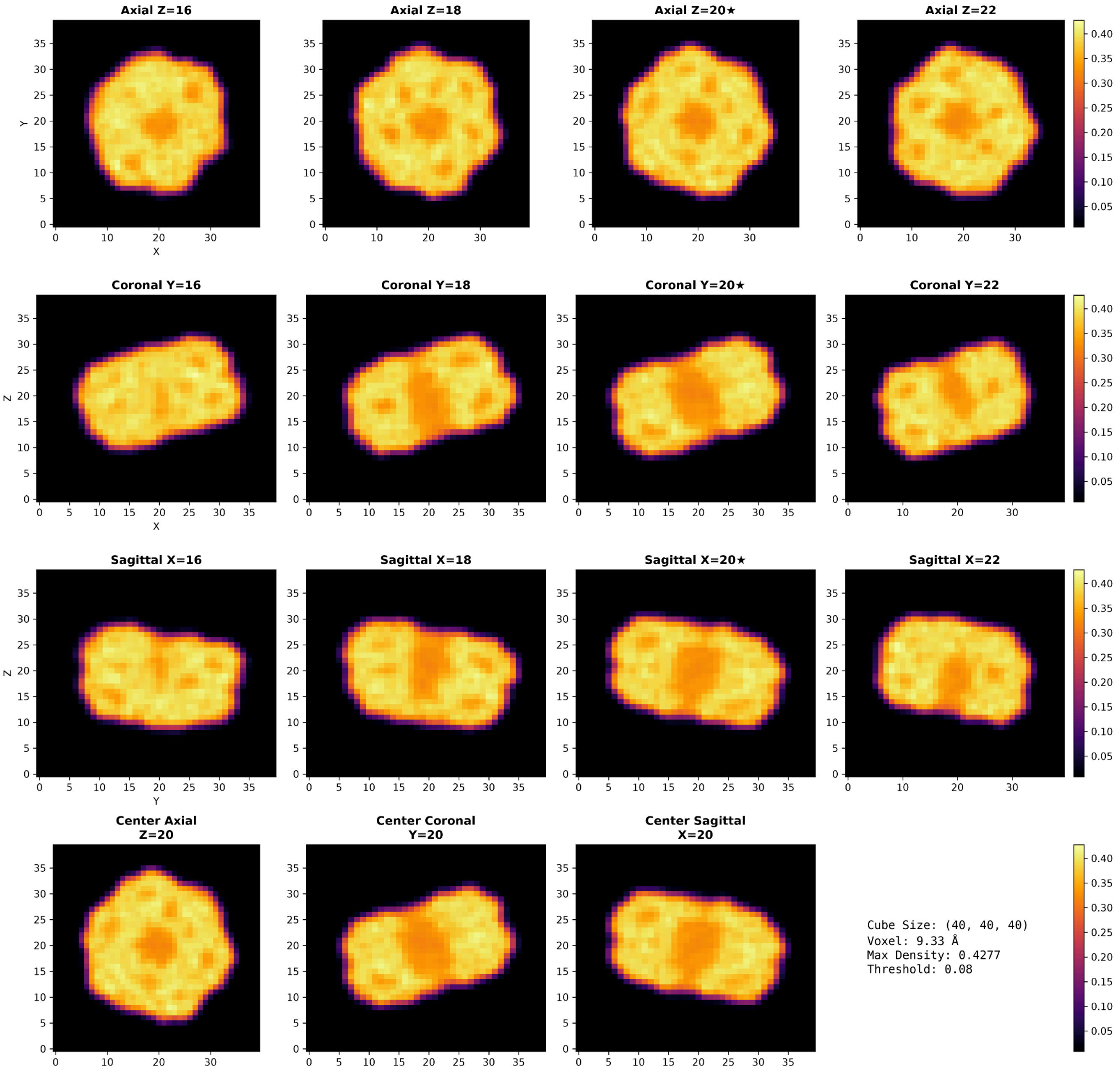

**Fig S25 |** Slices through the 500 ns gas-phase MD simulation molmap model with residual solvent at 28Å (closely matches our 2D and 3D reconstructed images).

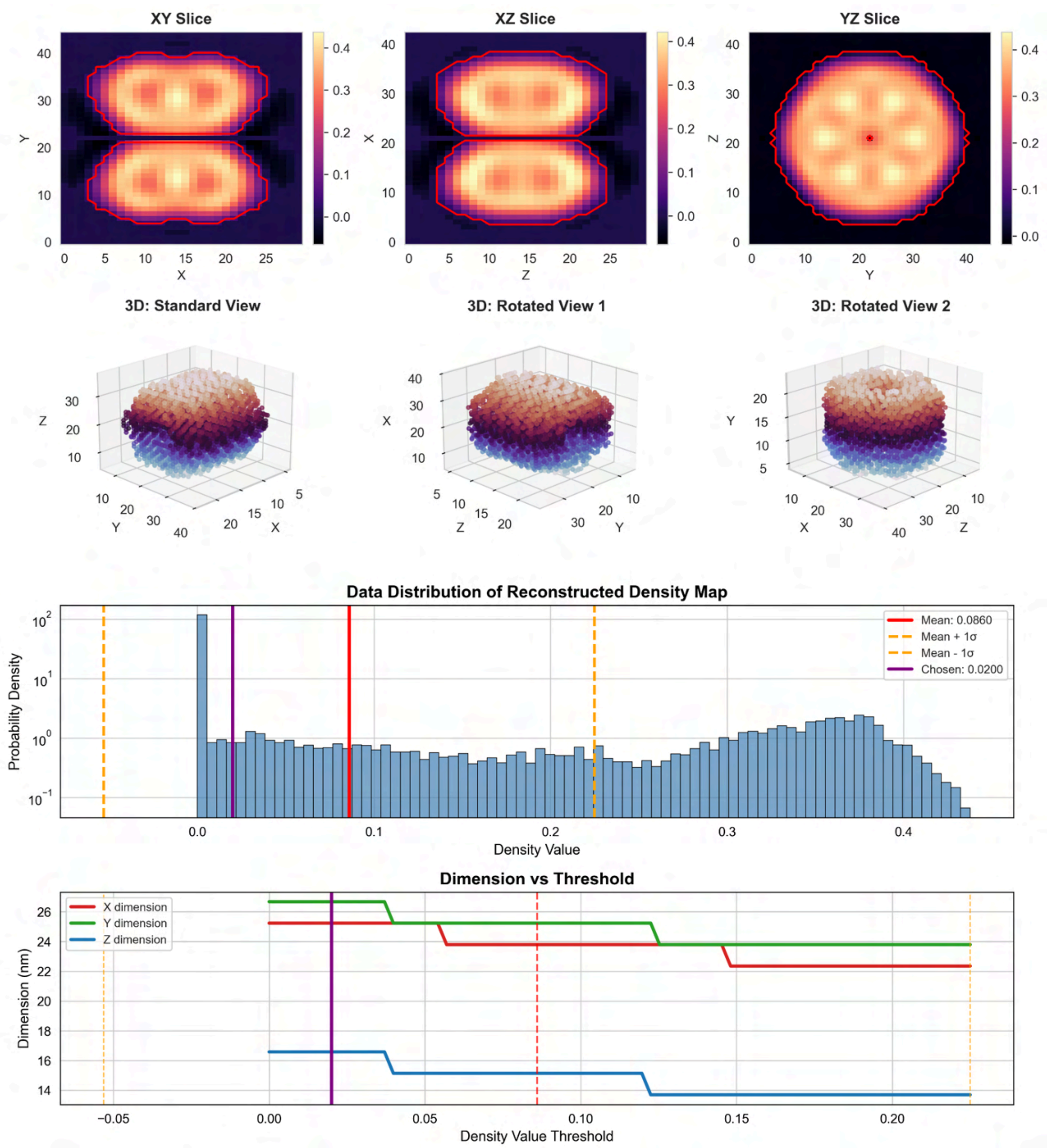

**Fig. S26 |** 3D view and data distribution histogram of the XFEL-SPI reconstructed density map (with low-q phasing method) and "threshold levels" (used as in ChimeraX) described in the text, illustrated for reference.

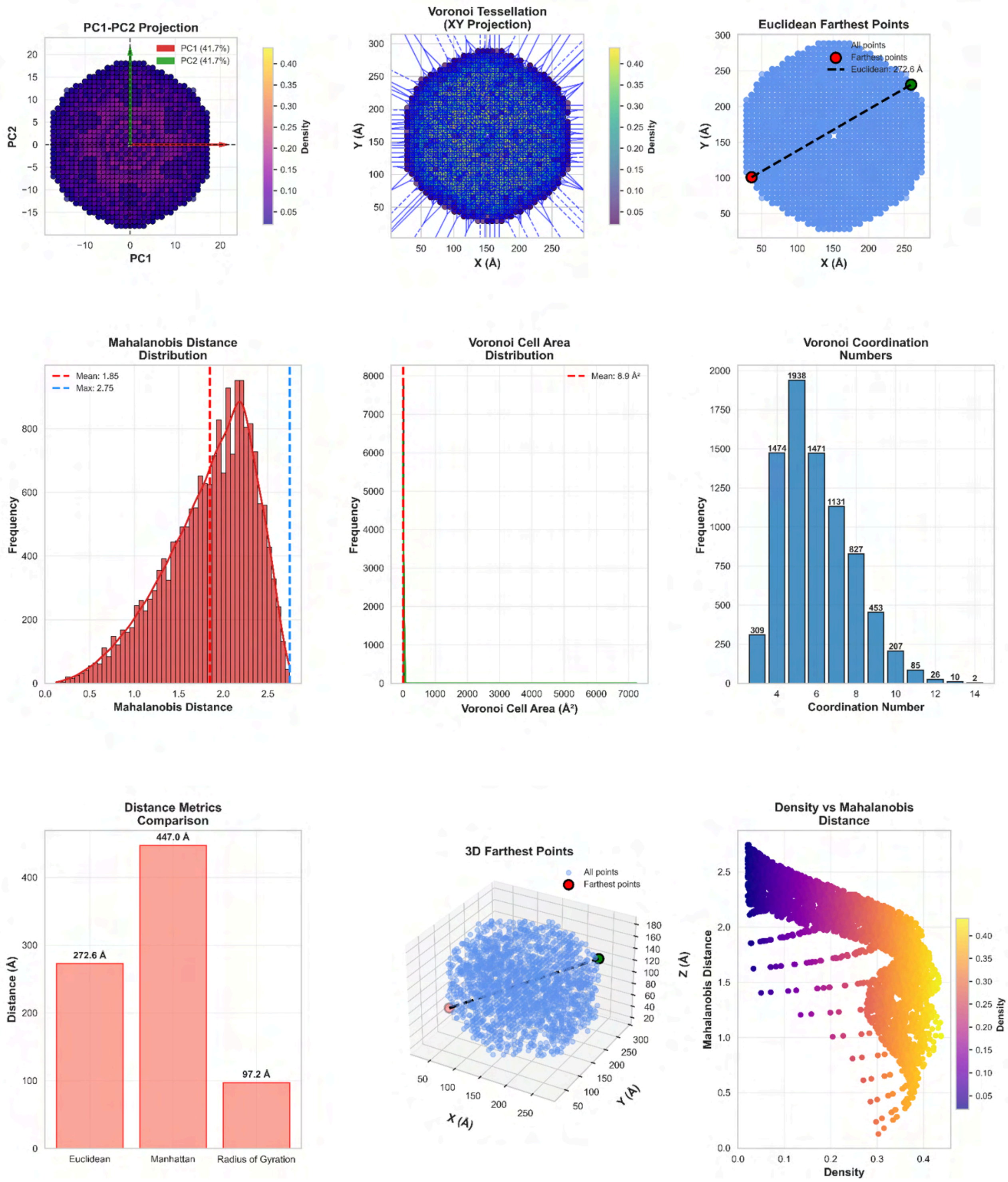

**Fig. S27 | Spatial analysis of the 3D reconstructed density map of Ery from aerosol XFEL-SPI at 6 keV.** Top right: PCA. threshold used 0.02

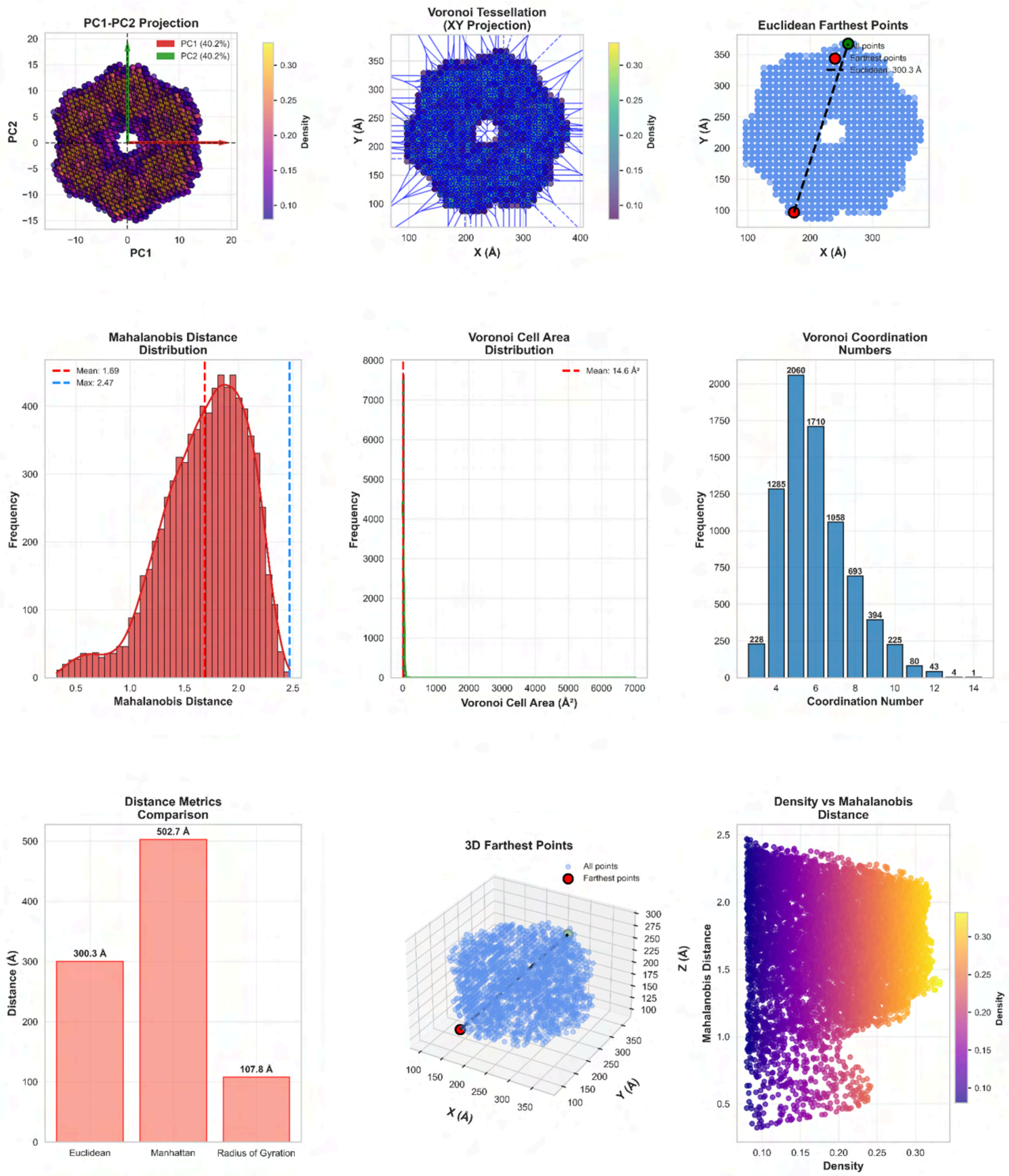

**Fig. S28 | Spatial analysis of Ery from the generated MolMap at 29Å from PDB for comparison.** Top right: PCA for comparison. Abs threshold used 0.08 to match the envelope size with ground truth PDB.

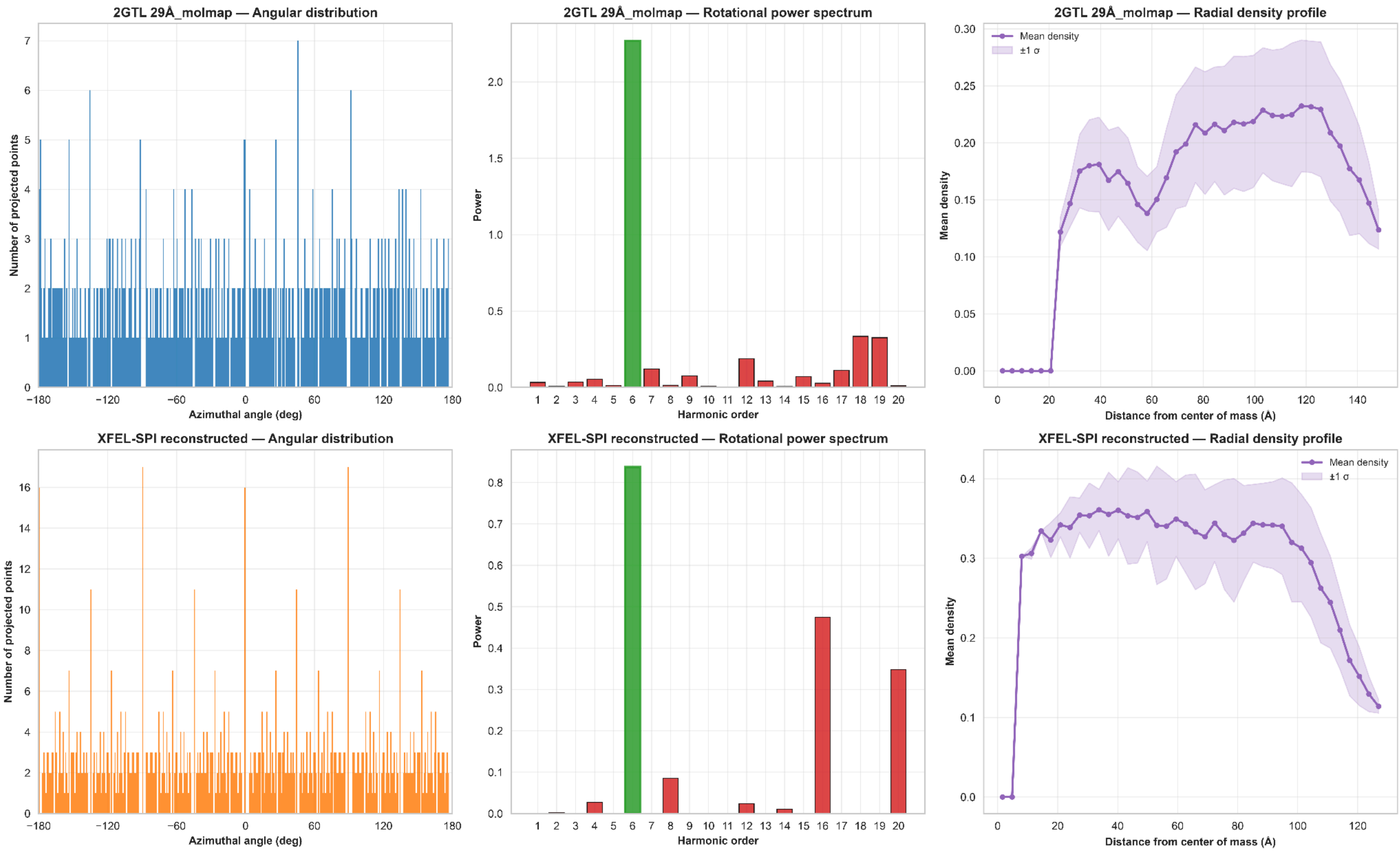

**Fig. S29 |** Six-fold symmetry and central cavity comparison between 2GTL PDB and XFEL-SPI map: **Left:** Angular distribution comparison between 29Å molmap from 2GTL PDB (top, blue) and XFEL-SPI reconstructed map (low-q phasing) (bottom, orange). **Middle:** Rotational power spectrum (harmonic order vs. power). The dominant peak at order k = 6 and the suppressed order k = 3 indicate the preservation of hexagonal order in both maps. A secondary peak at order 16 is observed in the reconstruction; its exact origin is unknown, but it may arise from the residual solvent or counterions that are not present in the pure PDB model. **Right:** Radial density profiles comparing central cavity. The 2GTL reference (top) exhibits a deep central cavity (ratio 0.40), whereas the gas-phase reconstruction (bottom) shows a shallower dip (ratio 0.83), indicating that the central pore is partially filled (with residual solvent and counter ions) or compacted.

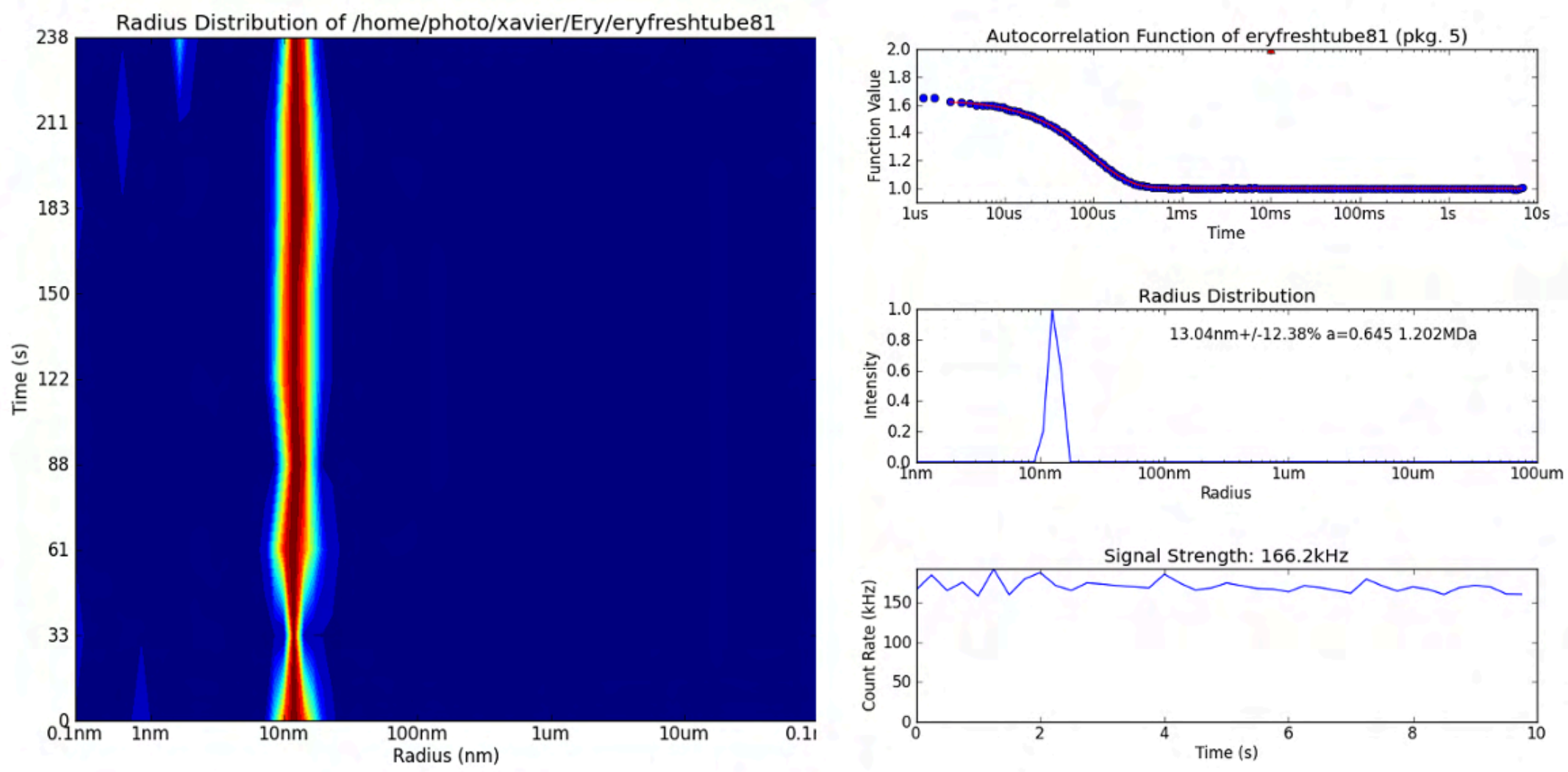

**Fig. S30 |** Dynamic Light Scattering measurement of Ery showing monodispersity and expected radius.

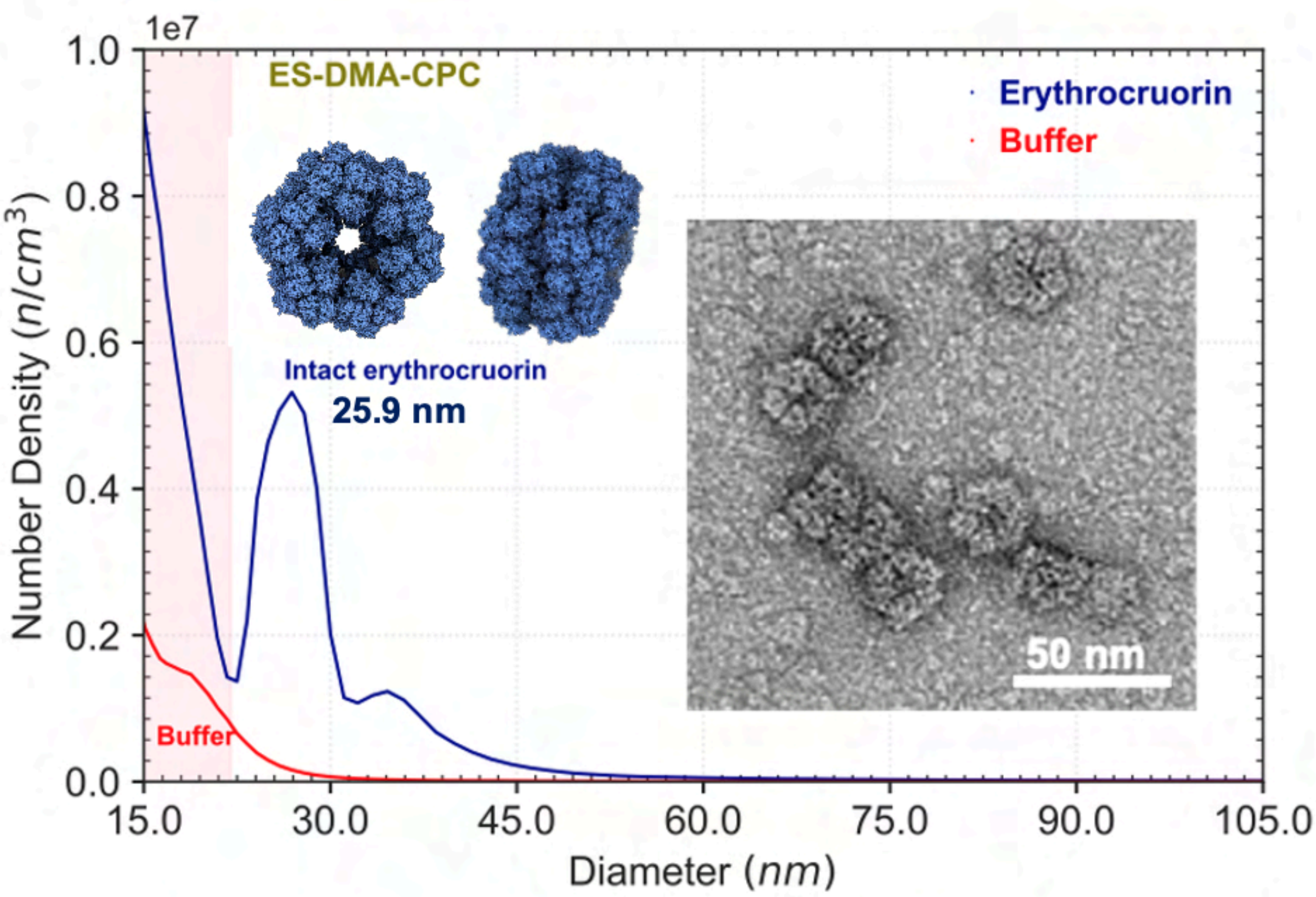

**Fig. S31 |** ES-DMA-CPC measurement showing intact Ery after electrospray aerosolization, with the buffer peak at ~10 nm we expect a caking layer of ~0.5 nm. Inset: TEM image of Ery, scalebar: 50 nm.

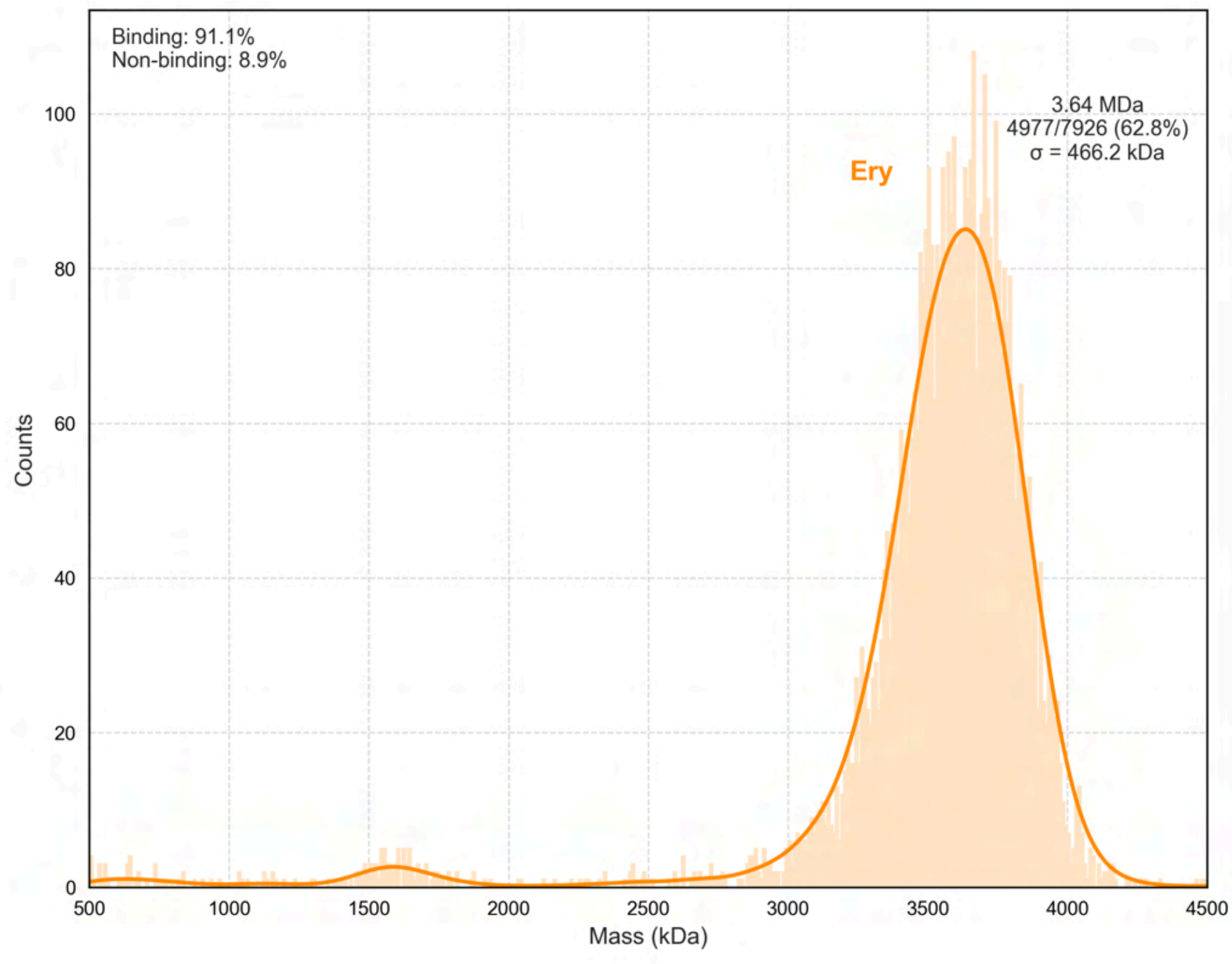

**Fig. S32 |** In-solution mass photometry characterization of diluted Ery (at ~2 nM concentration) indicating expected molecular weight ~3.6 MDa.

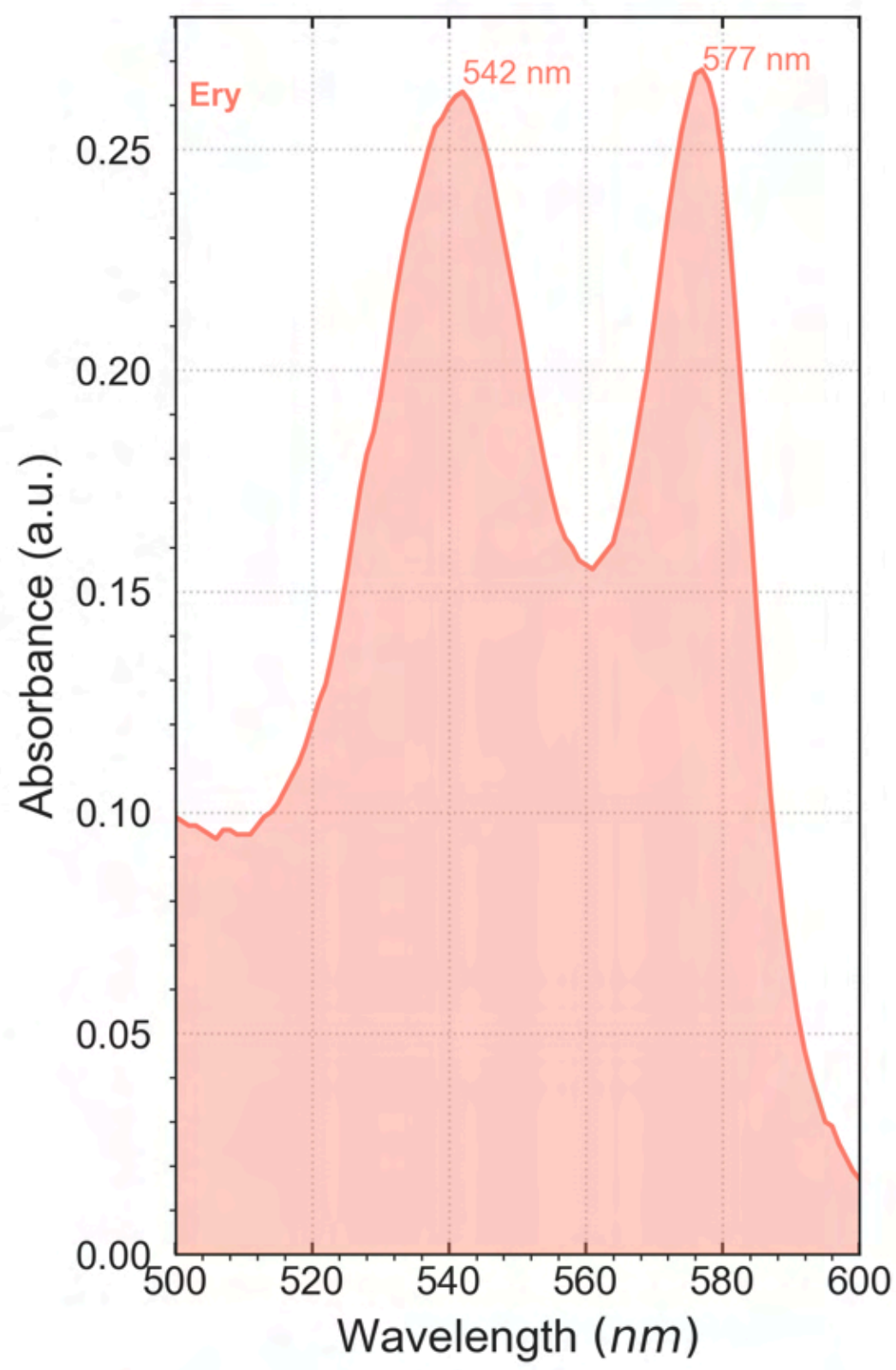

**Fig. S33 |** UV-Vis characterization of the Ery injected to XFEL beam shows the signature of oxy-hemoglobin state.

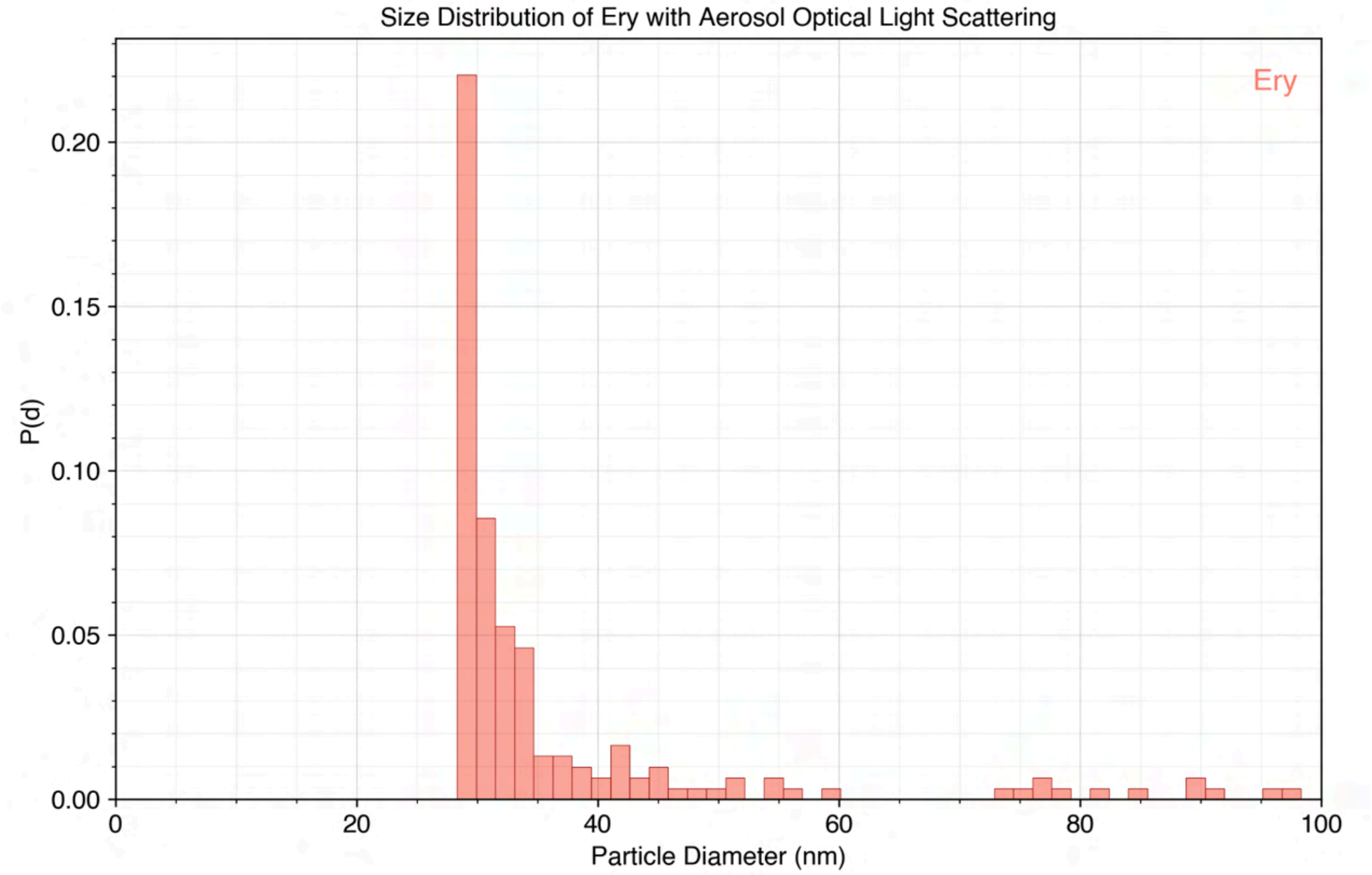

**Fig. S34 |** Size distribution of Ery based on aerosol optical laser light-scattering. Size distribution was estimated from the scattered intensities with polystyrene-spheres-calibration, as described elsewhere.

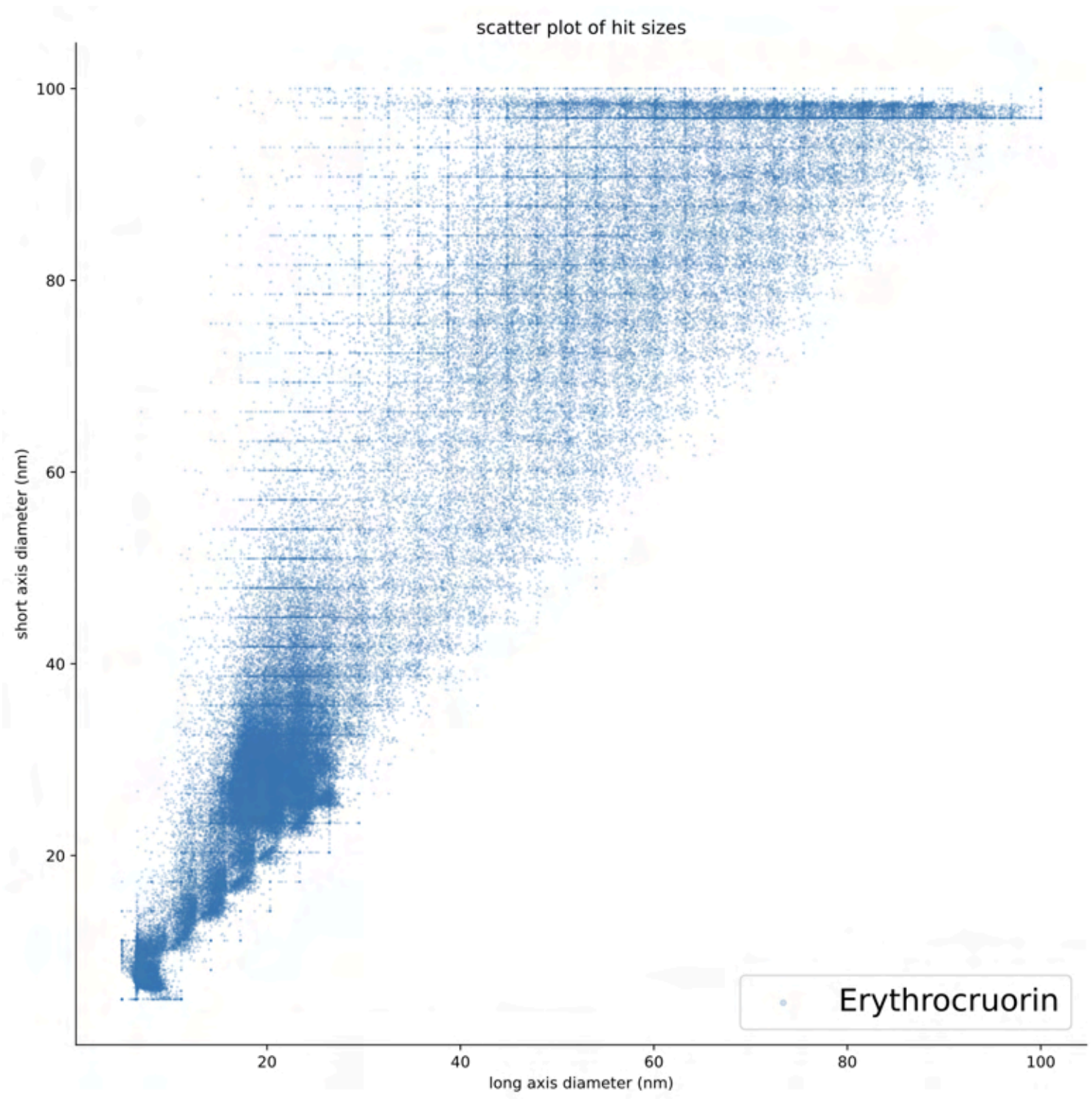

**Fig. S35 |** Size distribution from X-ray scattering based on ellipsoid fitting of all hits in Ery runs. Note that most of the hits of the right size are concentrated between 20 and 40 nm, as expected for intact Ery.

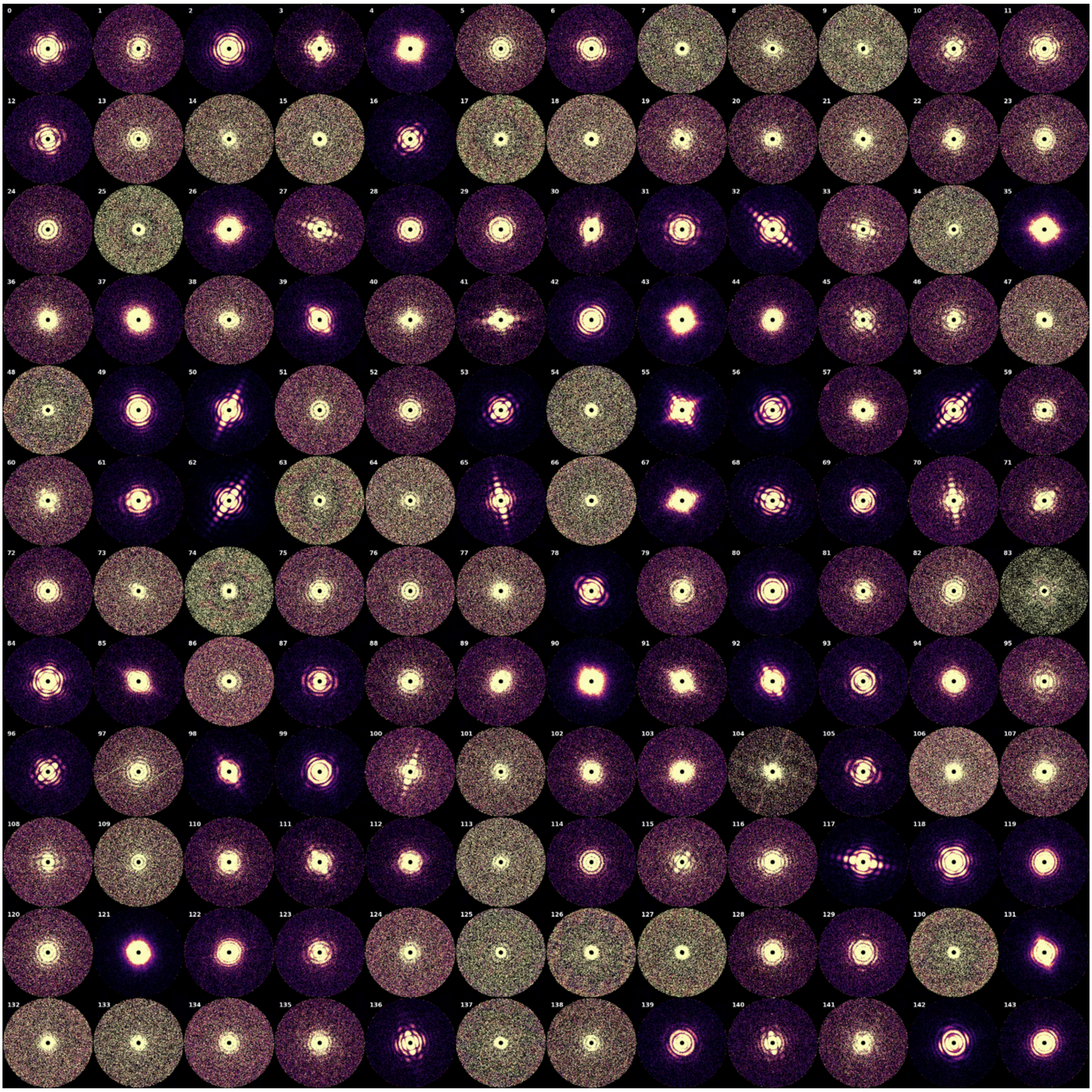

**Fig. S36 |** EMC based 2D class averages of hits in Ery runs, representing 144 classes from the 44,000 diffraction patterns identified as hits, covering several different orientations and also debris in the reciprocal space. Class average number on the top left corner. 2D phase retrieved single-protein structure in real space in Fig. S21.

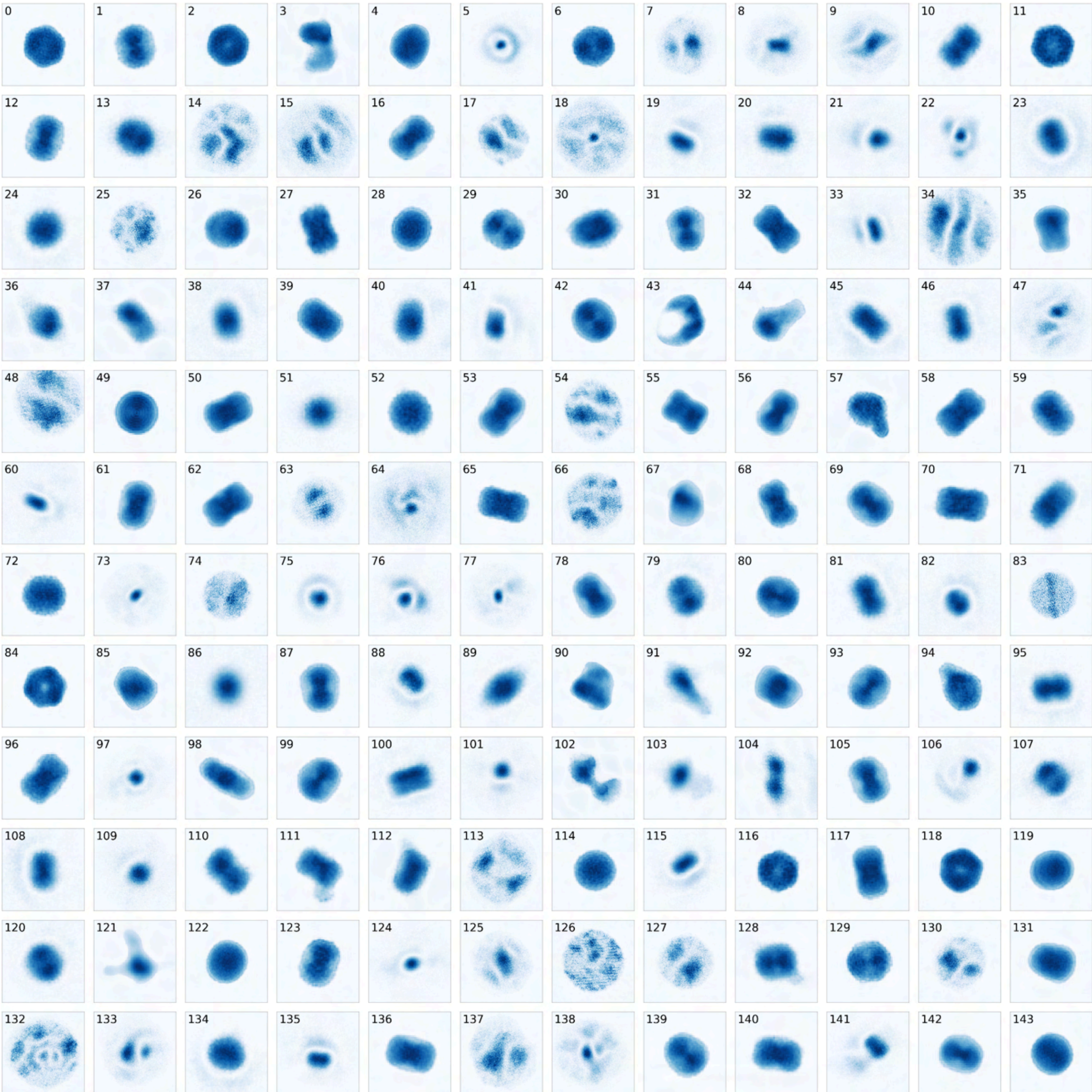

**Fig. S37 |** Collage of phased 2D class-averages from the 144 2D reciprocal space class averages shown in Fig. S35. Pixel size 0.72 nm, with an edge resolution of 1.5 nm. See corresponding PRTF plot in S22.

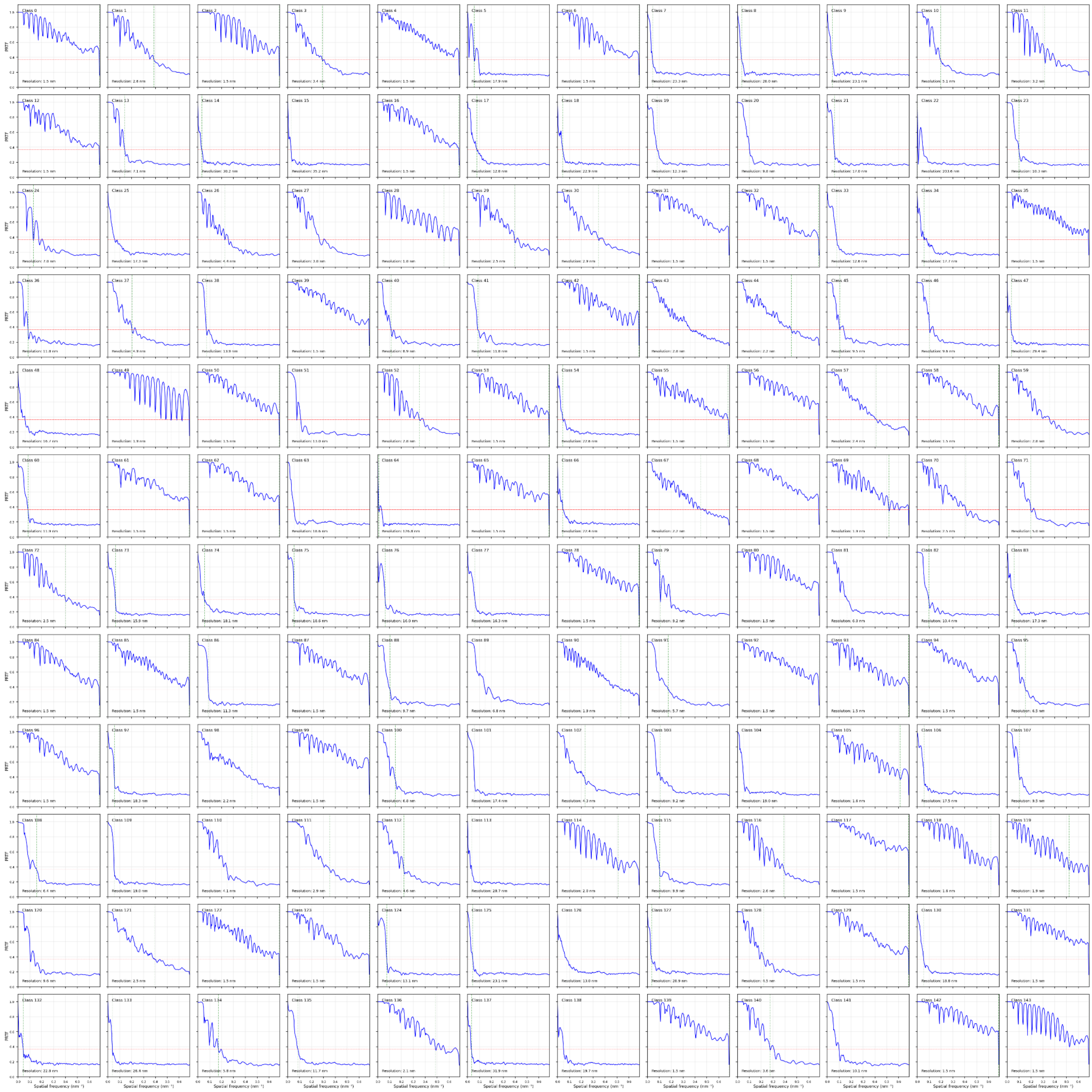

**Fig. S38 |** PRTF plot of all 144 phase retrieved classes as calculated in the same way as for Fig. S19. Edge resolution of 15 Å.

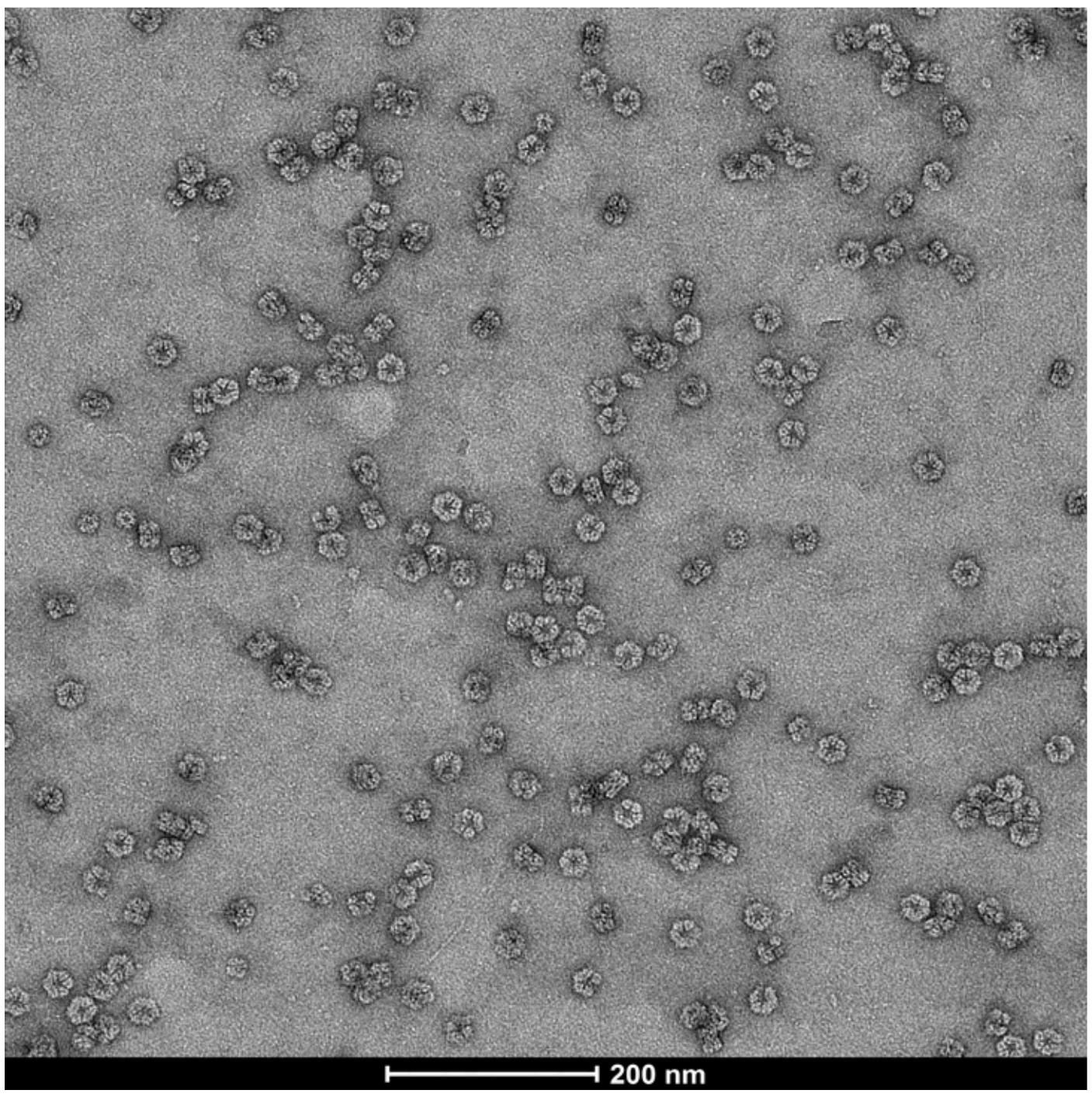

**Fig S39 |** Negative-stain transmission electron microscopy image of Ery. The scale bar is 200 nm.

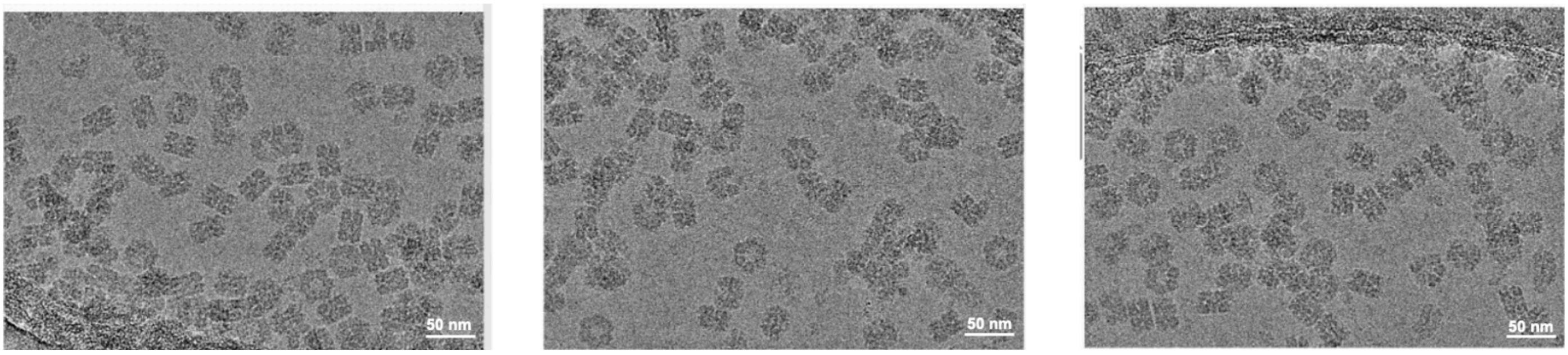

**Fig. S40 |** Representative Cryo-EM micrographs of Ery in ammonium acetate buffer as in the XFEL-SPI experiment. The scale bar is 50 nm.

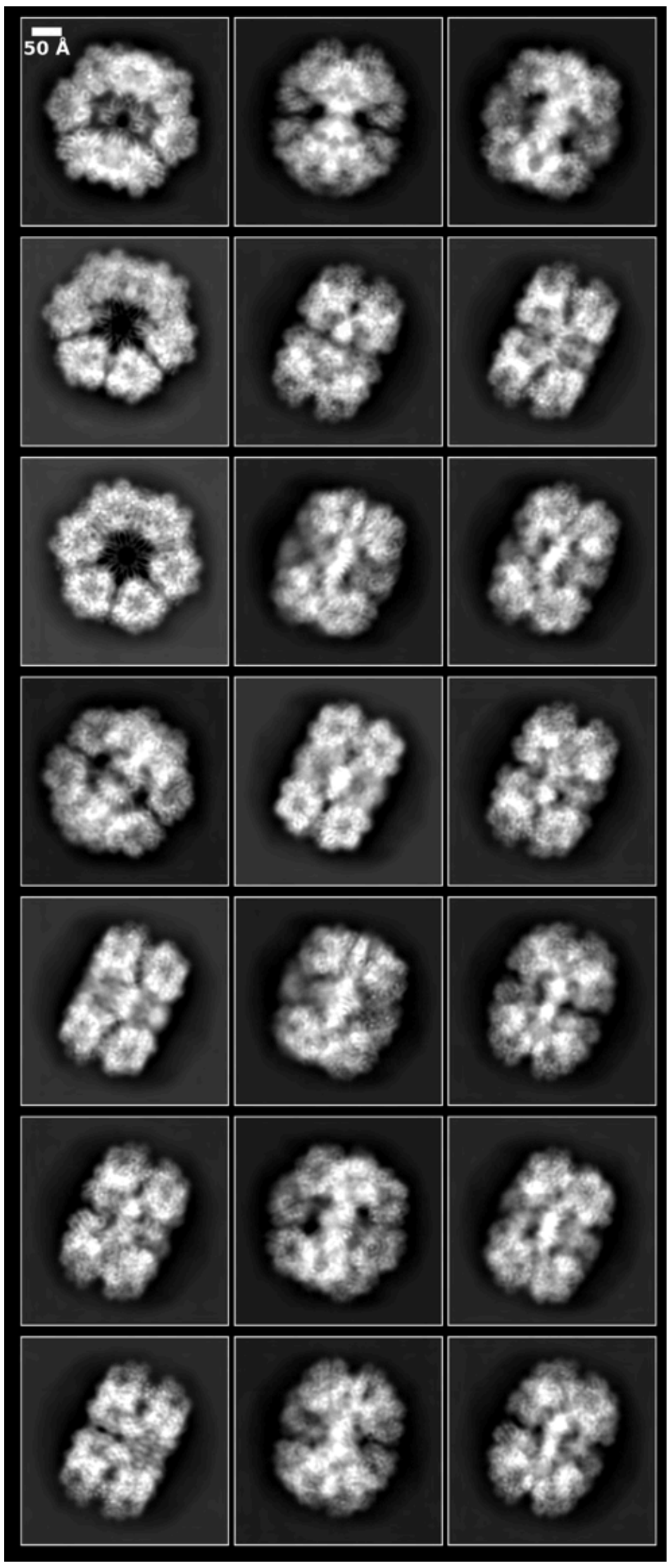

**Fig. S41 |** Representative cryoEM-SPI 2D class-averages of Ery in ammonium acetate buffer used for XFEL-SPI measurements from about 380k particles. Further 2D classifications and selections were done before 3D reconstruction. Scale bar: 50 Å

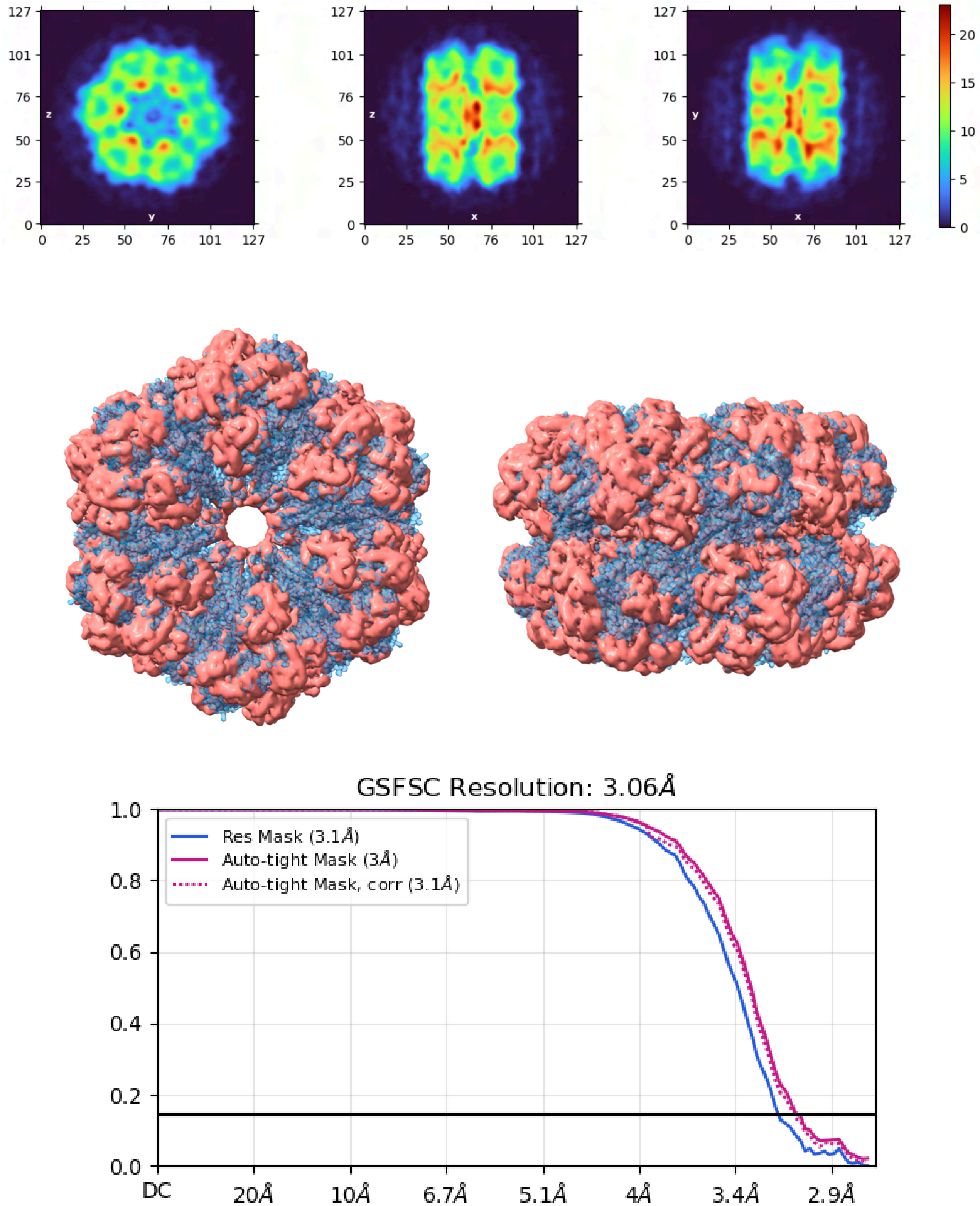

**Fig. S42** | Cryo-EM 3D reconstruction of Ery in the ammonium acetate buffer used for electrospray injection. **Top:** Initial *ab initio* reconstruction of Ery **Middle:** CryoEM reconstruction with D6 symmetry constraints (salmon) and the reference molmap (blue-transparent) from 2GTL PDB overlaid for comparison. **Bottom:** FSC resolution indicating 3.06 Å resolution.

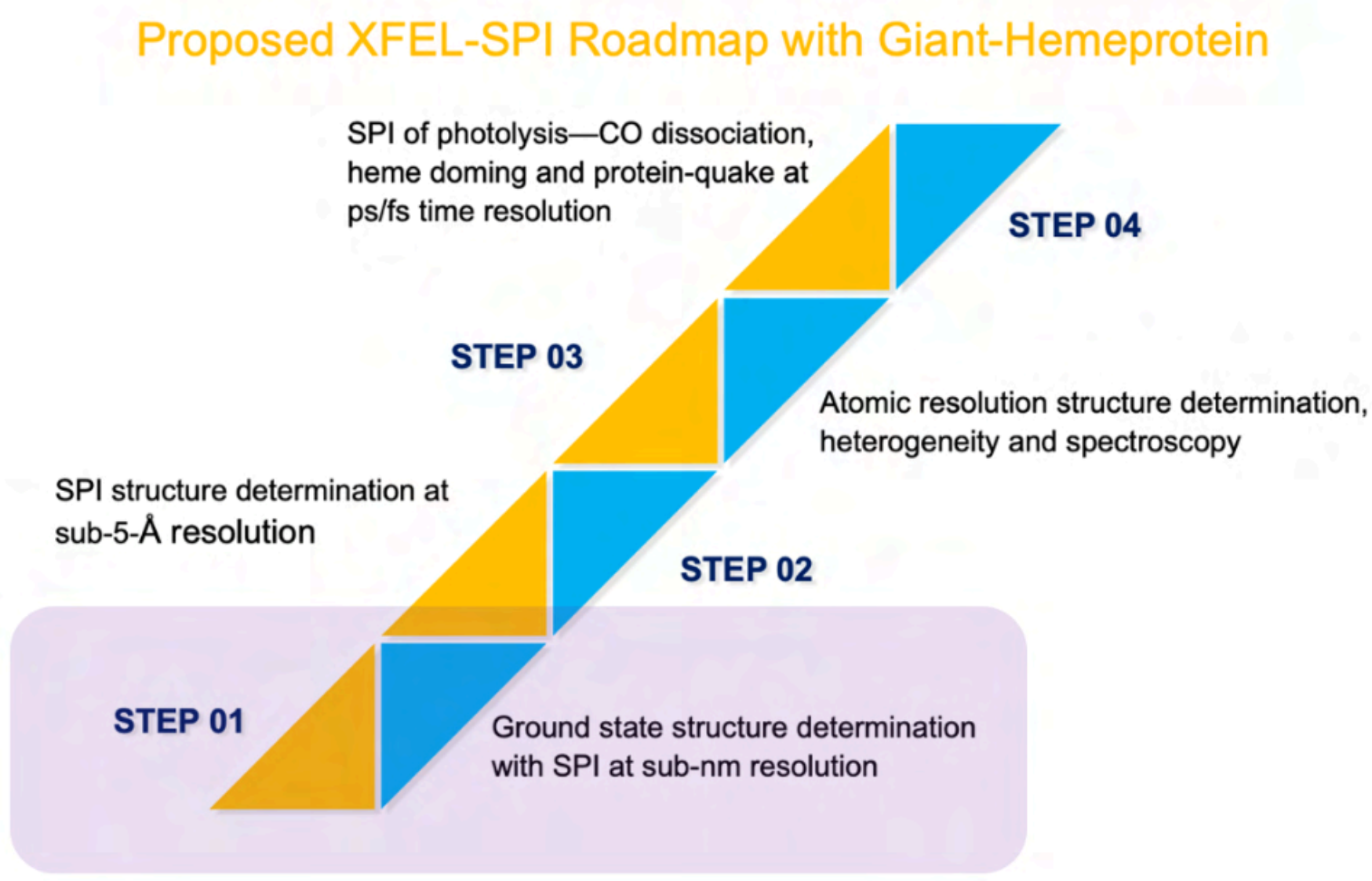

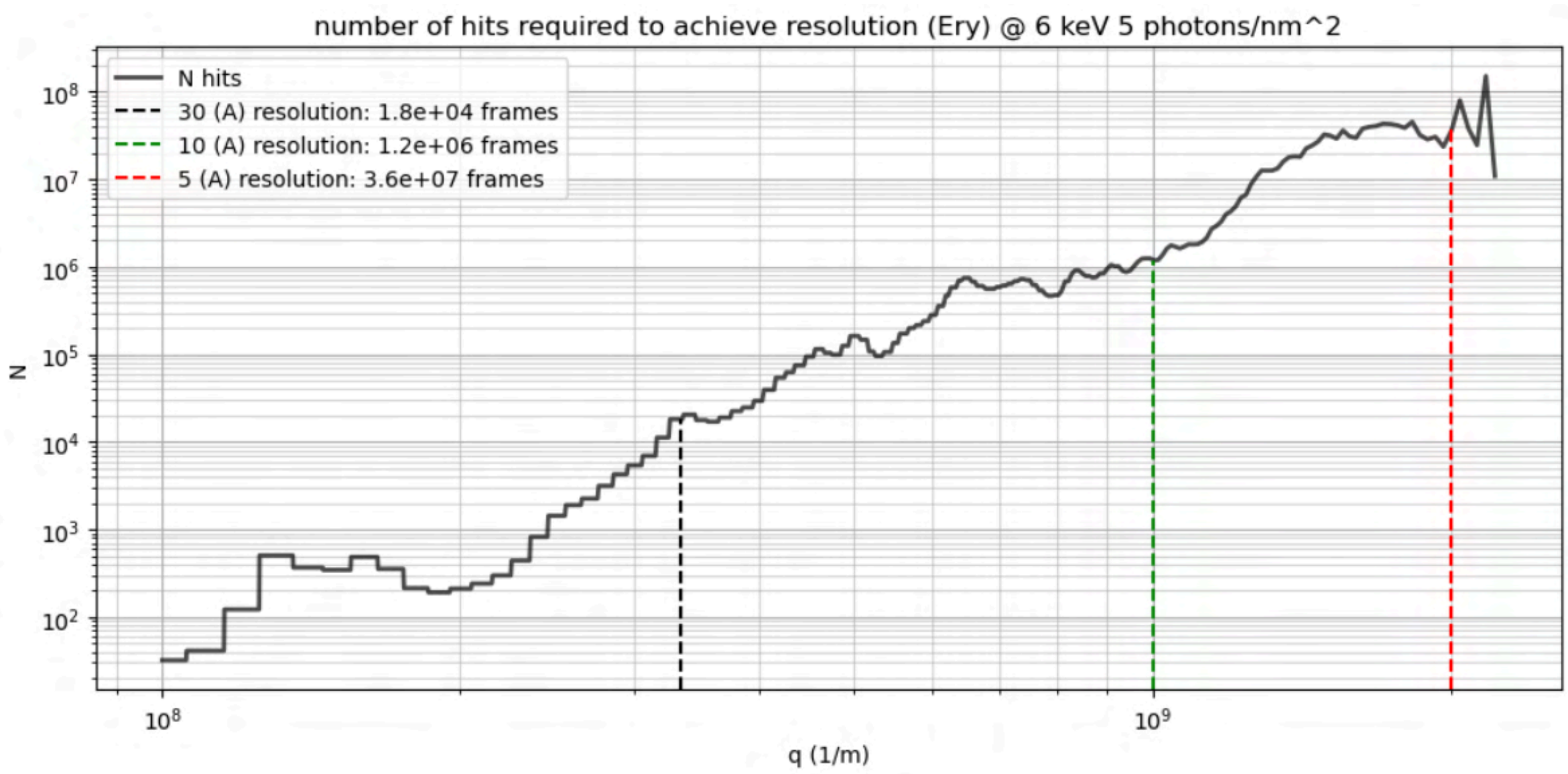

**Fig. S43 | Top:** Proposed roadmap towards high-resolution time-resolved ultrafast single-particle imaging with the giant-hemeprotein Ery (top).[1] And achieving high-resolution (near-atomic and atomic) requires three orders of magnitude more diffraction patterns from the current tens of thousand diffraction patterns we report here. **Bottom:** Plot indicating the number of hits required with the current SNR, and for even reaching sub-5-Å resolution (second step in the road-map) we already need a few million hits at least—this demands improved sample-delivery and high hit-rate. We are still at step 1 (highlighted) in the roadmap towards reaching sub-nm resolution.

## Observation of diffraction from intact Ery in a preparatory experiment at SQS with soft XFEL

### Experimental set-up

The preliminary screening experiment was performed at SQS endstation at 1.2 keV with a micron focus beam with an average pulse energy of 6.5 mJ. The focal spot was optimized with a mini-ion-ToF near the interaction region (based on the ionization of Argon), and the diffraction images were recorded with a DSSC detector[2] at a distance of 536 mm with a vertical-center-edge resolution of ~4.1 nm. Due to shadowing from the DSSC flange, corner-edge resolution was inaccessible. The sample was delivered with the similar electrospray-aerosol-injection set-up as described in the methods section which was used at SPB/SFX. Initially, when Ery was screened additional He supply was not used (hence the higher background). By the end of the SQS beamtime, the usage of additional He-supply after the skimmer was employed and we determined that it reduces background (see Fig. S44), which was then later implemented at SPB/SFX to image single-proteins in the hard x-ray regime (Extended Data Fig 9).

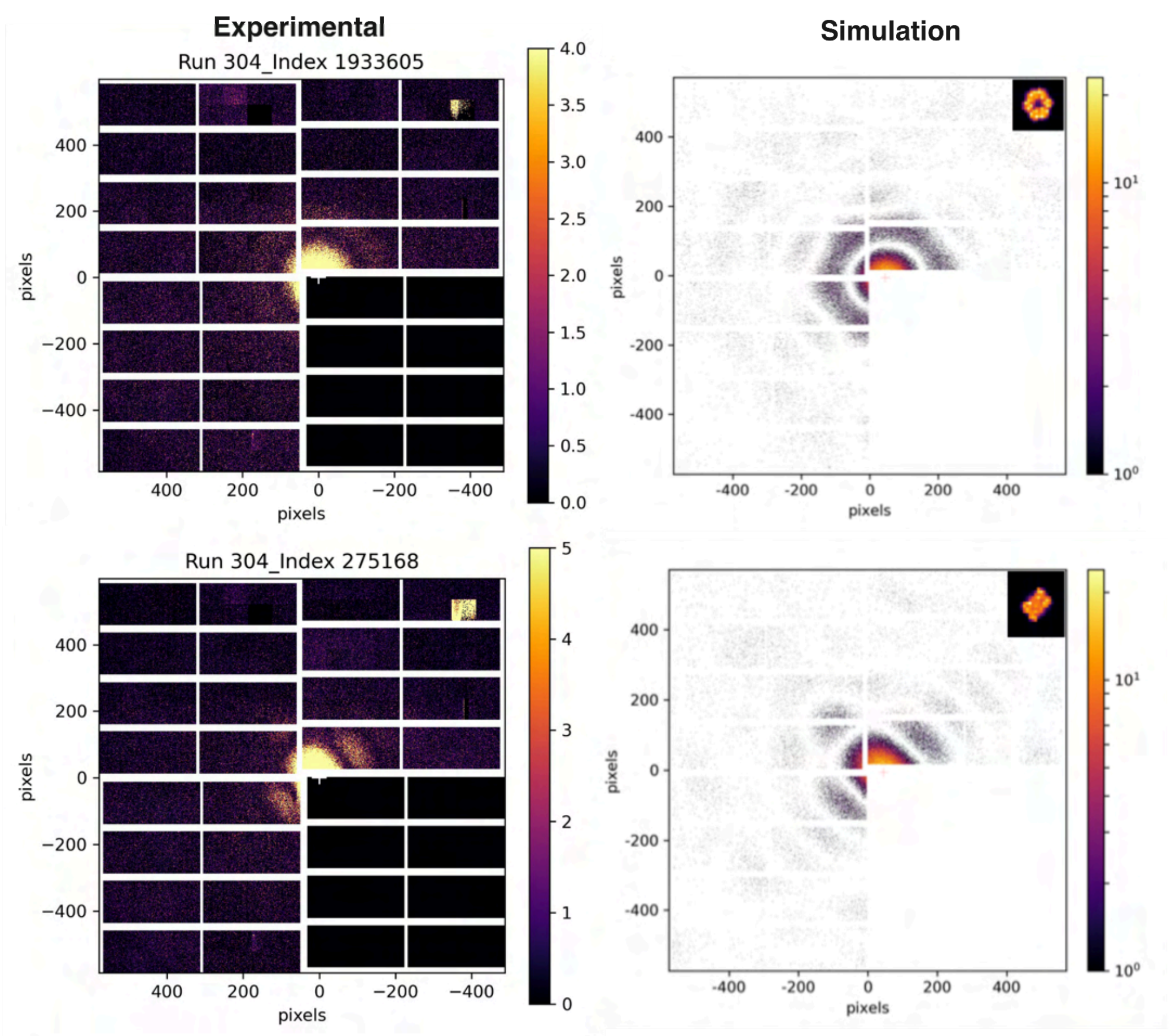

**Fig. S44 |** Single-protein diffraction patterns of Ery from screening at SQS in p3004 (02Nov2023) with DSSC detector at 1.2 keV without additional-He-supply (before using He to reduce background as in Fig. S44). The lower right quadrant in the DSSC detector was non-functional, hence was turned off, and only the rest of the three quadrants were used. These single-shot diffraction patterns of Ery represent the first diffraction snapshots recorded of a protein system with photoactive moieties exhibiting light-induced ps/fs bio-functional dynamics. The top row represents a diffraction showing the six-fold symmetry of Ery (face-on hit) and the bottom row shows a hit along the side-edge of Ery. The left column is experimental and the right column shows closely matching simulations and projected real space image of Ery as inset.

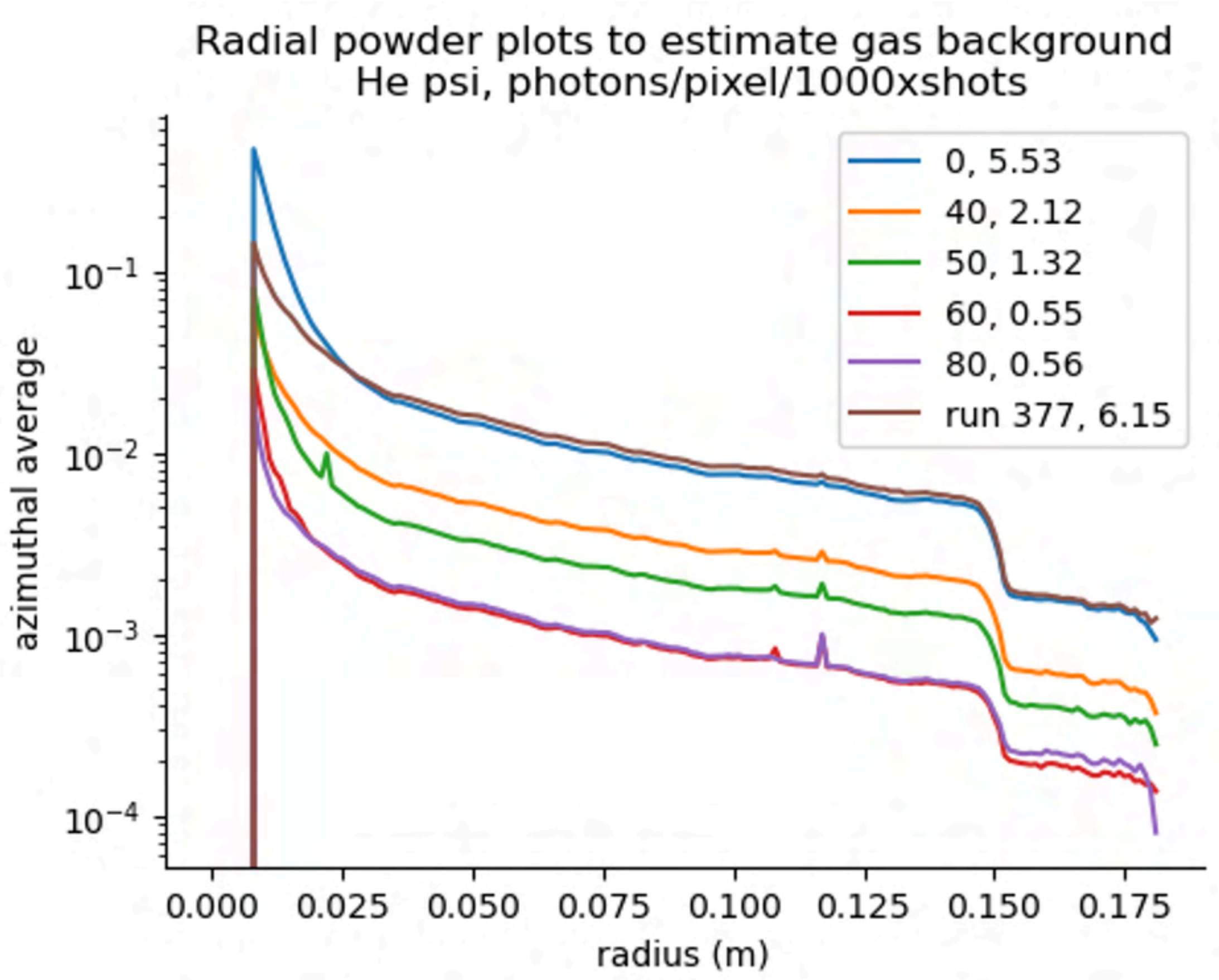

**Fig. S45 | Gas background scattering signal showing 90% background reduction with the supply of He after the skimmers** (while running DNA-origami[2] samples (manuscript in preparation P. L. Xavier, A. Morgan et al.) in the experiment p3004 at SQS in Nov 2023, and we used the same scheme at SPB/SFX in 7927 to image Ery. He supply ranged from 0 to 80 psi. In the legend, He used and photons/pixel/1000 shot is given next to the corresponding color-coded line. Run 377 was without additional He supply, with He-electrospray nozzle, which produced a similar background as in He = 0 psi.

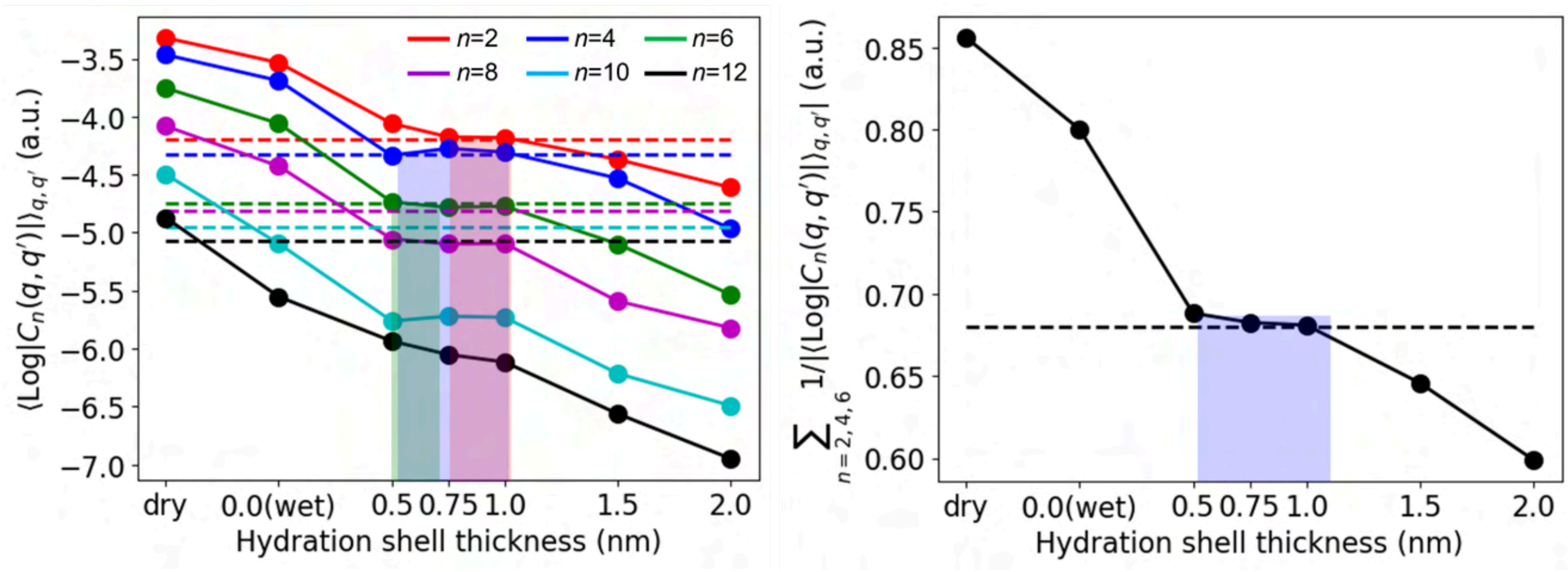

**Fig. S46 |** Hydration-shell estimation using angular cross-correlation analysis of diffraction patterns. (left) Circular markers represent the values of $\langle Log\left(\left|C_n(q,q')\right|\right)\rangle_{q,q'}$ determined for simulated Ery structures for different hydration states; different colors correspond to different harmonic orders $n$. The experimental values are indicated by dashed lines in the corresponding colors. The intersections regions between the simulated and experimental curves of the same colour—highlighted as shaded areas for the dominant orders $n$=2,4, and 6—provide an estimate of the hydration-shell thickness in the range of 0.5-1.0 nm. (right) Integrated hydration metric $\sum_{n=2,4,6} 1/\left|\langle Log\left(\left|C_n(q,q')\right|\right)\rangle_{q,q'}\right|$, where it indicates the estimated hydration-shell thickness is in the range of approximately 0.5-1 nm.

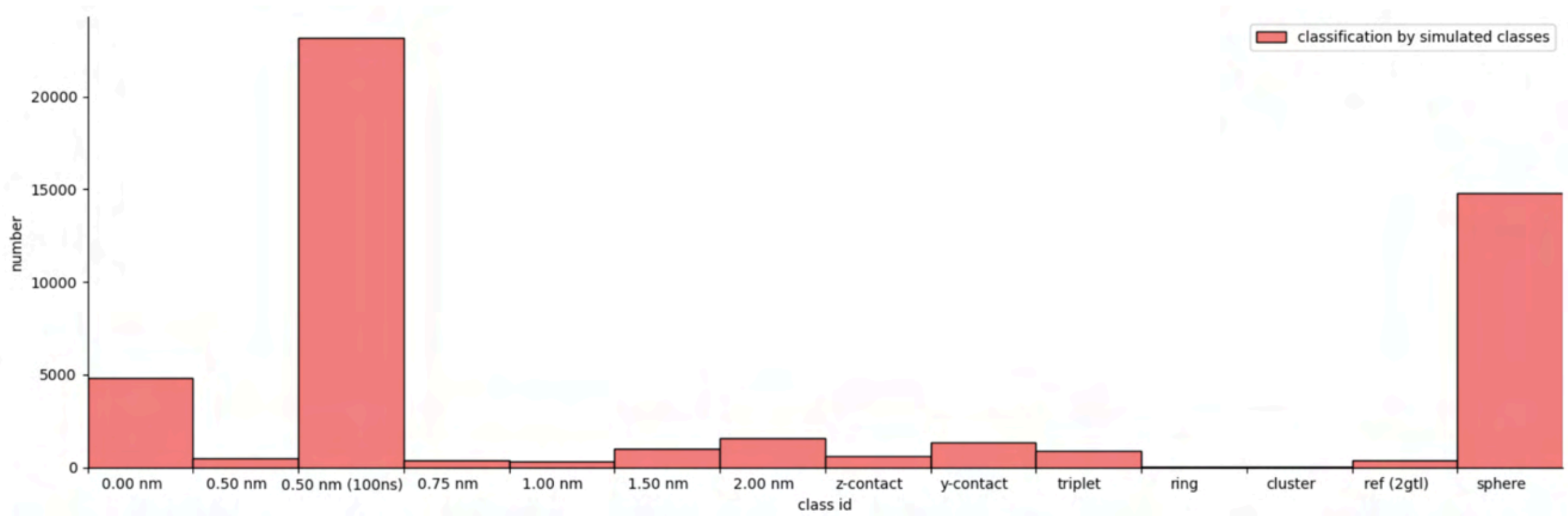

**Fig. S47 |** Molecular dynamics model-based EMC classification of diffraction patterns, indicate most diffraction patterns occupy the 0.5 nm hydration shell with 100 ns in vacuum.

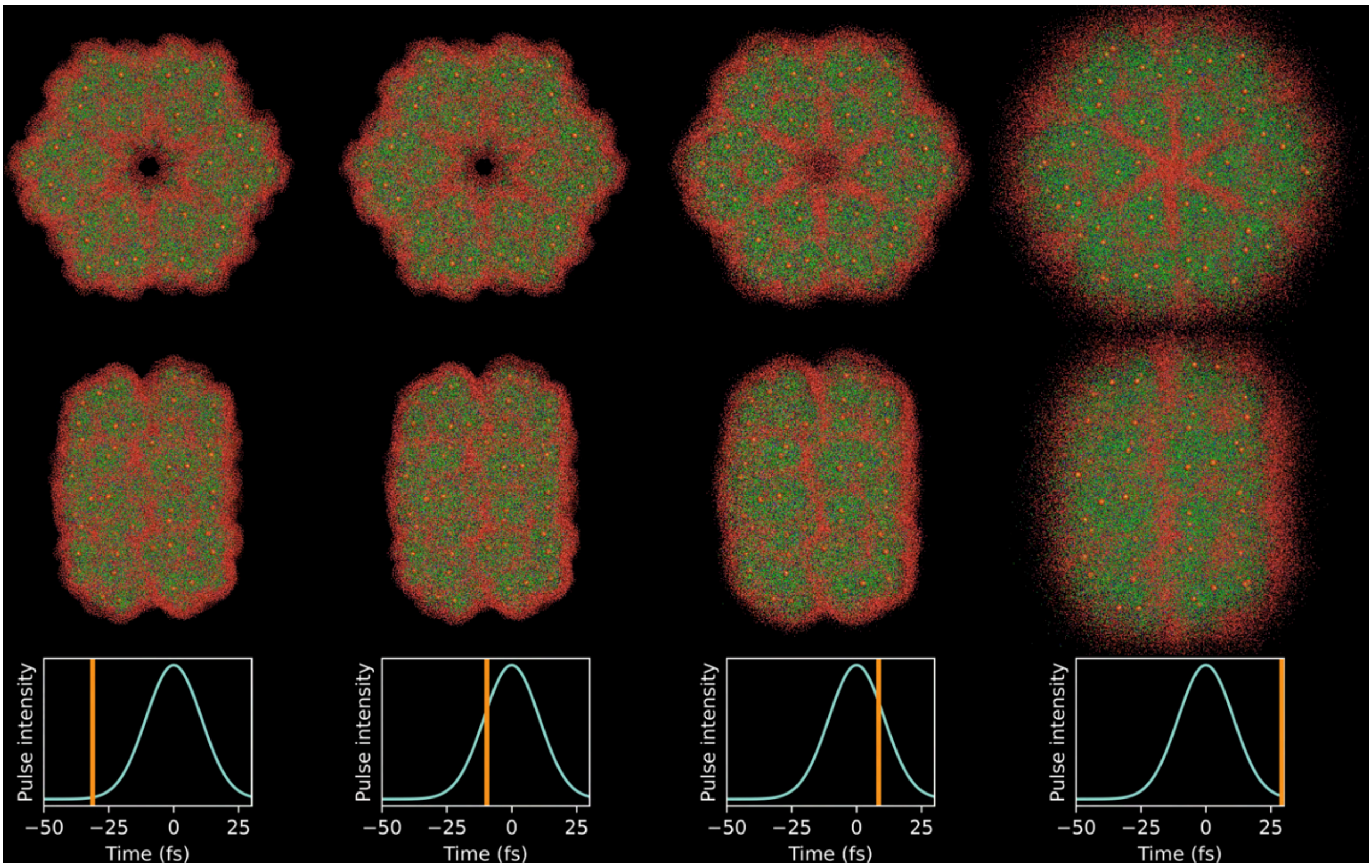

**Fig. S48 |** Snapshots of Ery solvated with 1 nm water shell during simulation under a 25 fs FDHM pulse, with a fluence of 8.3 x $10^{12}$ photons/$um^2$. Hydrogen atoms are not shown. The red spheres correspond to oxygen. The Fe ions are shown in orange with an exaggerated size.

**Fig. S49 |** For the animated-movie version of Fig S47 @ ery_explosion.mp4

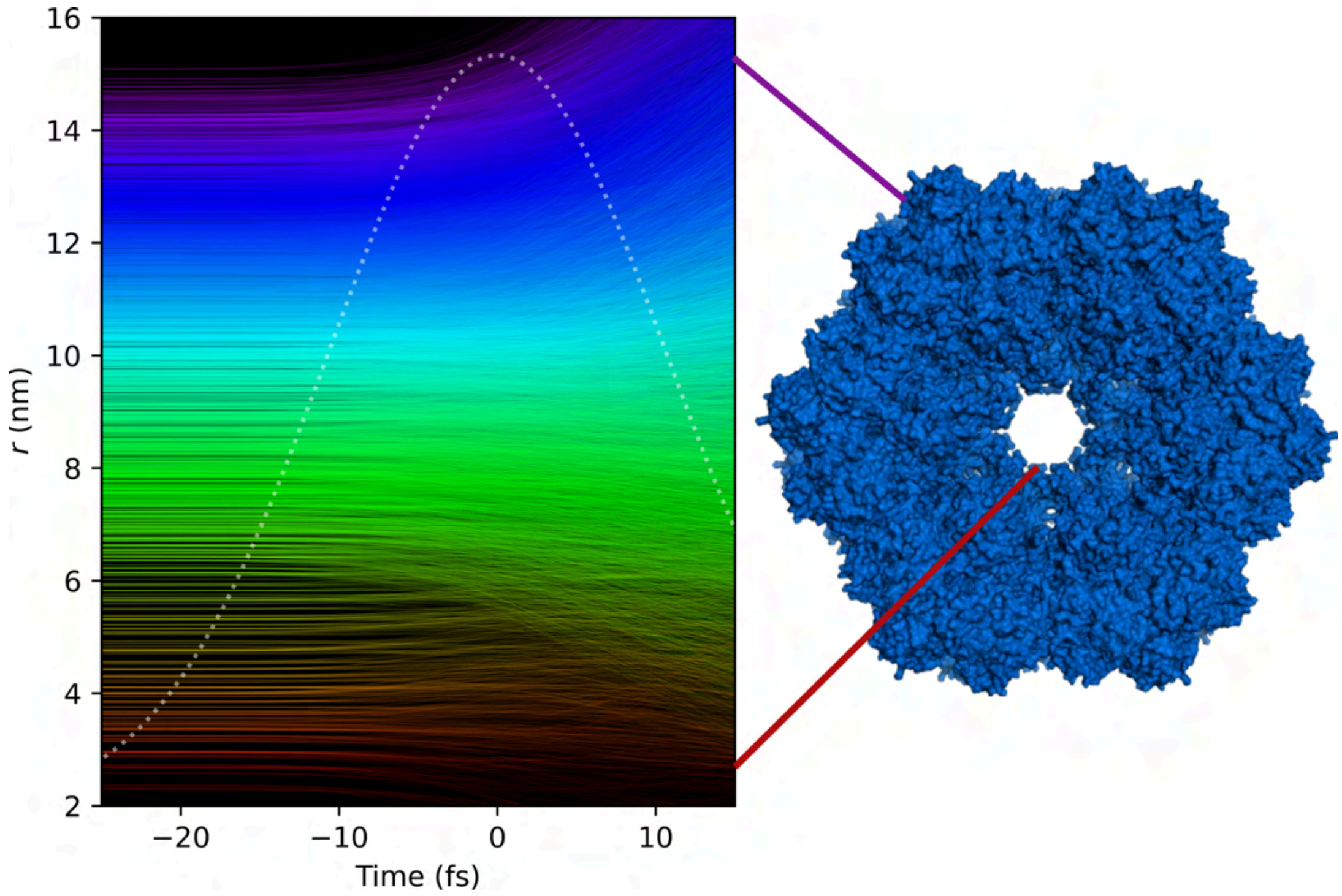

**Fig. S50 |** Radial movement of individual residues for ultrafast radiation damage simulation of Ery with 1 nm water shell. Each line traces the mean distance of an individual residue backbone from the centre of the structure. Lines are colored according to the distances in the initial, damage-free structure.

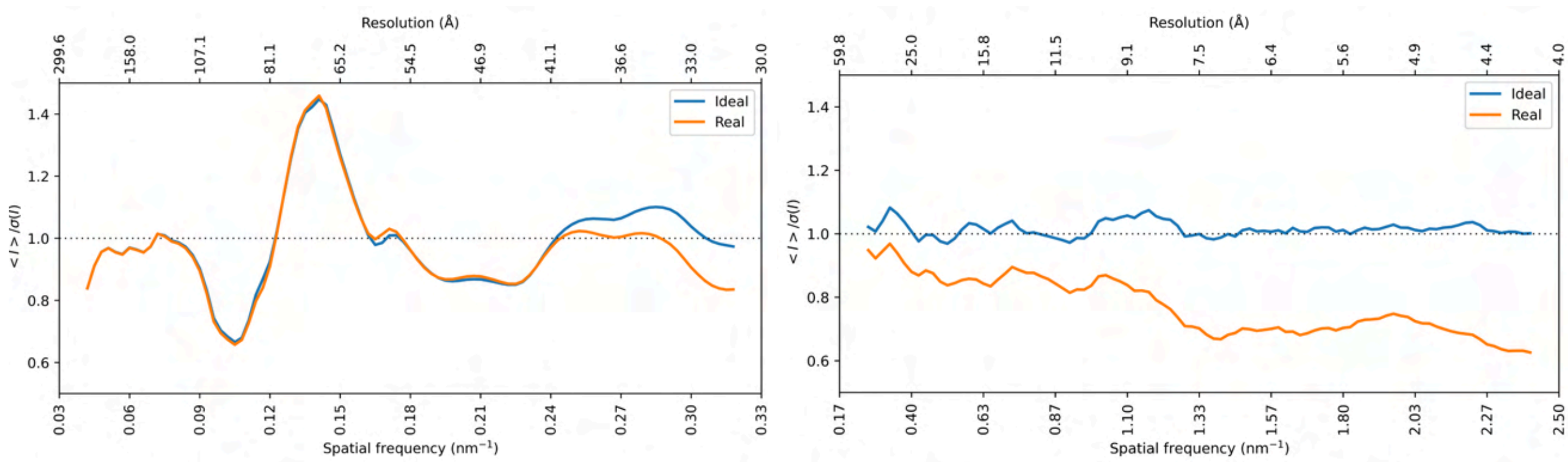

**Fig. S51 |** Simulated effect of radiation damage on 2D diffraction pattern speckle contrast for detector edge resolutions of 30 Å (left) and 4 Å (right), showing the 8-point moving average of <I>/σ(I) for 100 bins of pixel intensities with equal intervals in reciprocal space (spatial frequency). Sample orientation was randomized for each pattern. The blue curves (labelled "ideal") correspond to the diffraction intensities 1 FWHM before the temporal midpoint of the pulse and the orange curves (labelled "real") correspond to the pulse-integrated diffraction intensities.

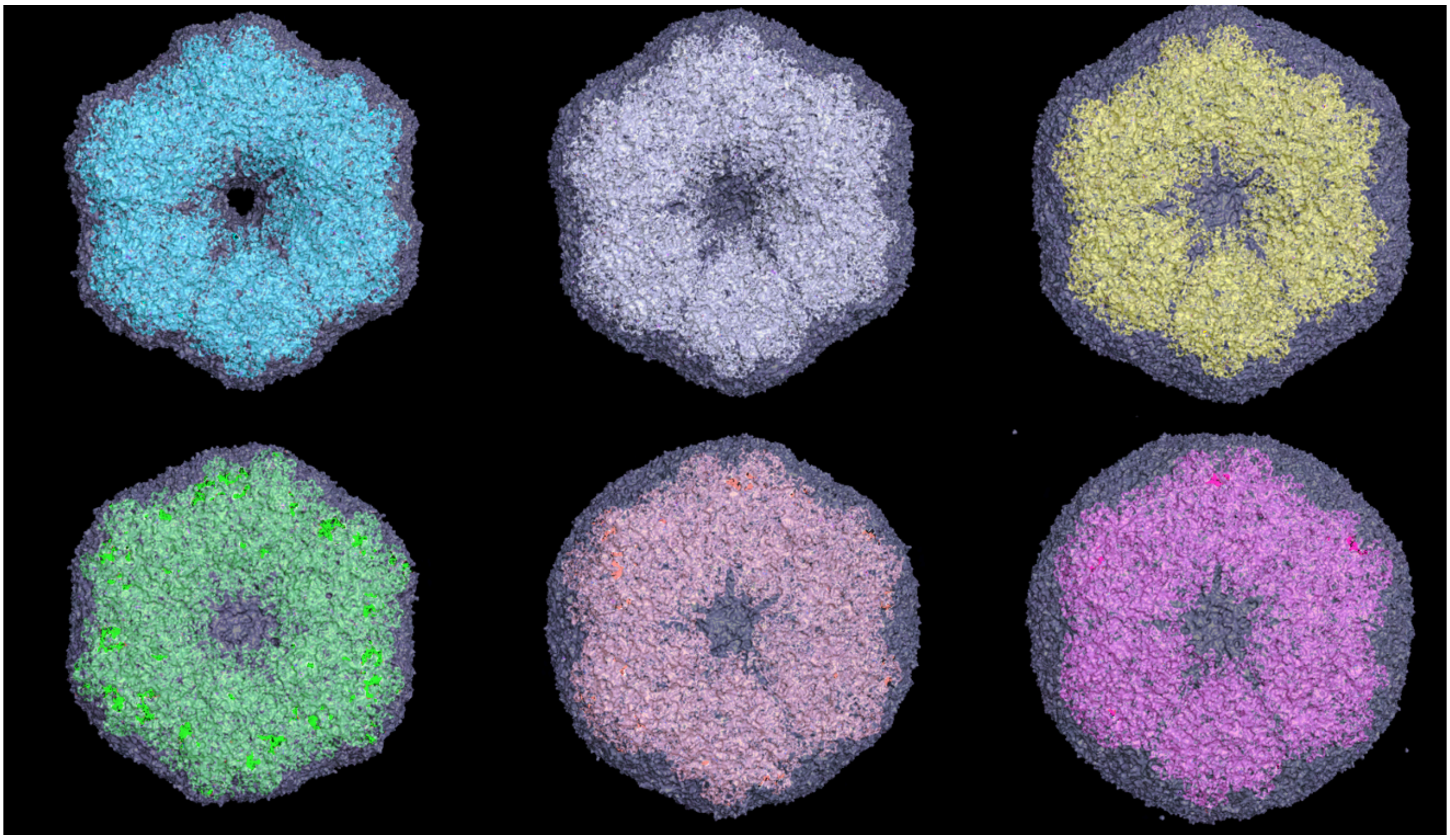

**Fig. S52 |** Cartoon representation of MD simulations with varying water-shell to avoid gas-phase compaction. Hydration layer column-wise: column 1: 1 nm, column 2: 1.5 nm, and column 3: 2 nm. Top row: t = 0 (before MD) and bottom row: after t = 3 ns MD simulations.

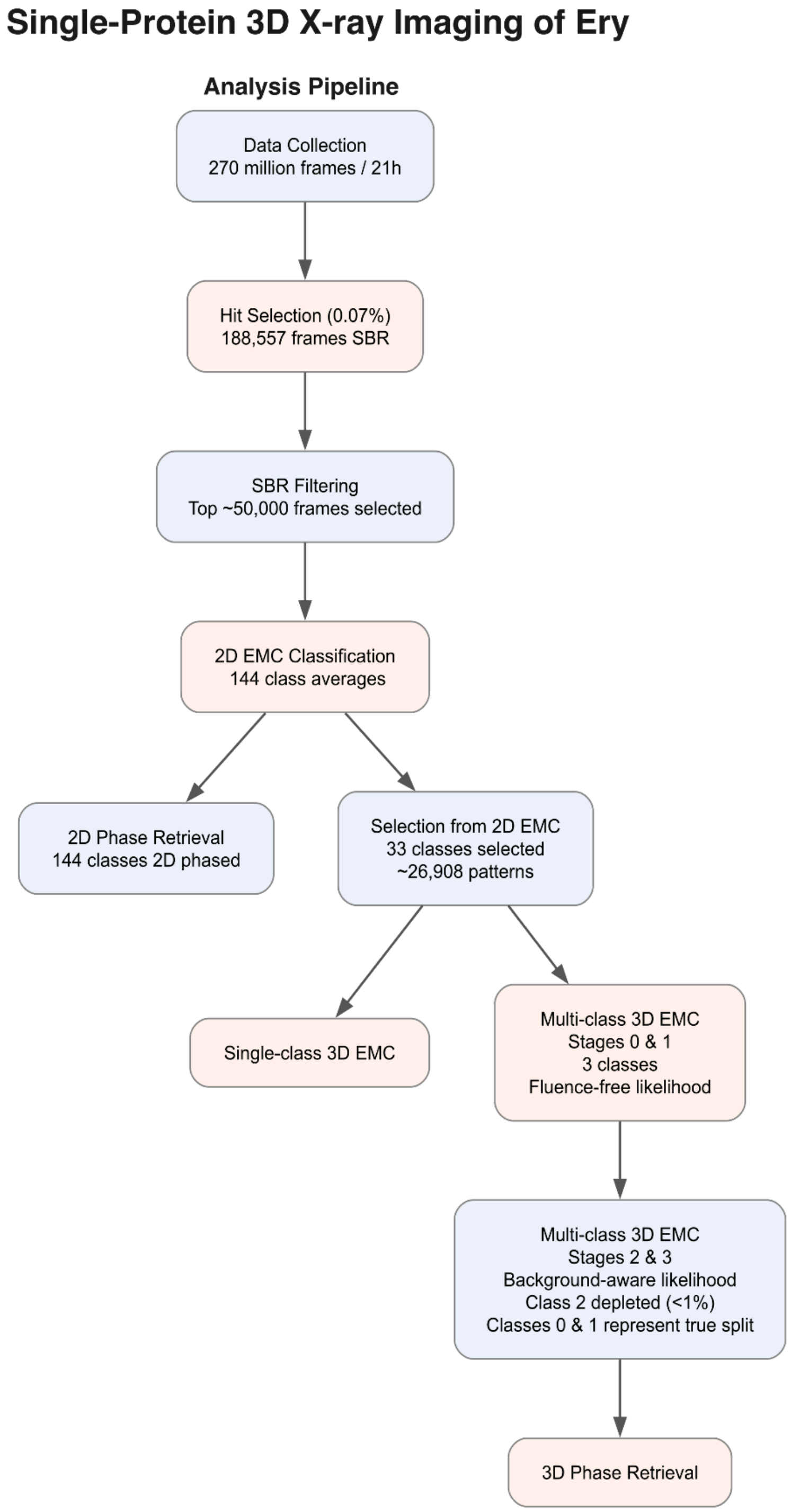

**Fig. S53 |** Flowchart summary of the data analysis pipeline of single-protein 3D X-ray imaging of Ery

**SI References**